\documentclass[fleqn,usenatbib]{mnras}

\usepackage{newtxtext,newtxmath}

\usepackage[T1]{fontenc}

\DeclareRobustCommand{\VAN}[3]{#2}
\let\VANthebibliography\thebibliography
\def\thebibliography{\DeclareRobustCommand{\VAN}[3]{##3}\VANthebibliography}

\usepackage{graphicx}	
\usepackage{amsmath}	
\usepackage{xcolor}
\usepackage{caption}
\usepackage{subcaption}

\title[Impact of spatial variability for LSST]{Impact of survey spatial variability on galaxy redshift distributions and the cosmological $3\times2$-point statistics for the Rubin Legacy Survey of Space and Time (LSST)}

\author[Q. Hang et al.]{
Qianjun Hang,$^{1}$\thanks{E-mail: e.hang@ucl.ac.uk}
Benjamin Joachimi,$^{1}$
Eric Charles,$^{2,3}$
John Franklin Crenshaw,$^{4,5}$
Patricia Larsen,$^{6}$ 
\newauthor
Alex I. Malz,$^{7}$
Sam Schmidt,$^{8}$ 
Ziang Yan,$^{9}$ 
Tianqing Zhang,$^{3,7,10}$ 
and the LSST Dark Energy Science 
\newauthor
Collaboration
\\
$^{1}$Department of Physics \& Astronomy, University College London, Gower Street, London WC1E 6BT, UK\\
$^{2}$Kavli Institute for Particle Astrophysics and Cosmology (KIPAC), Stanford University, Stanford, CA 94305, USA\\ 
$^{3}$SLAC National Accelerator Laboratory, 2575 Sand Hill Road, Menlo Park, CA 94025, USA\\ 
$^{4}$Department of Physics, University of Washington, Seattle, WA 98195, USA\\
$^{5}$DIRAC Institute, University of Washington, Seattle, WA 98195, USA\\
$^{6}$CPS Division, Argonne National Laboratory, 9700 S. Cass Ave., Lemont, IL 60439, USA\\
$^{7}$McWilliams Center for Cosmology, Department of Physics, Carnegie Mellon University, Pittsburgh, PA, USA\\
$^{8}$Department of Physics, University of California, One Shields Avenue, Davis, CA 95616, USA\\
$^{9}$Ruhr University Bochum, Faculty of Physics and Astronomy, Astronomical Institute (AIRUB), German Centre for Cosmological Lensing, 44780 Bochum, \\
Germany\\
$^{10}$Department of Physics and Astronomy and PITT PACC, University of Pittsburgh, Pittsburgh, PA 15260, USA\\
}

\date{Accepted XXX. Received YYY; in original form ZZZ}

\pubyear{2024}

\begin{document}
\label{firstpage}
\pagerange{\pageref{firstpage}--\pageref{lastpage}}
\maketitle

\begin{abstract}
We investigate the impact of spatial survey non-uniformity on the galaxy redshift distributions for forthcoming data releases of the Rubin Observatory Legacy Survey of Space and Time (LSST). Specifically, we construct a mock photometry dataset degraded by the Rubin OpSim observing conditions, and estimate photometric redshifts of the sample using a template-fitting photo-$z$ estimator, BPZ, and a machine learning method, FlexZBoost. We select the Gold sample, defined as $i<25.3$ for 10 year LSST data, with an adjusted magnitude cut for each year and divide it into five tomographic redshift bins for the weak lensing lens and source samples. We quantify the change in the number of objects, mean redshift, and width of each tomographic bin as a function of the coadd $i$-band depth for 1-year (Y1), 3-year (Y3), and 5-year (Y5) data. In particular, Y3 and Y5 have large non-uniformity due to the rolling cadence of LSST, hence provide a worst-case scenario of the impact from non-uniformity. We find that these quantities typically increase with depth, and the variation can be $10-40\%$ at extreme depth values. Using Y3 as an example, we propagate the variable depth effect to the weak lensing $3\times2$pt analysis, and assess the impact on cosmological parameters via a Fisher forecast. We find that galaxy clustering is most susceptible to variable depth, and non-uniformity needs to be mitigated below 3\% to recover unbiased cosmological constraints. There is little impact on galaxy-shear and shear-shear power spectra, given the expected LSST Y3 noise.
\end{abstract}



\section{Introduction}

Observational cosmology enters the era of high-precision measurements. For example, weak gravitational lensing, which probes the small distortion of distant galaxy shapes due to the gravity of foreground large-scale structures, is particularly sensitive to the clustering parameter $S_8=\sigma_8\sqrt{\Omega_{\rm m}/0.3}$. Current weak lensing surveys have measured this parameter to be $S_8=0.759^{+0.024}_{-0.021}$ by the Kilo-Degree Survey \citep[KiDS-1000,][]{2021A&A...645A.104A}, $S_8=0.759^{+0.025}_{-0.023}$ by the Dark Energy Survey \citep[DES-Y3,][]{2022PhRvD.105b3514A}, and $S_8=0.760^{+0.031}_{-0.034}$ ($S_8=0.776^{+0.032}_{-0.033}$) using the shear power spectra (two-point correlation function) by the Hyper Suprime-Cam \citep[HSC-Y3,][]{2023arXiv230400702L, 2023arXiv230400701D}. The constraints are comparable to that measured by \cite{2020A&A...641A...6P} from the primary cosmic microwave background (CMB), $S_8=0.830\pm0.013$, and the recent result from CMB lensing \citep{2023arXiv230405203M}, $S_8=0.840\pm0.028$, but are interestingly lower by $2-3\sigma$. 
The uncertainties of these measurements are already dominated by systematic errors - without a careful treatment of various systematic effects, the cosmological results can be biased up to a few sigma \citep[e.g.][]{2022MNRAS.511.2665R}.
The forthcoming Stage IV surveys such as the Rubin Observatory Legacy Survey of Space and Time (LSST) will achieve a combined Figure of Merit ten times as much as the Stage III experiments as mentioned above \citep{thelsstdarkenergysciencecollaboration2021lsst}. 
While the high statistical power enables pinning down the nature of such tensions, systematic error needs to be controlled down to sub-percent level to ensure that our results are not biased.

One major systematic uncertainties come from survey non-uniformity. Galaxy samples detected at different survey depth, for example, will have different flux errors and number of faint objects near the detection limit. This could propagate down to systematic errors in redshift distribution and number density fluctuation.
The majority of the LSST footprint will follow the wide-fast-deep (WFD) observing strategy, which means that a large survey region will be covered before building up the survey depth.
At early stages of the survey, fluctuations in observing conditions, such as sky brightness, seeing, and air mass, are expected to be significant across the footprint. These can change the per-visit $5\sigma$ limiting magnitude, $m_5$, leading to depth non-uniformity in the early LSST data \citep{2019ApJ...873..111I}.
The survey strategy later on could also affect uniformity. LSST will adopt a `rolling cadence', which means that during a fixed period, more frequent revisits will be assigned to a particular area of the sky, whereas the rest of the regions are deprioritized by up to 25\% of the baseline observing time. The high- and low-priority regions continue to swap, such that the full footprint is covered with the same exposure time after ten years. 
This can result in different limiting magnitudes across the sky at intermediate stages of rolling.
This strategy greatly advances LSST's potential for time domain science for e.g. denser sampling in light curves. However, it also poses challenges to the analysis of large-scale structure (LSS) probes, which normally prefers a uniform coverage.

Changes in $m_5$ can change the detected sample of galaxies and its photometric redshifts in two ways. Firstly, a larger $m_5$ means that fainter, higher redshift galaxies will pass the detection limit. 
This increases the sample size, and could shift the ensemble mean redshift higher. 
These faint galaxies also contain large photometric noise, resulting in larger scatter with respect to the true redshift, hence broadening the redshift distribution. 
Secondly, at fixed magnitude, the signal-to-noise is larger given a larger $m_5$. This means that, contrary to the previous effect, the scatter in spec-z vs photo-$z$ will be reduced due to the reduced noise.
These effect has been studied previously in a similar context.
The density fluctuation is quantified in \cite{2016ApJ...829...50A} via $1+\delta_{\rm o}=(1+\delta_{\rm t})(1+\delta_{\rm OS})$, where $\delta_{\rm o}$ is the observed density contrast, $\delta_{\rm t}$ is the true density, and $\delta_{\rm OS}$ is the fluctuation in the observing condition.
The effects on photo-$z$ have been investigated in \cite{2018AJ....155....1G} in the context of LSST. They showed that the photo-$z$ quality can change significantly with respect to different observing conditions, although they did not consider tomographic binning. 
\cite{2020A&A...634A.104H,2021A&A...646A.129J} also quantified the effects for KiDS-1000 data, where the depth varies significantly between different pointings. They showed that by varying the $r$-band limiting magnitude, a significant amount of high redshift objects can be included in the sample, such that the mean number density can double between the deepest and shallowest pointings, and the average redshift for a tomographic bin can shift by as much as $\Delta \langle z\rangle \sim 0.2$. 
Understanding these effects are important, because weak lensing is particularly sensitive to the mean redshift of the lens and source galaxies. \cite{2020A&A...634A.104H} demonstrated that this effect is similar to a spatially varying multiplicative bias, and for cosmic shear analysis in configuration space, constraints in the $\Omega_{\rm m}-\sigma_8$ plane can shift up to $\sim 1\sigma$ for a KiDS-like survey with the same area as LSST.
\cite{2023JCAP...07..044B} also derived an analytic expression for anisotropic redshift distributions for galaxy and lensing two-point statistics in Fourier space. They showed that, assuming a spatial variation of scale $\ell_{z}$, the effects are at percent and sub-percent level for the current and forthcoming galaxy surveys, and converge to the uniform case at $\ell \gg \ell_{z}$.

In this paper, we investigate how survey non-uniformity can affect the redshift distribution of tomographic bins for LSST 1-year, 3-year, and 5-year observation (hereafter Y1, Y3, and Y5 respectively).
The LSST Dark Energy Science Collaboration (DESC) Science Requirements Document \citep[][hereafter DESC SRD]{thelsstdarkenergysciencecollaboration2021lsst} states that the photometric redshifts needs to achieve a precision of $\langle \Delta z \rangle = 0.002(1+z)$ ($0.001(1+z)$) for Y1 (Y10) weak lensing analysis, and $\langle \Delta z \rangle = 0.005(1+z)$ ($0.003(1+z)$) for Y1 (Y10) large-scale structure analysis. 
Here, using these numbers as a bench mark, we quantify changes in the mean redshift ($\langle z \rangle$) and width ($\sigma_z$) of \textit{tomographic bins}, as depth varies.
\footnote{Notice that the DESC SRD also provides requirements on the photometric redshift scatter of the full, unbinned sample, $\sigma_{\Delta z}$. For weak lensing, this is $\sigma_{\Delta z}=0.006(1+z)$ ($0.003(1+z)$) for Y1 (Y10); for large-scale structure analysis, this is $\sigma_{\Delta z}=0.1(1+z)$ ($0.03(1+z)$) for Y1 (Y10). Because we do not try to optimize the photometric redshift estimation in this paper, we do not compare our results with the DESC SRD $\sigma_{\Delta z}$ values.}
We use the up-to-date LSST observing strategy and the simulated 10-year observing conditions for Rubin Observatory \citep[OpSim,][]{2016SPIE.9911E..25R, 2016SPIE.9910E..13D} to quantify the survey non-uniformity, and generate a mock catalogue of true galaxy magnitude in $ugrizy$, redshift, and ellipticity based on the Roman-Rubin (DiffSky) simulations \citep{Troxel_2023}. 
The degradation of photometry and photo-$z$ estimation relies on the public software, Redshift Assessment Infrastructure Layers\footnote{\url{https://github.com/LSSTDESC/RAIL}} \citep[RAIL;][]{sam_schmidt_2023_7927358}, which will also be used in the LSST analysis pipeline.
Finally, we propagate these effects to the clustering and weak lensing two-point statistics.

This paper is organized as follows. We describe our simulation datasets in Section~\ref{sec: Simulations} and introduce our methods in Section~\ref{sec: Methods}. The results are presented in Section~\ref{sec: Results}.
We show the variation of the angular power spectra with varying depth effects in Section~\ref{sec: cosmology}.
Finally, we conclude in Section~\ref{sec: Conclusions}.

\section{Simulations}
\label{sec: Simulations}

This section provides an overview of the simulations used in this work, namely, the Rubin Operation Simulator (OpSim; Section~\ref{sec: opsim}), which simulates the observing strategy and related properties for Rubin LSST, and the Roman-Rubin simulation (DiffSky; Section~\ref{sec: roman-rubin}), which provides a truth catalogue complete up to $z=3$ with realistic galaxy colours.

\subsection{Rubin Operations Simulator (OpSim)}
\label{sec: opsim}

\begin{figure}
     \centering
     \begin{subfigure}[b]{0.47\textwidth}
         \centering
         \includegraphics[width=\textwidth]{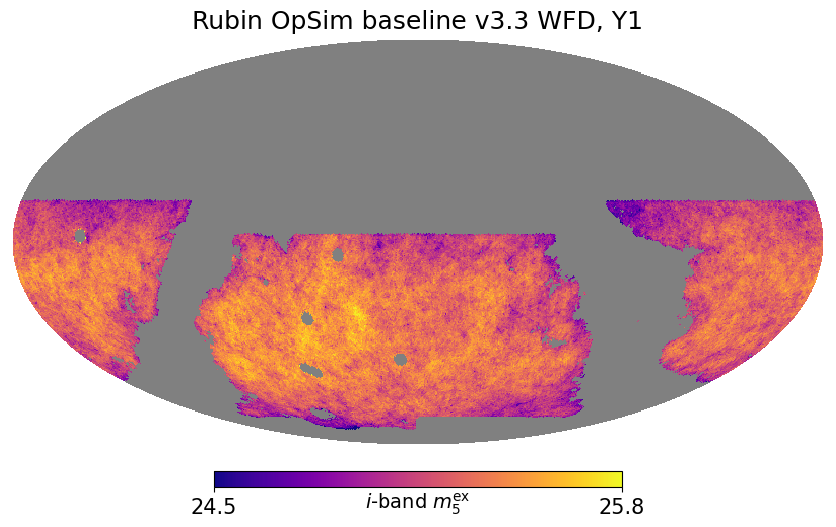}
     \end{subfigure}
      \hfill
     \begin{subfigure}[b]{0.47\textwidth}
         \centering
         \includegraphics[width=\textwidth]{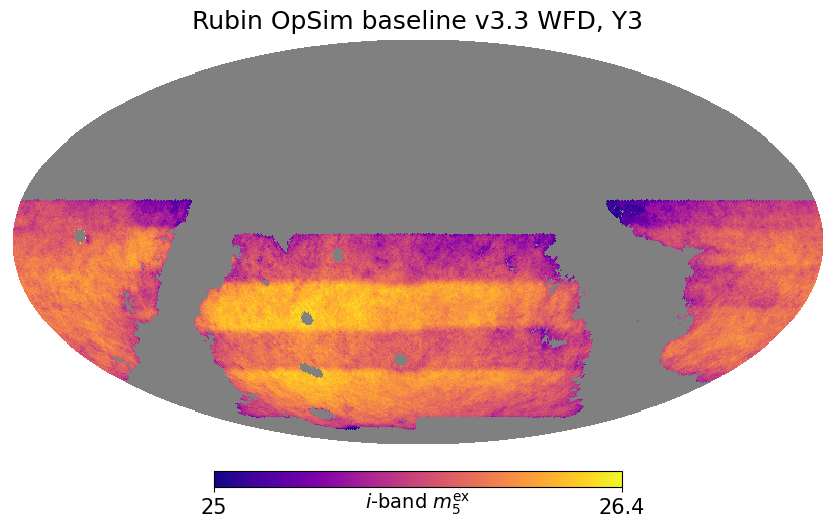}
     \end{subfigure}
     \hfill
     \begin{subfigure}[b]{0.47\textwidth}
         \centering
         \includegraphics[width=\textwidth]{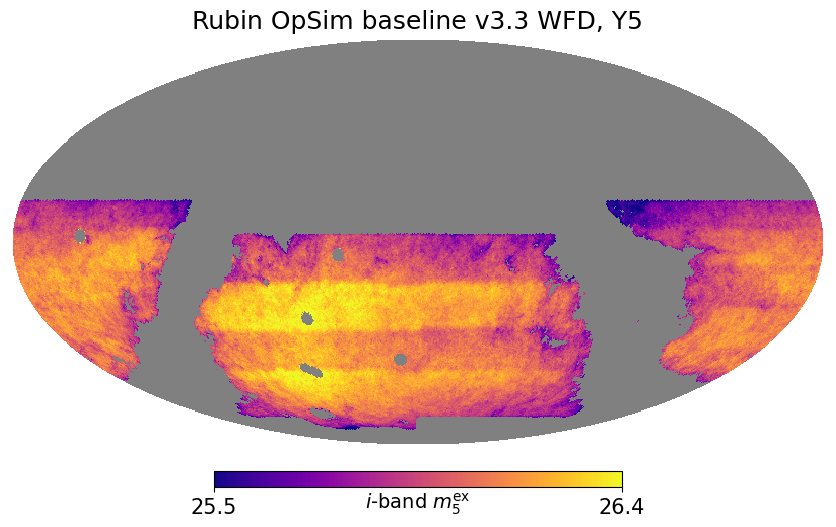}
     \end{subfigure}
        \caption{The simulated $i$-band coadd $5\sigma$ depth accounting for Galactic extinction, $m_5^{\rm ex}$, from the Rubin observatory OpSim baseline v3.3 over the LSST wide-fast-deep (WFD) footprint, for 1-year (upper), 3-year (middle), and 5-year (lower) observations. Notice the stripy patterns visible from the 3-year and 5-year observations are the result of rolling cadence. $i$-band is shown here because it is the detection band for LSST.}
        \label{fig: opsim coaddm5 i}
\end{figure}

\begin{figure*}
   \centering
   \includegraphics[width=\textwidth]{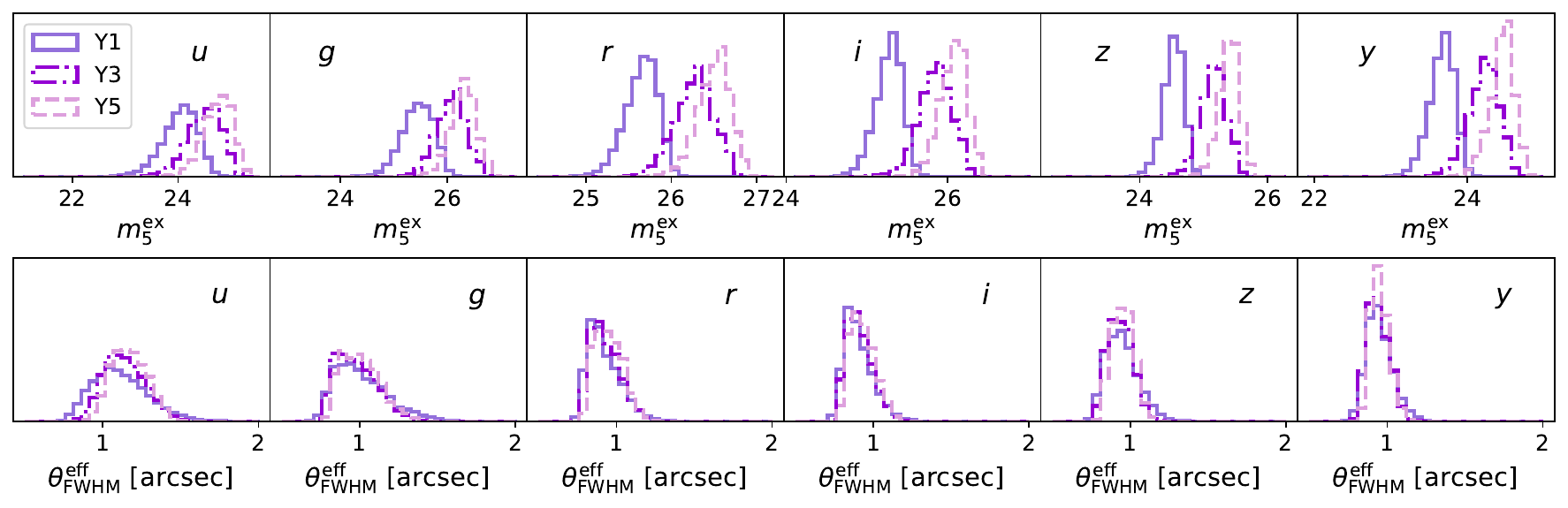}
    \caption{Distribution of the extinction-corrected coadd depth ($m_5^{\rm ex}$) and the median effective seeing ($\theta_{\rm FWHM}^{\rm eff}$) for the six LSST bands from the OpSim baseline v3.3. The different colours and line styles indicate 1-year, 3-years, and 5-years observations, as shown by the legend.}
    \label{fig: obscond_dist_y135}
\end{figure*}

The Operations Simulator\footnote{\url{https://rubin-sim.lsst.io/}} (OpSim) of the Rubin Observatory is an application that simulates the telescope movements and a complete set of observing conditions across the LSST survey footprint over the 10-year observation period, providing predictions for the LSST performance with respect to various survey strategies. 
OpSim uses a historical weather log from Cerro-Tololo Inter-American Observatory (CTIO), Chile from the ten year period 1996 to 2005, to simulate weather conditions. An observation is conducted when the weather log is no more than 42\% cloudy. This gives about the same amount of total weather downtime as Gemini South and SOAR. Realistic seeing values for each observation are generated using historical seeing logs from Cerror Pachón, Chile.
We utilise OpSim baseline v3.3, the most recent observing strategy. This strategy involves a rolling cadence that starts after the first year of observation. In subsequent years, parts of the sky will receive more visits than others, enabling higher resolution sampling for time domain science. At the end of the fiducial survey, uniformity will be recovered at the expected 10-year LSST depth. 
The output of OpSim is evaluated by the Metrics Analysis Framework (MAF), a software tool that computes summary statistics (e.g. mean and median of a particular observing condition over a given period) and derived metrics (e.g. coadd $5\sigma$ depth) that can be used to assess the performance of the observing strategy, in terms of survey efficiency and various science drivers. 
The \texttt{fieldRA} and \texttt{fieldDec} positions used in the MAF include the dithering that has been applied. The sky is first tessellated by the telescope field of view (a few degrees in diameter), and the orientation is then randomised at the start of each night. Visits are done in pairs to allow detection of moving solar system objects, so that within a night there is no dithering. The MAF loops over the HEALPix pixel centres, and for each one finds the observations that overlap with that point, including rejecting observations where the point falls on a chip gap.

For the purpose of this study, we obtain survey condition maps in HEALPix \citep{2005ApJ...622..759G} format using the MAF HEALPix slicer with $N_{\rm side}=128$ (corresponding to a pixel size of $755\, {\rm arcmin}^2$), using the (RA, Dec) coordinates. We do not choose a higher resolution for the map because we expect that survey conditions vary smoothly on large scales, and this choice of $N_{\rm side}$ is enough to capture the variation with the rolling pattern. 
For our purposes, we mainly consider the following quantities in each of the $ugrizy$ filters: extinction-corrected coadd $5\sigma$ point source depth (\texttt{ExgalM5}, hereafter $m_5^{\rm ex}$) and the effective full width half maximum seeing (\texttt{seeingFwhmEff}, hereafter $\theta_{\rm FWHM}^{\rm eff}$) in unit of arcsecond. 
The $m_5^{\rm ex}$ is different from the coadd depth, $m_5$, by the fact that it includes the lost of depth near galactic plane.
The effective seeing, $\theta_{\rm FWHM}^{\rm eff}$, has a wavelength dependence, with a poorer seeing at bluer filters from Kolmogorov turbulence. 
The MAF also takes into account for increase in PSF size with airmass, $X$, due to seeing, i.e. $\theta_{\rm FWHM}^{\rm eff} \propto X^{0.6}$. However, the MAF does not include the increase in PSF size along the zenith direction with zenith angle, due to differential chromatic refraction.
This quantity is used here to convert point-source depth to that for extended objects. 
We obtain maps of these quantities over the LSST footprint at the end of each full year of observation (e.g. Y3 for ${\rm nights}<1095$). 
The coadded depth in each band is computed by assessing the $5\sigma$-depth (in magnitudes) of each visit within each HEALPix pixel, then computing the `stacked' depth. The coadded depth calculation includes the airmass, seeing, and sky brightness of each visit. It is approximated that the whole field of view has values similar to the centre, so that vignetting or sky brightness gradients are not included. For the most part these gradients should be small and average out over many visits.
Maps of $\theta_{\rm FWHM}^{\rm eff}$ contain the median over all visits in a particular HEALPix pixel.

Throughout the paper, we will use Y1, Y3, and Y5 as examples to showcase the impact of spatial variability on photometric redshifts. 
Notice that the choice of Y3 and Y5 are a pessimistic one, because the survey strategy is close to uniformity in Y4 and Y7 where cosmological analysis are expected to be conducted. Hence, this paper provides a worst-case scenario of the severity of the impact from spatial variability.
Also, the Rubin observing strategy is still being decided, and the rolling cadence may move to different times during the survey. 
There are ongoing efforts on recommendations about the observing strategy, and hence the results shown here should be interpreted in light of this particular strategy and years chosen. 
We will focus on the wide-fast-deep (WFD) survey program footprint, and exclude areas with high galactic extinction $E(B-V)>0.2$ for cosmological studies. Notice that, in practice, additional sky cuts could also be applied (e.g. a depth cut that removes very shallow regions).
Specifically, we will focus on the variation with respect to $i$-band, the detection band of LSST. 
Figure~\ref{fig: opsim coaddm5 i} shows the spatial variation of the extinction-corrected coadd $i$-band depth for OpSim baseline v3.3 in Y1, Y3, and Y5. The stripes visible across the footprint in Y3 and Y5 are the characteristics of the rolling cadence. 
The distribution of all OpSim variables are shown in Fig.~\ref{fig: obscond_dist_y135} for each of the six filters and for selected years of observation. One can see that the coadd depths build up in each band over the years, whereas the distributions of the median effective seeing per visit are relatively unchanged. One can also see a strong skewness in these distributions.


\subsection{Roman-Rubin simulation (DiffSky)}
\label{sec: roman-rubin}

In order to investigate the impact of varying survey conditions on photo-$z$ for LSST, we need a simulated truth catalogue that is complete to beyond the LSST 10-year depth and realistic in colour-redshift space. For this purpose, we use the joint Roman-Rubin simulation v1.1.3. This simulation is an extension of the effort in \cite{Troxel_2023}, but with many improvements, including self-consistent, flexible galaxy modelling. The simulation is based on its precursor, CosmoDC2 \citep{2019ApJS..245...26K}, a synthetic sky catalogue out to $z=3$ built from the `Outer Rim' N-body cosmological simulation \citep{2019ApJS..245...16H}. The N-body simulation contains a trillion particles with a box size of $(4.225 {\rm Gpc})^2$.
The galaxies are simulated with Diffsky\footnote{\url{https://github.com/LSSTDESC/lsstdesc-diffsky}}, based on two differentiable galaxy models: Diffstar \citep{2023MNRAS.518..562A} and Differentiable Stellar Population Synthesis \citep[DSPS; ][]{2023MNRAS.521.1741H}. 
Using Diffstar, one can build a parametric model that links galaxy star formation history with physical parameters in halo mass assembly. Then, with DSPS, one can calculate the SED and photometry of a galaxy as a function of its star formation history, metallicity, dust, and other properties. 
The advantage of this galaxy model is that the distribution in colour-redshift is smooth and more realistic compared to that in CosmoDC2. This is thanks to the separate modeling for different galaxy components, i.e. bulge, disk, and star-forming regions. The SEDs built from these different components with different stellar populations makes the colours more realistic for photo-$z$ estimation. 
The calibration of the Roman-Rubin simulation galaxy colours as a function of redshift matches that of the COSMOS2020 sample \citep{Weaver_2022}, although some evidence of a low amount of variance in the NIR colors at $z>1$ is obvious. For more details of the Roman-Rubin DiffSky simulation, see the DESC Note by Troxel et al. \textit{In prep}.

We randomly subsample the full simulated catalogue to $N=10^6$ objects complete to $i<26.5$ as our truth sample. For each object, we obtain its magnitude in the six LSST bands, true redshift, bulge size $s_b$, disk sizes $s_d$, bulge-to-total ratio $f_b$, and ellipticity $e$.
We obtain the galaxy semi-major and semi-minor axes, $a,b$ via $a=s/\sqrt{q}$ and $b= s\sqrt{q}$, where $s$ is the weighted size of the galaxy, $s= s_bf_b + s_d(1-f_b)$, and $q$ is the ratio between the major and minor axes, related to ellipticity via $q=(1-e)/(1+e)$.

One caveat of the current sample is that, at $z>1.5$, there is an exaggerated bimodal distribution in the $g-r$ colour and redshifts, which is not found in real galaxy data. As a result, the bluest objects in the sample are almost always found at high redshifts. This could be due to the high-redshift SPS models being less well constrained. One direct consequence of this is that, when training a machine learning algorithm to estimate the photo-$z$, the high-redshift performance may be too optimistic due to this colour-space clustering.

\section{Methods}
\label{sec: Methods}

This section describes our methodology for generating a mock LSST photometry catalogue for Y1, Y3, and Y5, applying photometric redshift estimation algorithms, and defining metrics to assess the impact of variable depth. Specifically, we describe the degradation process using the LSST error model in Section~\ref{sec: degradation}, the two photo-$z$ estimators, BPZ and FlexZBoost, in Section~\ref{sec: photo-z estimators}, the tomographic binning strategy in Section~\ref{sec: Tomographic bins}, and the relevant metrics Section~\ref{sec: Metrics for impacts of varying depth}.

\subsection{Degradation of the truth sample}
\label{sec: degradation}


Given a galaxy with true magnitudes $m_{\rm t}=\{ugrizy\}$ falling in a HEALPix pixel within the footprint, we `degrade' its magnitude with observing conditions associated with that pixel, and assign a set of `observed' magnitudes $m_{\rm o}$ and the associated magnitude error $\sigma_{m,{\rm o}}$, using the following procedure: (1). Apply galactic extinction. (2). Compute the point-source magnitude error for each object in each filter, using the LSST error model detailed in \cite{2019ApJ...873..111I}. (3). Compute the correction to obtain the extended-source magnitude errors. (4). Sample from the error and add it to the true magnitudes. Steps (2) - (4) are carried out using the python package \texttt{photerr}\footnote{\url{https://github.com/jfcrenshaw/photerr/tree/main}} \citep{crenshaw2024probabilistic}. We detail each step below.

Firstly, we apply the galactic extinction to each band with the $E(B-V)$ dust map \citep{2018JOSS....3..695M} via:
\begin{equation}
    m_{\rm dust}=m+\left[\frac{A_{\lambda}}{E(B-V)}\right]E(B-V),
    \label{eq: MW redden}
\end{equation}
where for each of the six LSST filters we adopt $[A_{\lambda}/E(B-V)]=\{4.81,3.64,2.70,2.06,1.58,1.31\}$.

Then, we utilize the LSST error model \citep{2019ApJ...873..111I}  to compute the expected magnitude error, $\sigma_m$, per band. 
The magnitude error is related to the noise-to-signal ratio, nsr, via:
\begin{equation}
    \sigma_m = 2.5\log_{10}(1+{\rm nsr}). 
    \label{eq: low snr}
\end{equation}
The total nsr consists of two components:
\begin{equation}
    {\rm nsr}^2 = {\rm nsr}_{\rm sys}^2+ {\rm nsr}_{\rm rand, ext}^2,
    \label{eq: mag err 1}
\end{equation}
where ${\rm nsr}_{\rm sys}$ is the systematic error from the instrument read-out and ${\rm nsr}_{\rm rand}$ is the random error arising from observing conditions on the sky, for extended objects. 
Notice that in the high signal-to-noise limit where ${\rm nsr}\ll 1$, $\sigma_m\sim {\rm nsr}$, and Eq.~\ref{eq: mag err 1} recovers the form in \cite{2019ApJ...873..111I}. Throughout the paper, we set ${\rm nsr_{\rm sys}}\approx \sigma_{\rm sys}=0.005$, which corresponds to the maximum value allowed from the LSST requirement. 
For point sources, the random component of nsr is given by
\begin{equation}
    {\rm nsr}_{\rm rand, pt}^2=(0.04-\gamma)x+\gamma x^2, 
    \label{eq: mag err 2}
\end{equation}
where $\gamma$ is a parameter that depends on the system throughput. We adopt the default values from \cite{2019ApJ...873..111I}, $\gamma=\{0.038, 0.039, 0.039, 0.039, 0.039, 0.039\}$ for $ugrizy$. $x$ is a parameter that depends on the magnitudes of the object, $m$, and the corresponding coadd $5\sigma$ depth, $m_5$, in that band:
\begin{equation}
    \log_{10} x \equiv 0.4 \left(m-m_5\right). 
    \label{eq: x}
\end{equation}
For extended sources, we adopt the expression in \cite{2020A&A...642A.200V, Kuijken_2019}, where the nsr receives an additional factor related to the ratio between the angular size of the object and that of the PSF:
\begin{equation}
    {\rm nsr}_{\rm rand, ext} = {\rm nsr}_{\rm rand, pt} \sqrt{{A_{\rm ap}}/{A_{\rm psf}}}.
    \label{eq: extended err}
\end{equation}
Here, 
\begin{equation}
    A_{\rm psf}=\pi \sigma_{\rm psf}^2, \quad \sigma_{\rm psf} = \theta_{\rm FWHM}^{\rm eff}/2.355, 
\end{equation}
where $\theta_{\rm FWHM}^{\rm eff}$ is the effective FWHM seeing (it is linked to the seeing by $\theta_{\rm FWHM}^{\rm eff} = \theta_{\rm FWHM} X^{0.6}$, where $X$ is the airmass) for a given LSST band. The AP angular size of the object is given by
\begin{align}
    A_{\rm ap}&=\pi a_{\rm ap}b_{\rm ap},\nonumber\\
    a_{\rm ap} & = \sqrt{\sigma_{\rm psf}^2+(2.5a)^2}, \nonumber\\
    b_{\rm ap} & = \sqrt{\sigma_{\rm psf}^2+(2.5b)^2},
\end{align}
where $a, b$ are the galaxy semi-major and minor axis. 
We make one modification to Eq.~\ref{eq: extended err}, where we replace the denominator by the \textit{mean} PSF area, $\sqrt{\langle A_{\rm psf}\rangle}$, averaged over pixels in the $i$-band quantiles which we will elaborate shortly. 
In the approximation that ${\rm nsr}_{\rm rand,pt}\propto x$, the point-source noise is then proportional to $\theta_{\rm FWHM}^{\rm eff}$ (see Eq.~\ref{eq: mag err}), and so for the extended-source noise, $\theta_{\rm FWHM}^{\rm eff}$ cancels and Eq.~\ref{eq: extended err} effectively changes the dependence of $m_5$ on PSF size to that on the extended aperture size. However, in this work, we utilize the \textit{median} seeing, for which the cancellation may not be exact. 
Naively taking Eq.~\ref{eq: extended err} could lead to unrealistic cases, where, at fixed depth, ${\rm nsr}_{\rm rand,ext}$ increases with a better seeing. We have tested both scenarios, i.e., using individual $A_{\rm psf}$ or the mean $\langle A_{\rm psf} \rangle$ in Eq.~\ref{eq: extended err}, and find negligible difference for our main conclusion in the $i$-band quantiles. However, it does make a significant difference if one were to bin the samples by quantiles of seeing, as investigated in Appendix~\ref{sec: Variation with other survey properties}.

To obtain the observed magnitudes $m_{\rm o}$, we degrade in flux space, $f_{\rm o}$, by adding a random noise component $\Delta f$ drawn from a normal error distribution, $\Delta f\sim \mathcal{N}(0,{\rm nsr})$, to the reddened flux $f_{\rm dust}$ of the object. Here, nsr is computed by setting $m=m_{\rm dust}$ in Eq.~\ref{eq: x}. The flux and magnitude are converted back and forth via
\begin{equation}
    m_k=-2.5\log_{10}f_k,\quad k=\{{\rm dust, o}\}.
\end{equation}
Negative fluxes are set as `non-detection' in that band. The corresponding magnitude error $\sigma_{m,{\rm o}}$ is computed using Eq.~\ref{eq: low snr} and setting $m=m_{\rm o}$ in Eq.~\ref{eq: x}, such that the error de-correlates with the observed magnitude.

\begin{table}
    \centering
   \caption{The mean and standard deviation of the $i$-band extinction-corrected coadd depth, $m_5^{\rm ex}$, split in 10 quantiles, from the Rubin OpSim baseline v3.3 map with $N_{\rm side}=128$, for year 1, 3, and 5, respectively.}
    \label{tab: i-band quantiles}
   \begin{tabular}{cccc} 
    \hline
    qtl ($i$-band $m_5^{\rm ex}$) &  Y1 & Y3 & Y5\\
    \hline
    0  & $24.95\pm0.10$ & $25.46\pm0.12$ & $25.75\pm0.10$ \\
    1  & $25.10\pm0.03$ & $25.64\pm0.03$ & $25.89\pm0.02$ \\
    2  & $25.17\pm0.02$ & $25.72\pm0.02$ & $25.96\pm0.02$ \\
    3  & $25.22\pm0.01$ & $25.78\pm0.02$ & $26.01\pm0.01$ \\
    4  & $25.27\pm0.01$ & $25.83\pm0.02$ & $26.06\pm0.01$ \\
    5  & $25.31\pm0.01$ & $25.88\pm0.01$ & $26.10\pm0.01$ \\
    6  & $25.35\pm0.01$ & $25.93\pm0.01$ & $26.14\pm0.01$ \\
    7  & $25.39\pm0.01$ & $25.99\pm0.02$ & $26.18\pm0.01$ \\
    8  & $25.44\pm0.02$ & $26.06\pm0.03$ & $26.23\pm0.02$ \\
    9  & $25.53\pm0.05$ & $26.18\pm0.05$ & $26.33\pm0.04$ \\
    \hline
   \end{tabular}
\end{table}

To focus on the trend in the depth variation in the detection band, we subdivide pixels in the survey footprint into 10 quantiles in $i$-band $m_5^{\rm ex}$, where the first quantile (${\rm qtl}=0$) contains the shallowest pixels, and the last quantile (${\rm qtl}=9$) contains the deepest.
Table~\ref{tab: i-band quantiles} shows the mean and standard deviation of each $i$-band depth quantile. We also show in Table~\ref{tab: sub-prop} the mean and standard deviation of all other survey condition maps used in the analysis in each of the $i$-band depth quantiles.
Within each quantile, we randomly assign each galaxy to a HEALPix pixel in that quantile, with its associated observing conditions $\{ E(B-V), m_5^{\rm ex}, \theta_{\rm FWHM}^{\rm eff}\}$ on that pixel for each LSST band, from the OpSim MAF maps. Then, we carry out the above degradation process to our truth sample.
On average, each pixel within each quantile is assigned 121 galaxies.
Notice that there are many other parameters that could affect the photometric errors, e.g. sky background, exposure time, and atmospheric extinction. Following \cite{2019ApJ...873..111I}, because these quantities only contribute towards $m_5$, we do not include them otherwise in the degradation, and assume that $m_5^{\rm ex}$ completely captures their variation. 
Additionally, we explore the relation between $m_5$ and these extended quantities using OpSim in Appendix~\ref{sec: Calibration of LSST error model}, and we explore the galaxy redshift distribution dependence with other survey properties in Appendix~\ref{sec: Variation with other survey properties}.

Finally, we apply an $i$-band magnitude cut corresponding to the LSST Gold sample selection on the degraded catalogue. For the full 10-year sample this is defined as $i<25.3$. For data with an observation period of $N_{\rm yr}$ years, we adjust the gold cut to $i_{\rm lim}=25.3 + 2.5\log_{10}(\sqrt{N_{\rm yr}/10})$.
Thus for Y1, Y3, and Y5, we adopt the following gold cuts respectively: $i_{\rm lim}=24.0, 24.6, 24.9$. 
Notice that this is slightly shallower than the definition in the DESC SRD, where the Gold cut is defined as one magnitude shallower than the median coadd $m_5$. This is due to the fact that OpSim baseline v3.3 has a slightly deeper $i$-band depth in early years compared to previous expectations. For Y1, the median $i$-band $m_5^{\rm ex}$ is $\sim 25.2$, giving a DESC SRD Gold cut to be $0.2$ mag deeper than what we adopt here. 
Additionally, for our fiducial sample, we also apply a signal-to-noise cut in $i$-band: ${\rm SNR}=1/{\rm nsr}\geq10$, although we also look at the case with the full sample. This cut is motivated by the selection of the source sample, where shape measurements typically require a high SNR detection in $i$-band. In this work, we apply this cut to both the weak lensing and clustering samples.

\subsection{Photo-\texorpdfstring{$z$}{z} estimators}
\label{sec: photo-z estimators}

\begin{figure*}
     \centering
     \begin{subfigure}[b]{0.47\textwidth}
         \centering
         \includegraphics[width=\textwidth]{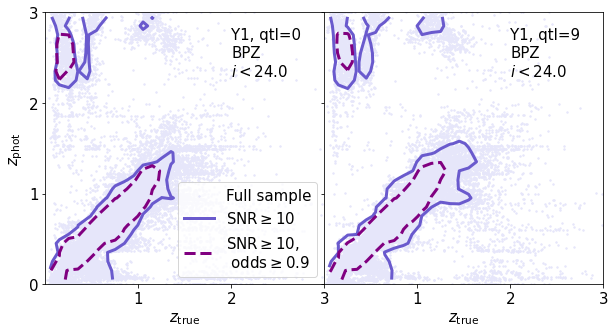}
     \end{subfigure}
      \hfill
     \begin{subfigure}[b]{0.47\textwidth}
         \centering
         \includegraphics[width=\textwidth]{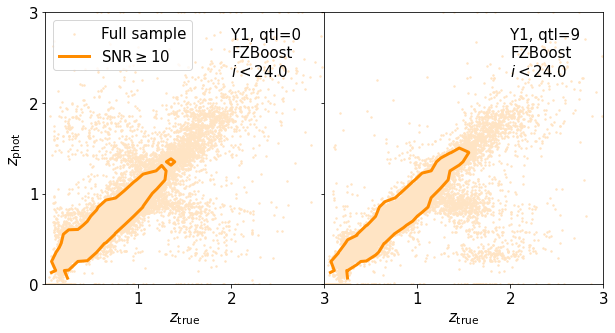}
     \end{subfigure}
     \hfill
     \begin{subfigure}[b]{0.47\textwidth}
         \centering
         \includegraphics[width=\textwidth]{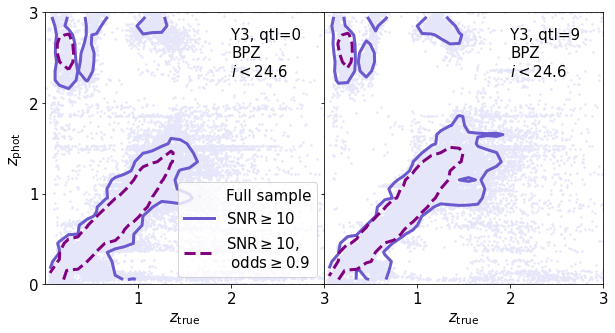}
     \end{subfigure}
     \hfill
     \begin{subfigure}[b]{0.47\textwidth}
         \centering
         \includegraphics[width=\textwidth]{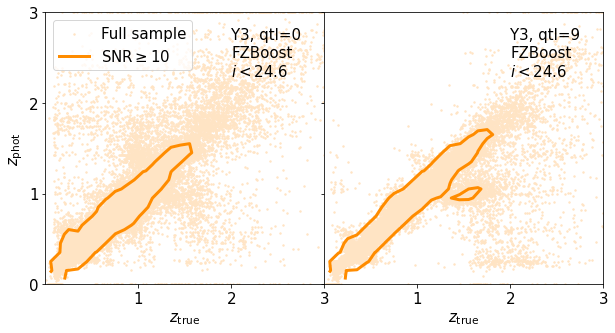}
     \end{subfigure}
     \hfill
     \begin{subfigure}[b]{0.47\textwidth}
         \centering
         \includegraphics[width=\textwidth]{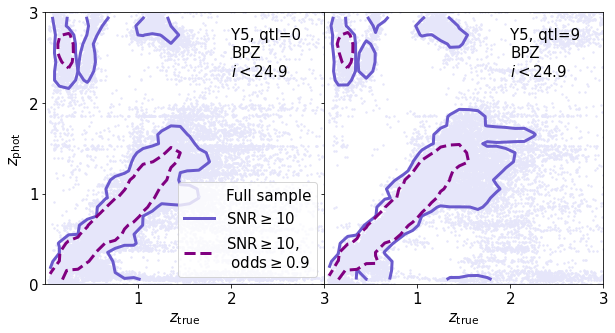}
     \end{subfigure}
     \hfill
     \begin{subfigure}[b]{0.47\textwidth}
         \centering
         \includegraphics[width=\textwidth]{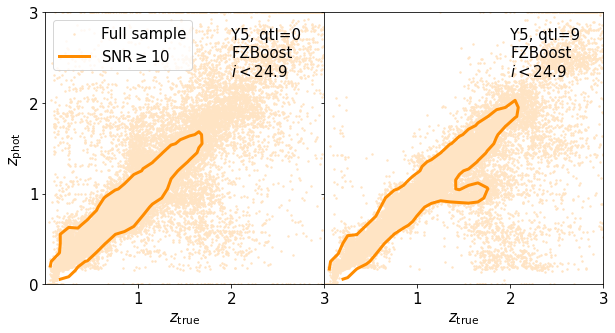}
     \end{subfigure}
        \caption{photo-$z$ vs. true redshifts for the sample degraded with Rubin OpSim baseline v3.3 observing conditions using BPZ (left two columns) and FZBoost (right two columns) mode as the photo-$z$ point estimator. For each photo-$z$ method, we show sample degraded with pixels containing the shallowest 10\% $i$-band Coadded depth with galactic extinction (${\rm qtl}=0$), and that from the deepest 10\% (${\rm qtl}=9$). This is repeated for the cases of Y1, Y3, and Y5 observing conditions with respective gold cut in $i$-band applied. The faint dots show all the samples included within the gold cut, whereas the solid contour shows the samples (90\% contour) with ${\rm SNR}\geq10$ (fiducial). In the BPZ case, the dashed lines show the 90\% contour for the sample with an additional selection of ${\rm odds}\geq 0.9$. In the FZBoost case, the model is trained on a perfectly representative sample for each observation year.}
        \label{fig: photo-$z$ vs spec-z}
\end{figure*}

Methods for photometric redshift estimation can be broadly divided into two main categories: template-fitting and machine learning. Template fitting methods assume a set of SED templates for various types of galaxies, and use these to fit the observed magnitudes of the targets. Machine Learning methods, on the other hand, use machine learning algorithms trained on a reference sample, to infer the unknown target redshifts. 
See \cite{2020MNRAS.499.1587S} for a review and comparison of the performance of various photo-$z$ estimators in the context of Rubin LSST. In this work, we adopt two algorithms with reasonable performance, a template fitting method, BPZ (Bayesian Photometric Redshifts), and a machine learning method, FlexZBoost. In this work, before applying these redshift estimators, all observed magnitudes are de-reddened, by applying the inverse of Eq.~\ref{eq: MW redden}.

\subsubsection{BPZ (Bayesian Photometric Redshifts)}

BPZ \citep[Bayesian Photometric Redshifts;][]{2000ApJ...536..571B, 2006AJ....132..926C} is a template-based photometric estimation code. Given a set of input templates $\mathbf{t}$, BPZ computes the joint likelihood $P(z,\mathbf{t})$ for each galaxy with redshift $z$. A prior $P(z,\mathbf{t}|m)$ is included based on the observed magnitude of the galaxy $m$. For example, the prior restricts bright, elliptical galaxies to lower redshifts. 
For each galaxy, a likelihood $P(z,\mathbf{t}|c, m)$ given the galaxy's colour $c$ and magnitude is computed, and by marginalising over the templates, one obtains the per-object redshift probability $P(z)$. 

We use the RAIL interface of the BPZ algorithm, with the list of spectral energy distribution (SED) templates adopted in \cite{2006AJ....132..926C}: the CWW+SB4 set introduced by \cite{2000ApJ...536..571B}, the El, Sbc, Scd \& Im from \cite{1980ApJS...43..393C}, the SB2 \& SB3 from \cite{1996ApJ...467...38K}, and the 25Myr \& 15Myr `SSP' from \cite{2003MNRAS.344.1000B}. We set the primary observing band set to $i$-band, and adopt the prior from the original BPZ paper \citep{2000ApJ...536..571B}, which was used to fit data from the Hubble Deep Field North \citep[HDF-N;][]{1996AJ....112.1335W}. Notice that these set of SEDs may be different from that in the Roman-Rubin simulation, and the prior distributions may not match exactly. 
The prior mismatch would only affect samples with low signal-to-noise and hence those posteriors are prior-dominated. For the Gold sample considered in this paper, the impact of the prior on the mean difference and scatter of the true and photometric redshifts is expected to be small, although galaxies with broad or bimodal posteriors may end up having a different point estimate (e.g. mode), hence the outlier rate could be slightly higher. We do not include extra SED templates here. The SED template colours are able to cover the range of colours in the Roman-Rubin simulation, as shown in Appendix~\ref{sec: bpz templates}.

Additionally we compute the odds parameter, defined as 
\begin{equation}
    {\rm odds} = \int_{z_{\rm mode}-\Delta z}^{z_{\rm mode}+\Delta z}P(z)\,dz,
\end{equation}
where $z_{\rm mode}$ is the mode of $P(z)$, and $\Delta z =\epsilon (1+z_{\rm mode})$ defines an interval around the mode to integrate $P(z)$. The maximum value of odds is 1, which means that the probably density is entirely enclosed within the integration range around the mode, whereas a small odds means that the probability density is diffuse given the range. 
Hence, odds denotes the confidence of the BPZ redshift estimation, and the choice of $\epsilon$ essentially sets the criteria. The $(1+z_{\rm mode})$ factor accounts for the fact that larger redshift errors are expected at higher redshifts. We choose $\epsilon=0.06$ as a nominal photo-$z$ scatter, and we use odds as a BPZ `quality control', where a subsample is selected with ${\rm odds}\geq0.9$, as comparison to the baseline sample.

\subsubsection{FlexZBoost}

FlexZBoost \citep[][; hereafter FZBoost]{2020A&C....3000362D} is a machine-learning photo-$z$ estimator based on FlexCode \citep{2017arXiv170408095I}, a conditional density estimator (CDE) that estimates the conditional probability density $p(y|\mathbf{x})$ for the response or parameters, $y$, given the features $\mathbf{x}$.
The algorithm uses basis expansion of univariate $y$ to turn CDE to a series of univariate regression problems. 
Given a set of orthonormal basis functions $\{ \phi_i(y) \}_i$, the unknown probability density can be written as an expansion:
\begin{equation}
    p(y|\mathbf{x})=\sum_j \beta_j(\mathbf{x})\phi_j(y).
\end{equation}
The coefficients $\beta_j(\mathbf{x})$ can be estimated by a training set $(\mathbf{x},y)$ using regression.
The advantage of FlexCode is the flexibility to apply any regression method towards the CDE.
The main hyper-parameters involved in training is the number of expansion coefficients and those associated with the regression. 
\cite{2020MNRAS.499.1587S} found that FZBoost was among the the strongest performing photo-$z$ estimators according to the established performance metrics.

In this paper, we utilise the RAIL interface of the FZBoost algorithm with its default training parameters. 
We construct the training sample by randomly drawing 10\% of the degraded objects from each of the deciles, and train each year separately. 
Notice that this training sample is fully representative of the test data, which is not true in practice. Spectroscopic calibration samples typically have a magnitude distribution that is skewed towards the brighter end, and the selection in colour space can be non-trivial depending on the specific dataset used. Although there are methods to mitigate impacts from this incompleteness, such as re-weighting in redshift or colours \citep{2008MNRAS.390..118L}, and, more recently, using training data augmentation from simulations \citep{2024arXiv240215551M}, the photo-$z$ performance is not comparable to having a fully representative sample, and one would expect some level of bias and increased scatter depending on the mitigation method adopted. Here, we are interested in whether our results on the non-uniformity impact changes significantly with an alternative photo-$z$ algorithm. We thus leave the more realistic and sophisticated case with training sample imperfection to future work.

\subsubsection{Performance}

For both photo-$z$ estimators, we use the mode of the per-object redshift probability, $P(z)$, as the point estimate, $z_{\rm phot}$.
Fig.~\ref{fig: photo-$z$ vs spec-z} shows the scatter in spec-$z$ and photo-$z$ for Y1, Y3, and Y5 with BPZ and FZBoost redshifts, for the shallowest (${\rm qtl}=0$) and the deepest (${\rm qtl}=9$) quantiles in the $i$-band $m_5^{\rm ex}$ respectively. 
The scatter is always larger for the shallower sample in the full sample case (faint purple dots). This is expected following Eqs.~\ref{eq: low snr} and~\ref{eq: mag err 2}, given that the coadd depths in each band are strongly correlated. At fixed magnitude, the larger the $m_5$, the smaller the photometric error, hence also the smaller the scatter in photo-$z$. The signal-to-noise cut at ${\rm SNR}\geq10$ removes some extreme scatter as well as objects from the highest redshifts. This is more obvious for the shallowest sample compared to the deepest, due to the better signal-to-noise for the deepest sample at high redshifts.

There is a significant group of outliers in the BPZ case that are at low redshifts but are estimated to be at $z>2$, highlighted by the blue contours. By examining individual BPZ posteriors for this group, we find that these objects tend to have very broad or bimodal redshift distributions. This could be a result of confusion between the Lyman break and the $4000 \text{\AA}$ Balmer break, and notice that the fraction of this population as well as its location can be influenced by the choice of the BPZ priors. Another possible cause is the spurious bimodal distribution in the colour-redshift space in the Roman-Rubin simulation, as mentioned in Section~\ref{sec: roman-rubin}.
We see that after applying a strict cut with ${\rm odds}\geq0.9$, shown by the purple dashed lines enclosing 90\% of the sample, the outlier populations are significantly reduced, as expected.
This cut retains 20.4\% (27.7\%), 25.7\% (44.4\%), 29.5\% (44.0\%) of the ${\rm SNR}\geq10$ sample in ${\rm qtl}=0$ (9) for Y1, Y3, and Y5, respectively. We see that this cut further reduces the scatter at $z_{\rm phot}\sim 1.5$. FZBoost in general shows a much better performance, given that the training data is fully representative of the test data.
Table~\ref{tab: stats on photoz} summarizes these findings for each sample via a few statistics of the distribution of the difference between photo-$z$ and true redshifts: $\Delta z=(z_{\rm phot}-z_{\rm true})/(1+z_{\rm true})$. Namely, the median bias ${\rm Median}(\Delta z)$, the standard deviation, the normalized Median Absolute Deviation (NMAD) $\sigma_{\rm NMAD}=1.48 {\rm Median}(|\Delta z|)$, and the outlier fraction with outliers defined as $|\Delta z|>0.15$. 

Notice that the odds cut could introduce bias to the galaxy distribution.
Given that the relation between photometry and the redshift PDF shape that influences odds is highly complex and non-linear, the odds can be correlated with both galaxy type and redshift. 
For cosmological analysis, the imposed selection in galaxy type is not a great concern as long as the $n(z)$ is accurately determined, and the galaxy sample is uniformly distributed spatially. 
A potential worry is that a spatial variation in the \textit{galaxy bias} is introduced, or that the bias evolution is changed, due to the odds cut. This would have to be tested out in a large cosmological simulation that includes both realistic photometry and clustering information, which we leave to future work.

\subsection{Tomographic bins}
\label{sec: Tomographic bins}

\begin{figure*}
   \centering
   \includegraphics[width=\textwidth]{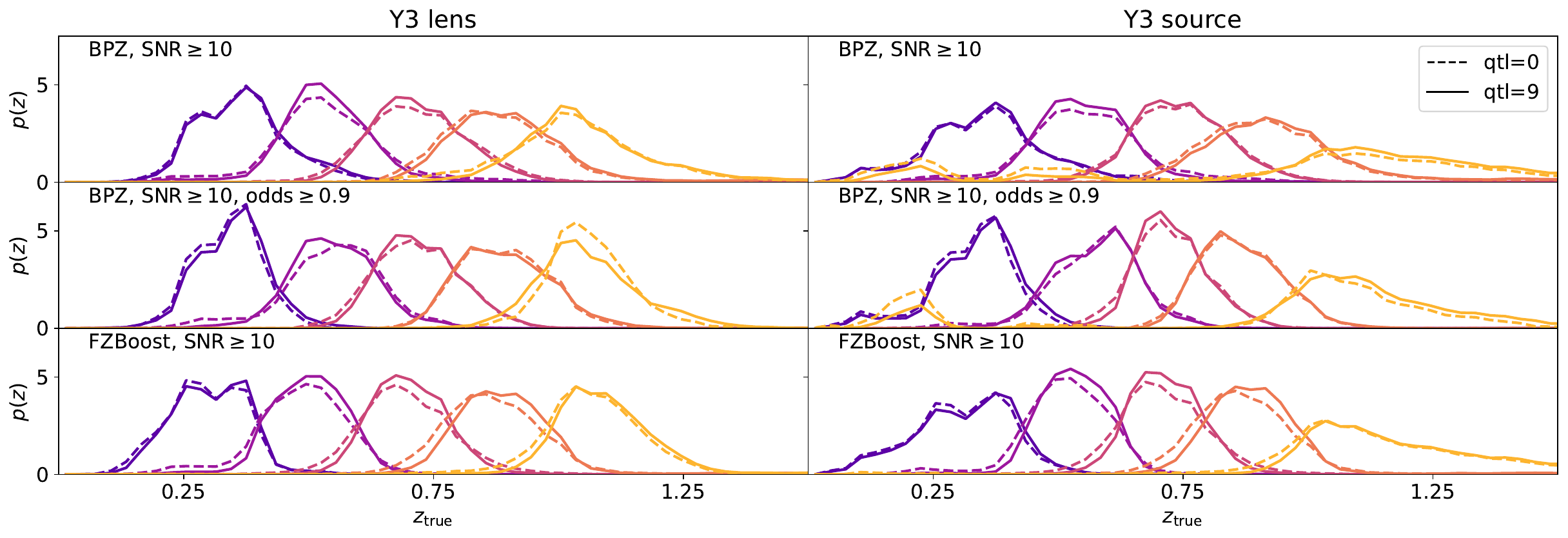}
    \caption{True redshift distribution for tomographic bins as defined in the DESC SRD for lens (left column) and source galaxies (right column) in Y3. The tomographic bins are determined using the mode of BPZ redshifts (first two rows) and FZBoost (last row). In all cases, the sample has been applied a gold cut $i<24.6$ and ${\rm SNR}\geq10$. The middle row shows the sample selected with an additional cut with ${\rm odds}\geq 0.9$. The dashed lines show samples degraded using the shallowest 10\% pixels in $i$-band coadd depth (${\rm qtl}=0$), and the solid lines show those from the deepest 10\% (${\rm qtl}=9$).}
    \label{fig: nz-baselinev2-y2-goldcut-snr-10}
\end{figure*}

In weak lensing analysis, the full galaxy catalogue is sub-divided into a `lens' sample and a `source' sample. The lens sample is often limited at lower redshifts, acting as a tracer of the foreground dark matter field which `lenses' the background galaxies. The source sample contains the background galaxies extending to much higher redshifts, whose shapes are measured precisely to construct the shear catalogue. The two samples together allow measurement of the so-called `3$\times$2pt statistics', including galaxy clustering from the lens sample, galaxy-galaxy lensing from the lens galaxies and source shapes, and cosmic shear from the source shapes alone.
Additionally, both the lens and source samples are divided into several tomographic bins, i.e., sub-samples separated with sufficient distinction in redshifts. This further includes evolution information that improves cosmological constraints. 

We adopt the Y1 tomographic bin definitions in the DESC SRD for all of our samples. The lens sample has 5 bins equally spaced in $0.2<z<1.2$, with bin width $\Delta z =0.2$, and bin edges defined using $z_{\rm phot}$. 
For source samples, the DESC SRD requires five bins with equal number of galaxies. To do so, we first combine the 10 depth quantiles, and then split the sample into five $z_{\rm phot}$ quantiles. 

Notice that in practice, tomographic binning can be determined in different ways, often with the aim of maximizing the signal-to-noise of the two-point measurements. In some cases, clustering algorithm, e.g. random forest, rather than a photo-$z$ estimator, is used to separate samples into broad redshift bins. We refer the interested readers to \cite{Zuntz_2021} for explorations of optimal tomographic binning strategies for LSST. 
Notice also that, following the DESC SRD, we do not apply additional magnitude cuts for the lens sample. This is done, for example, for the DES Y3 MagLim lens sample, where a selection of $i>17.5$ and $i<4z_{\rm phot}+18$ is applied \citep{2022PhRvD.106j3530P}. These cuts are applied to reduce faint, low-redshift galaxies in the lens sample, such that the photometric redshift calibration is more robust. 
Notice that if the lens samples are selected with a brighter cut, one would expect a different and likely reduced depth variation.
We explore this particular case in Appendix~\ref{sec: maglim}.

Fig.~\ref{fig: nz-baselinev2-y2-goldcut-snr-10} shows the normalized true redshift distribution, $p(z)$, of the lens and source tomographic bins for Y3 as an example, split by the BPZ redshifts (with or without odds selection) and the FZBoost redshifts. The dashed lines show the $p(z)$ measured from the shallowest samples, whereas the solid lines show that from the deepest samples. The BPZ case shows more extended tails in each tomographic bin compared to the FZBoost case, and for the source galaxies, a noticeable outlier population at low redshifts in the highest tomographic bin. We see that in most cases, there is a clear difference in $p(z)$ between the shallow and the deep samples: the deep samples seem to shrink the tails, making $p(z)$ more peaky towards the mean redshift (although this is not the case for the ${\rm odds}\geq0.9$ sample), and their $p(z)$ seems to shift towards higher redshift at the same time. To quantify these changes, we define metrics for the impact of variable depth below.

\subsection{Metrics for impact of variable depth}
\label{sec: Metrics for impacts of varying depth}

The first metric is the variation in the number of objects in each sample, $N_{\rm gal}$, as a function of the coadd $i$-band depth. This is the most direct impact of varying depth: deeper depth leads to more detection of objects within the selection cut. The result is that the galaxy density contrast, $\delta_g(\boldsymbol{\theta})=[N(\boldsymbol{\theta})-\bar{N}]/\bar{N}$, where $N(\boldsymbol{\theta})$ is the per-pixel number count at pixel $\boldsymbol{\theta}$, and $\bar{N}$ is the mean count over the whole footprint, fluctuates according to the depth variation, leading to a spurious clustering signal in the two-point statistics.
To quantify the relative changes, we measure the average number of objects per tomographic bin across all 10 depth quantiles, $\bar{N}_{\rm gal}=\sum_i N_{{\rm gal},i}w_i$, where $i=1,..,10$ denotes the depth bin, and $w_i\sim0.1$ is the weight proportional to the number of pixels in that quantile. We quote the change of object number in terms of $N_{\rm gal}/\bar{N}_{\rm gal}$.

The second metric quantifies the mean redshift of the tomographic bin as a function of depth:
\begin{equation}
    \langle z \rangle  = \int z\,p(z)\,dz , 
    \label{eq: meanz}
\end{equation}
where $p(z)$ is the true redshift distribution of the galaxy sample in the tomographic bin with normalization $\int p(z)\,dz=1$. Weak lensing is particularly sensitive to the mean distance to the source sample: the lensing kernel thus differs on patches with different depth. 
Here, we look at the difference between the mean redshift $\langle z \rangle_i$ of depth quantile $i$ and that of the full sample,  $\langle z \rangle_{\rm tot}$, i.e., $\Delta \langle z \rangle\equiv \langle z \rangle_i - \langle z \rangle_{\rm tot}$. More specifically, we look at the quantity $\Delta \langle z \rangle /(1+\langle z \rangle_{\rm tot})$, where the weighting accounts for the increase in photo-$z$ error towards higher redshifts. This format also allows us to compare with the DESC SRD requirements. 

The third metric quantifies the width of the tomographic bin. This is not a well-defined quantity because the $p(z)$ in many cases deviate strongly from a Gaussian distribution. One could use the variance, or the second moment of the redshift distribution:
\begin{equation}
    \sigma_z^2  = \int (z-\langle z \rangle)^2\,p(z)\,dz. \label{eq: sigmaz}
\end{equation}
However, this quantity is very sensitive to the tails of the distribution: larger tails of $p(z)$ increases $\sigma_z$, even if the bulk of the distribution does not change much.
In our case, the width of the tomographic bin is most relevant for galaxy clustering measurements: the smaller the bin width, the larger the clustering signal. 
Specifically, in the Limber approximation, the galaxy auto-correlation angular power spectrum is given by
\begin{equation}
    C_{\ell}^{\rm gg}=\int\frac{d\chi}{\chi^2(z)} \left[ \frac{H(z)}{c} p(z)\right]^2 P_{gg}\left(k=\frac{\ell+1/2}{\chi}, z\right),
\end{equation}
where $\ell$ is the degree of the spherical harmonics, $\chi$ is the comoving distance, $H(z)$ is the expansion rate at redshift $z$, $c$ is the speed of light, $k$ is the 3-dimensional wave vector, and $P_{gg}$ is the 3-dimensional galaxy power spectrum. 
Assuming that within the tomographic bin, the redshift evolution of galaxy bias is small, and all other functions can be approximated at the mean value at the centre of the bin, the clustering signal is proportional to the integral of the square of the galaxy redshift distribution, $p(z)$. This assumption breaks down if the tomographic bin width is broad, for instances, the combination of all five lens bins.
Hence, we define the following quantity:
\begin{equation}
    W_z \coloneqq \int \,p^2(z)\,dz
    \label{eq: wz}
\end{equation}
as the LSS diagnostic metric, which corresponds to changes of the two-point angular power spectrum kernel with respect to changes in $p(z)$.
 This is a useful complement to the second moment, $\sigma_z$, because $\sigma_z$ can be sensitive to the tails of the $p(z)$ distribution caused by a small population of outliers in photo-$z$; however, the impact of this population could be small for galaxy clustering, which is characterised by $W_z$. For both of these quantities, we look at the ratio with the overall sample combining all depth quantiles. 
We show all the mean metric quantities in each tomographic bin and each quantile for Y1, Y3, and Y5 in Table~\ref{tab: mean metrics bpz} for BPZ and Table~\ref{tab: mean metrics fzb} for FZBoost.

Notice that for the $p(z)$-related quantities, we have used the \textit{true} redshifts, but in practice, these are not accessible. Rather, unless one uses a Bayesian hierarchical model such as CHIPPR \citep{malzhogg}, one only has access to the \textit{calibrated} redshift distribution $p_c(z)$ against some calibration samples via, e.g. a Self-Organizing Map (SOM), which is itself associated with bias and uncertainties that can be impacted by varying depth.
The case we present here thus is idealized, where the calibration produces the perfect true $p(z)$. This allows us to propagate the actual impact of varying depth on $p(z)$ to the $3\times2$pt data vector, but does not allow us to assess the bias at the level of modelling due to using an `incorrect' $p_c(z)$ that is affected also by the varying depth. 
We leave this more sophisticated case to future work.

\section{Results}
\label{sec: Results}

This section presents our results on the impact of variable depth via three metrics: the number of objects (Section~\ref{sec: Number of objects}), mean redshift of the tomographic bin (Section~\ref{sec: Mean redshift}), and the width of the tomographic bin (Section~\ref{sec: Width of the tomographic bin}).

\begin{figure*}
     \centering
     \begin{subfigure}[b]{0.47\textwidth}
         \centering
         \includegraphics[width=\textwidth]{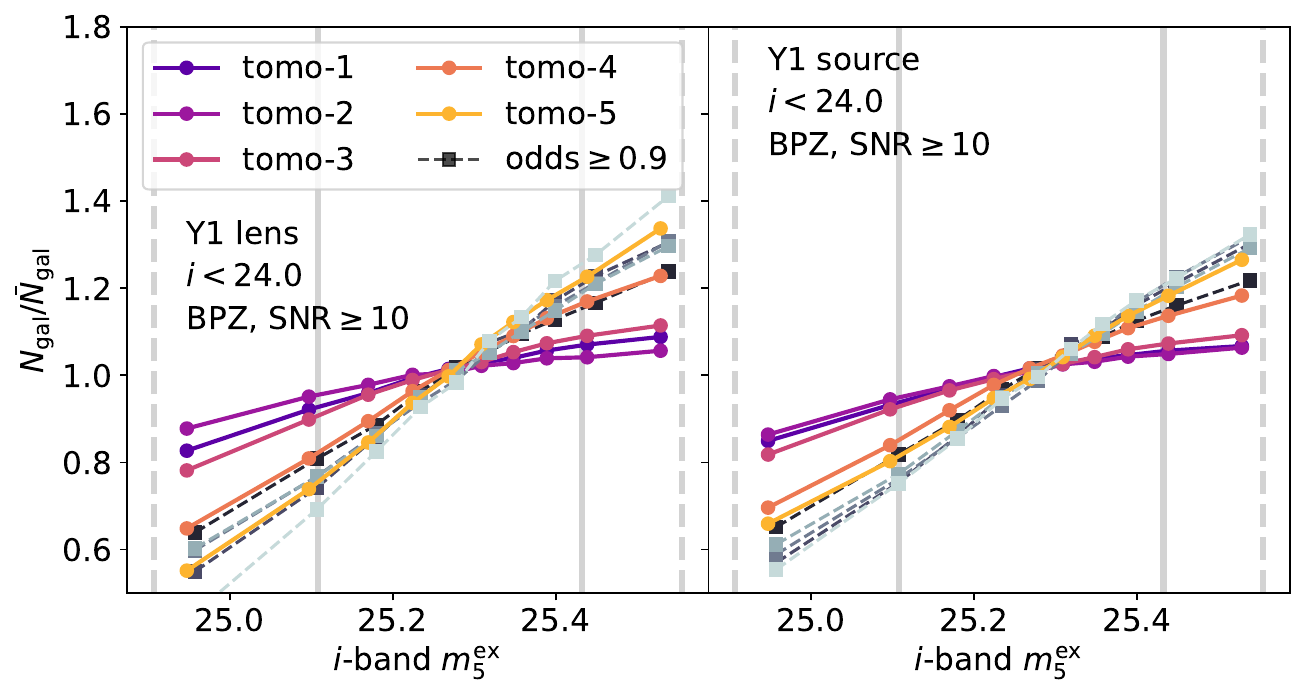}
     \end{subfigure}
      \hfill
     \begin{subfigure}[b]{0.47\textwidth}
         \centering
         \includegraphics[width=\textwidth]{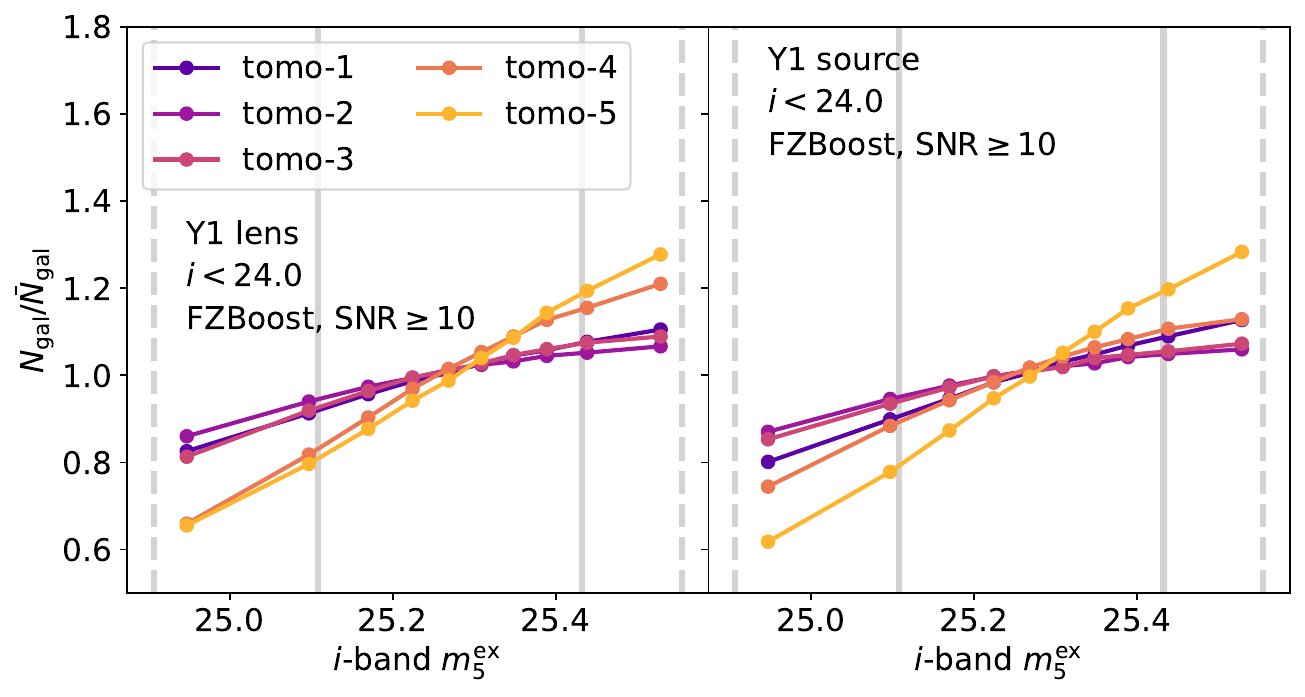}
     \end{subfigure}
     \hfill
     \begin{subfigure}[b]{0.47\textwidth}
         \centering
         \includegraphics[width=\textwidth]{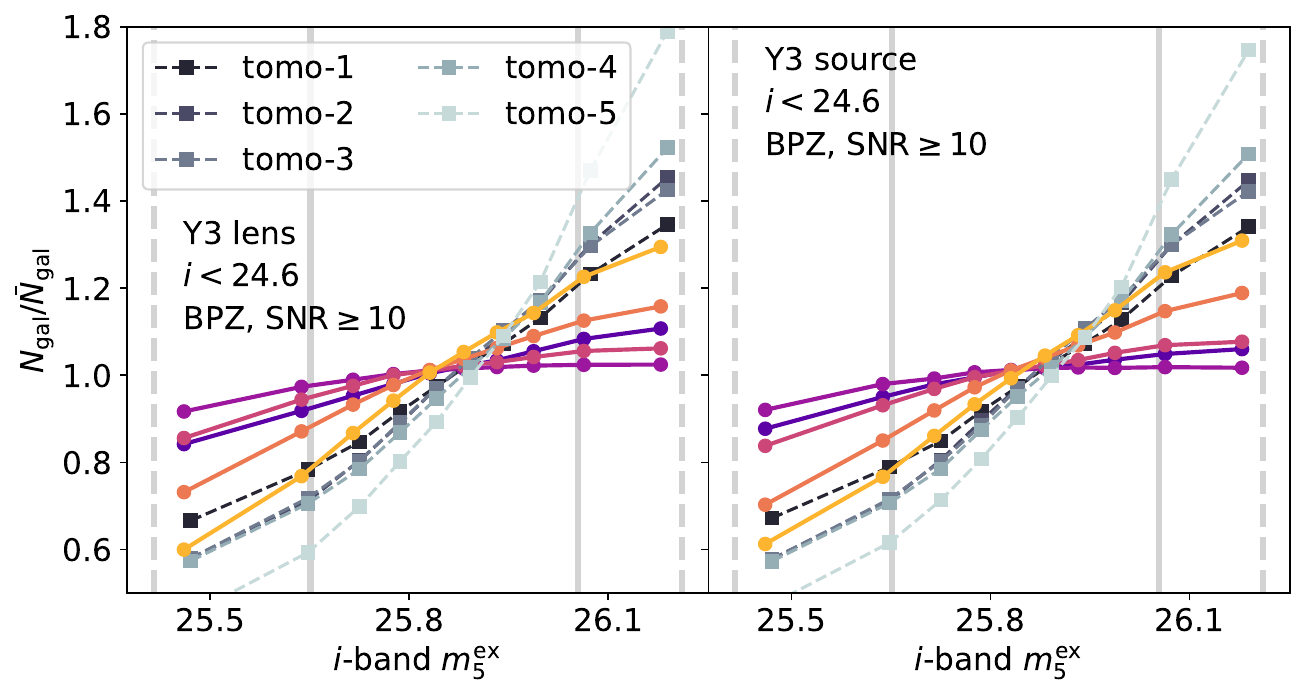}
     \end{subfigure}
     \hfill
     \begin{subfigure}[b]{0.47\textwidth}
         \centering
         \includegraphics[width=\textwidth]{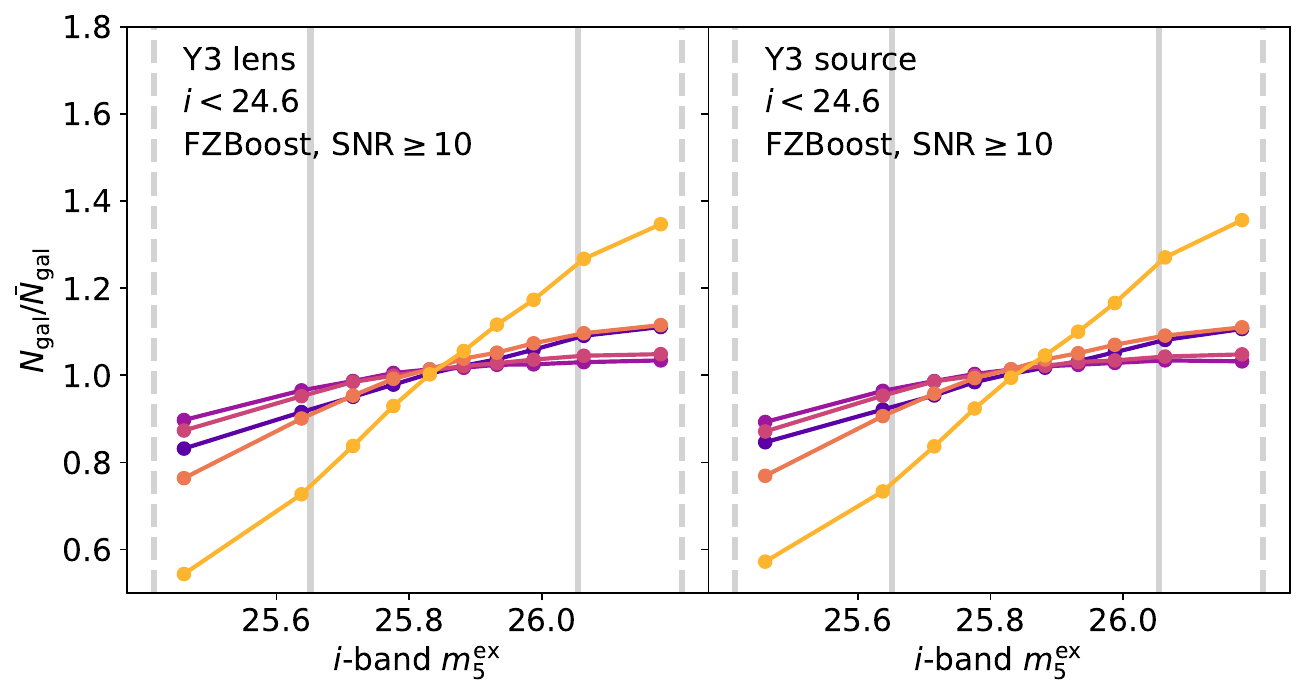}
     \end{subfigure}
     \hfill
     \begin{subfigure}[b]{0.47\textwidth}
         \centering
         \includegraphics[width=\textwidth]{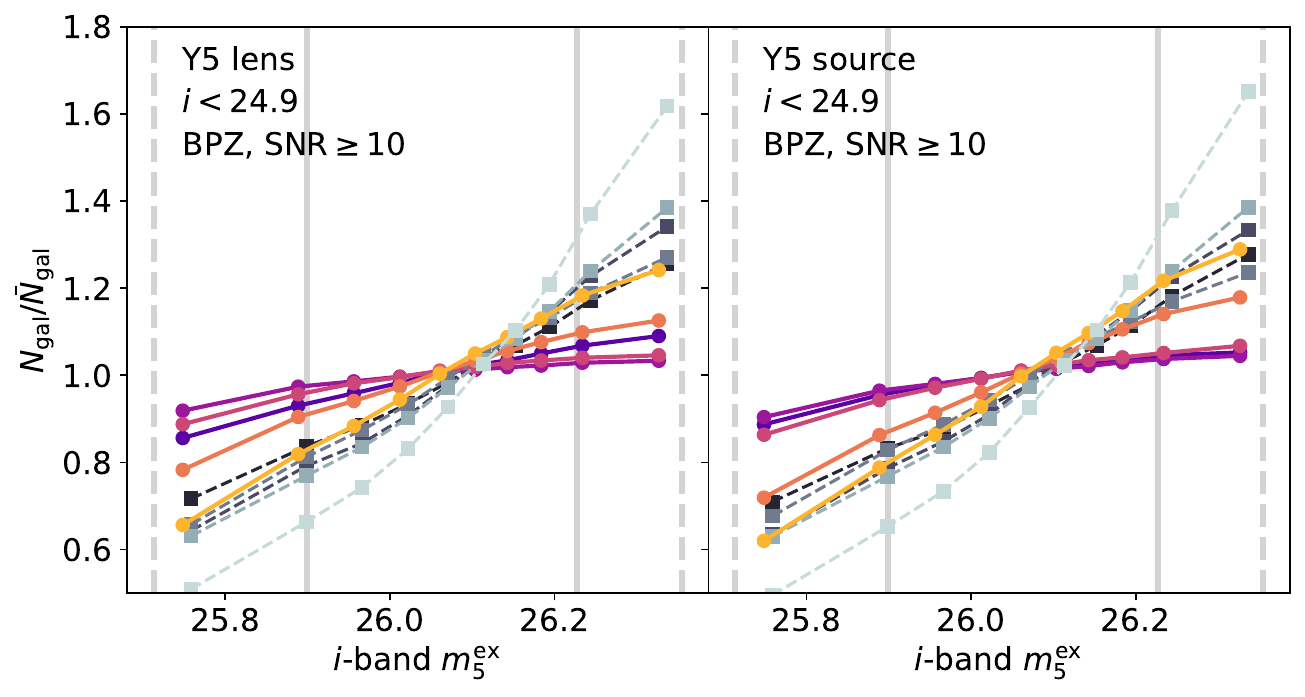}
     \end{subfigure}
     \hfill
     \begin{subfigure}[b]{0.47\textwidth}
         \centering
         \includegraphics[width=\textwidth]{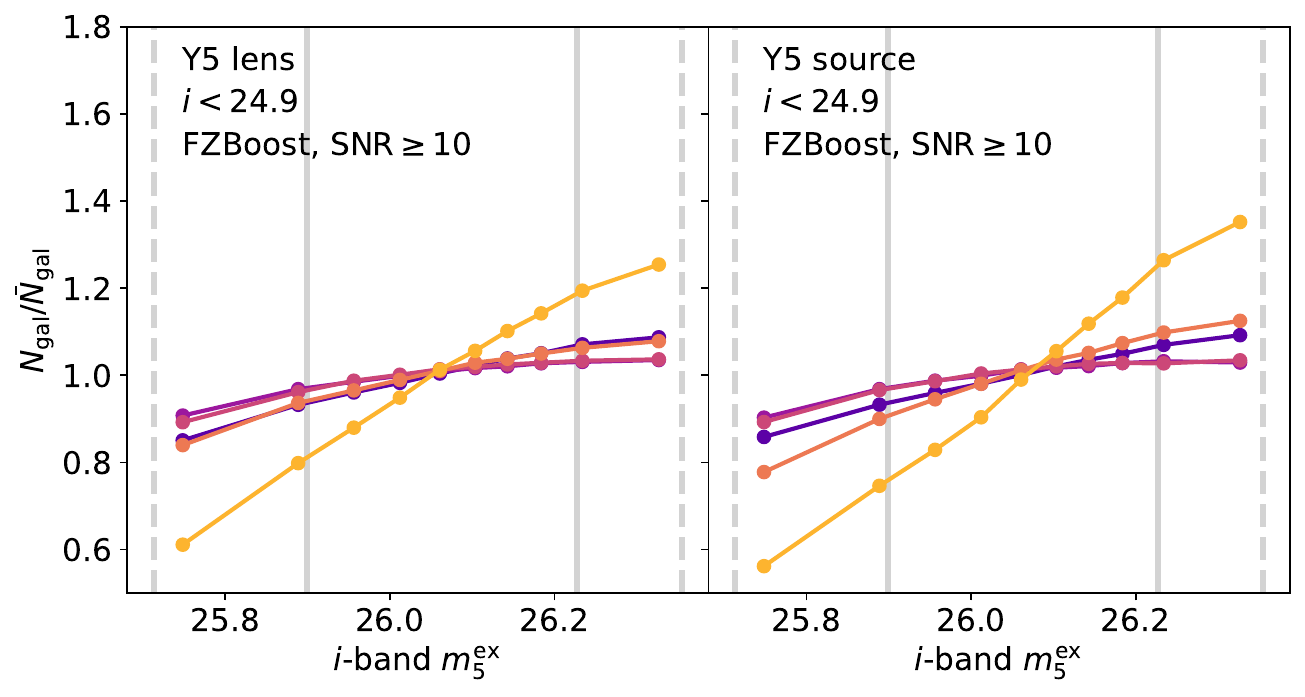}
     \end{subfigure}
        \caption{The number of galaxies in tomographic bins as a function of the $i$-band extinction-corrected coadd depth, $m_5^{\rm ex}$, for Y1, Y3, and Y5. The number is normalised by the average number of objects combing all quantiles for each tomographic bin, $\bar{N}_{\rm gal}$. The tomographic bins are determined using the mode of BPZ redshifts (left two columns) and FZBoost (right two columns). For each redshift estimator, both lens and source galaxy samples are shown, with the gold cut and ${\rm SNR}\geq10$. In the BPZ case, we also show the sample with ${\rm odds}\geq 0.9$ in squares with dashed lines. The vertical solid and dashed lines marks the $1\sigma$ and $2\sigma$ regions of the depth distribution.}
        \label{fig: ngal-baselinev2-y1-y5-goldcut-snr-10}
\end{figure*}

\begin{figure*}
     \centering
     \begin{subfigure}[b]{0.47\textwidth}
         \centering
         \includegraphics[width=\textwidth]{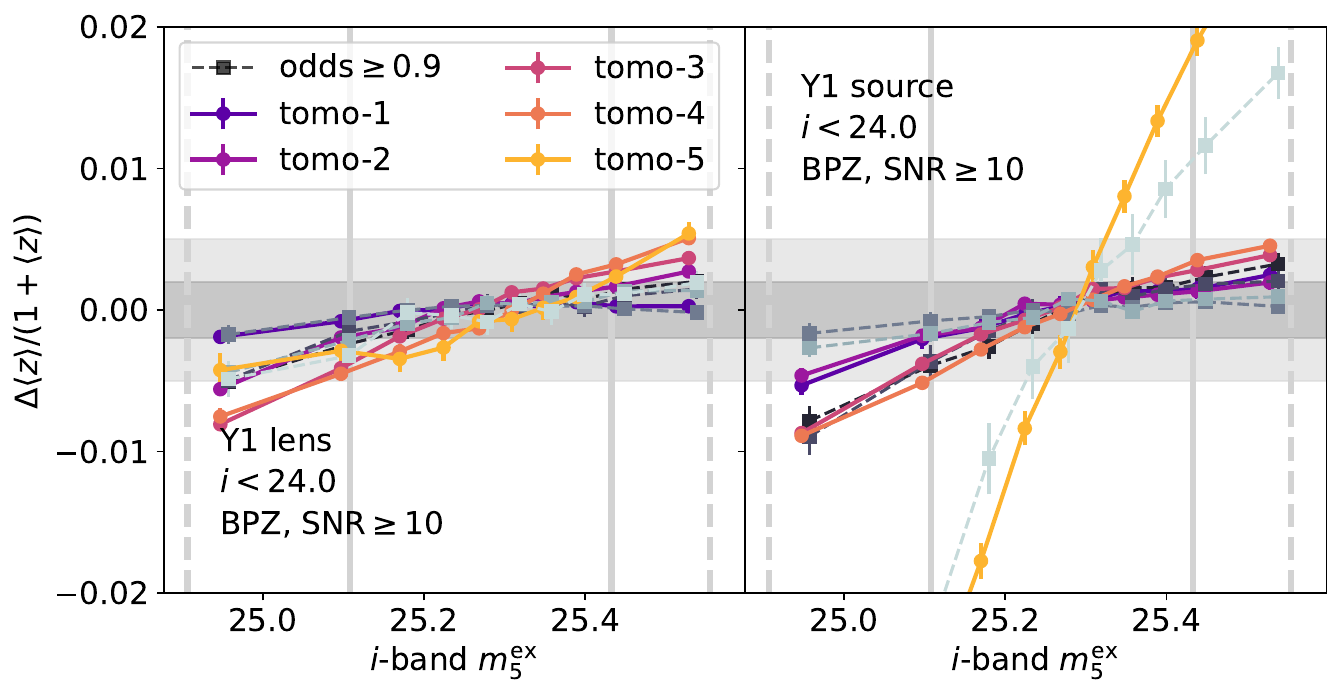}
     \end{subfigure}
      \hfill
     \begin{subfigure}[b]{0.47\textwidth}
         \centering
         \includegraphics[width=\textwidth]{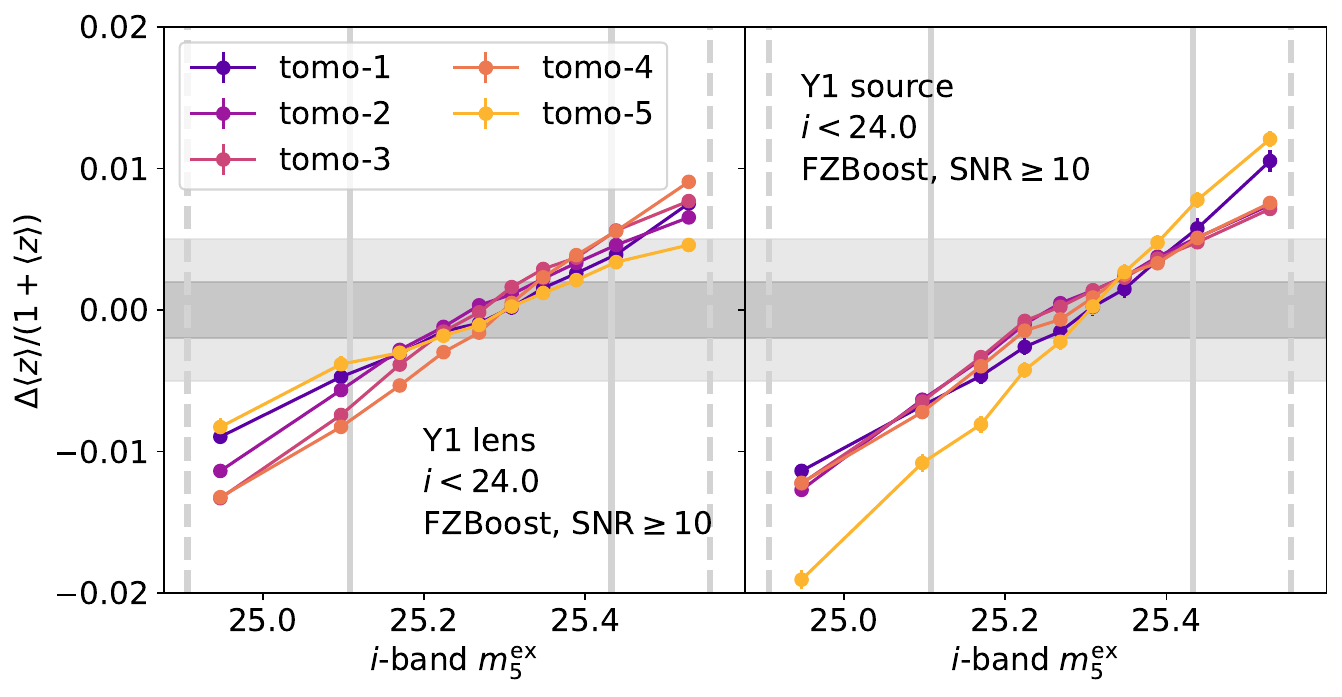}
     \end{subfigure}
     \hfill
     \begin{subfigure}[b]{0.47\textwidth}
         \centering
         \includegraphics[width=\textwidth]{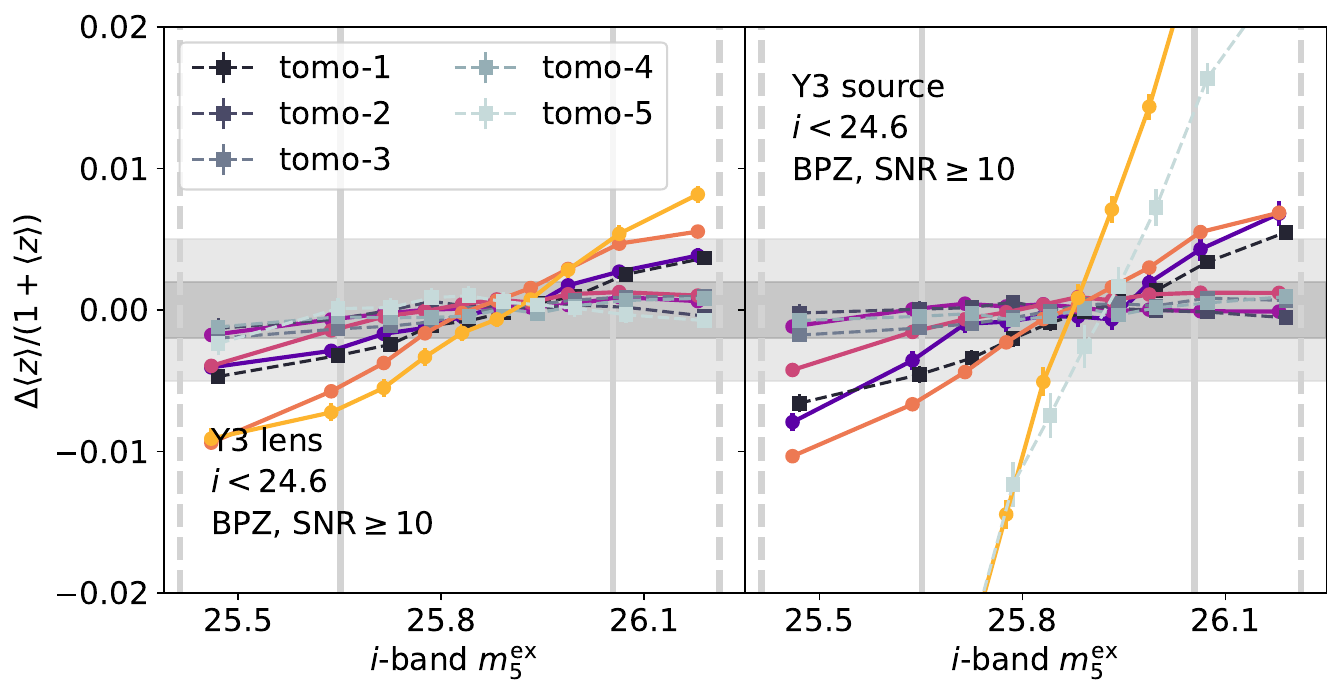}
     \end{subfigure}
     \hfill
     \begin{subfigure}[b]{0.47\textwidth}
         \centering
         \includegraphics[width=\textwidth]{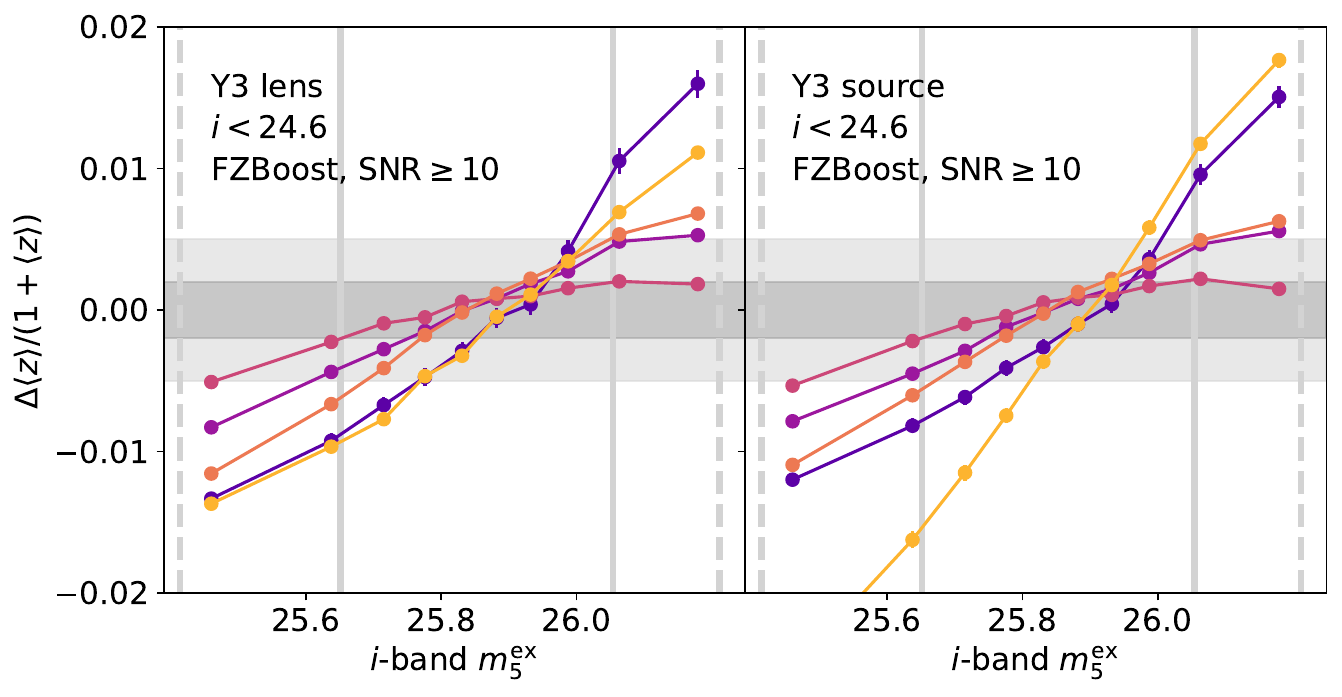}
     \end{subfigure}
     \hfill
     \begin{subfigure}[b]{0.47\textwidth}
         \centering
         \includegraphics[width=\textwidth]{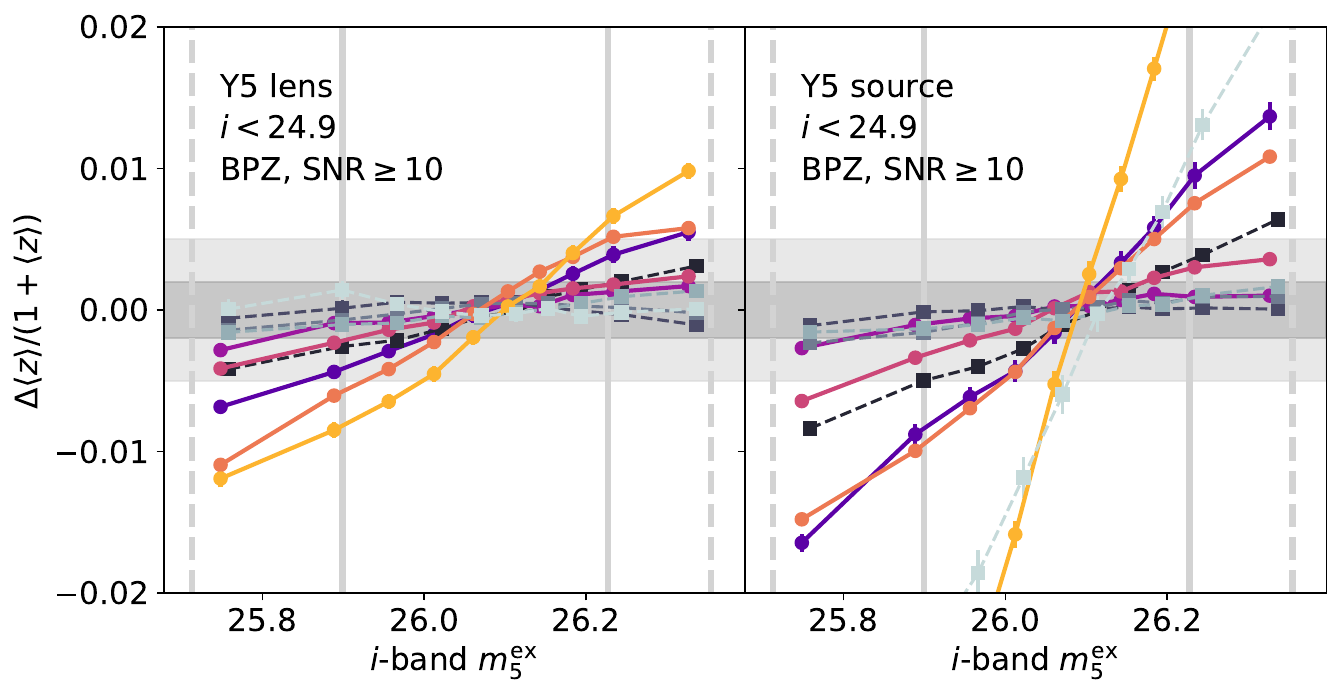}
     \end{subfigure}
     \hfill
     \begin{subfigure}[b]{0.47\textwidth}
         \centering
         \includegraphics[width=\textwidth]{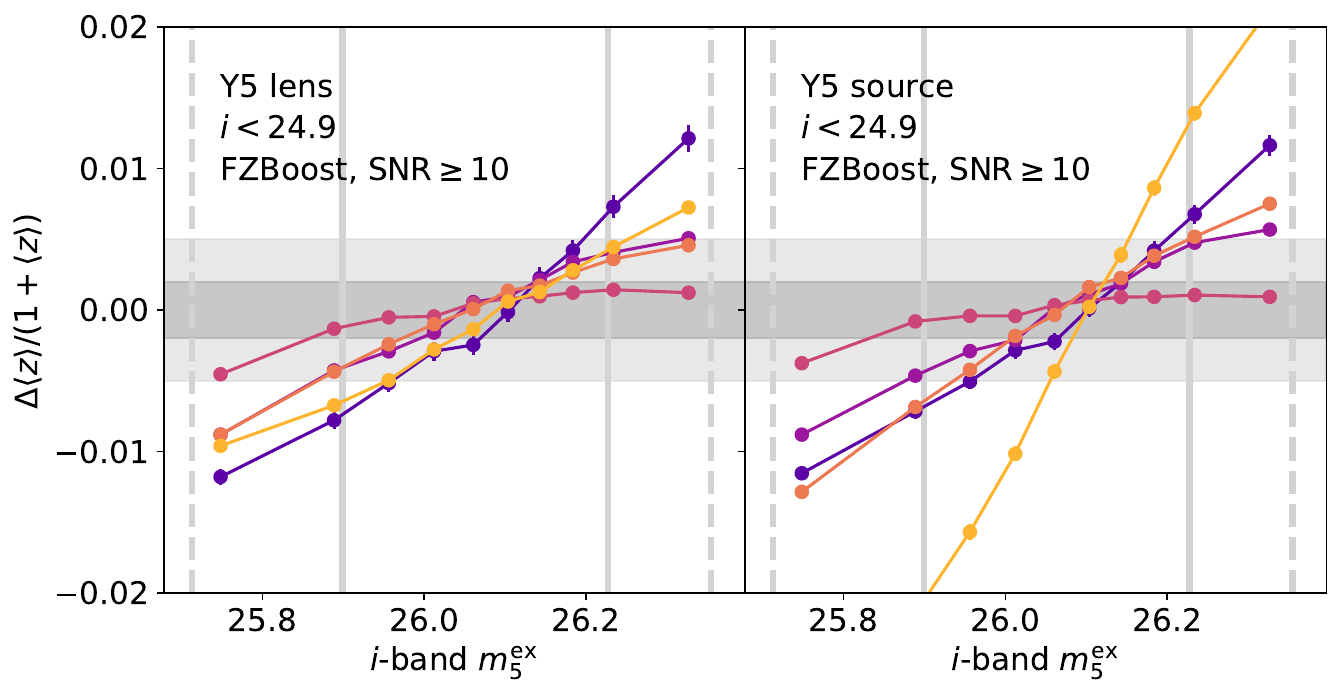}
     \end{subfigure}
        \caption{The change in mean redshift in each tomographic bin as a function of the $i$-band extinction-corrected coadd depth, $m_5^{\rm ex}$, for Y1, Y3, and Y5. The difference in mean redshift, $\Delta z$, between a given quantile and the combined sample $\langle z \rangle$, is normalised by $1/(1+\langle z \rangle)$ to account for expected larger uncertainties at higher redshifts. The fainter and darker grey bands marks $\pm 0.005$ and $\pm 0.002$, corresponding to the DESC SRD requirements for Y1 large-scale structure and weak lensing science. The tomographic bins are determined using the mode of BPZ redshifts (left two columns) and FZBoost (right two columns). For each redshift estimator, both lens and source galaxy samples are shown, with the gold cut and ${\rm SNR}\geq10$. In the BPZ case, we also show the sample with ${\rm odds}\geq 0.9$ in squares with dashed lines. The vertical solid and dashed lines marks the $1\sigma$ and $2\sigma$ regions of the depth distribution.}
        \label{fig: delta-meanz-baselinev2-y1-y5-goldcut-snr-10}
\end{figure*}

\begin{figure*}
     \centering
     \begin{subfigure}[b]{0.47\textwidth}
         \centering
         \includegraphics[width=\textwidth]{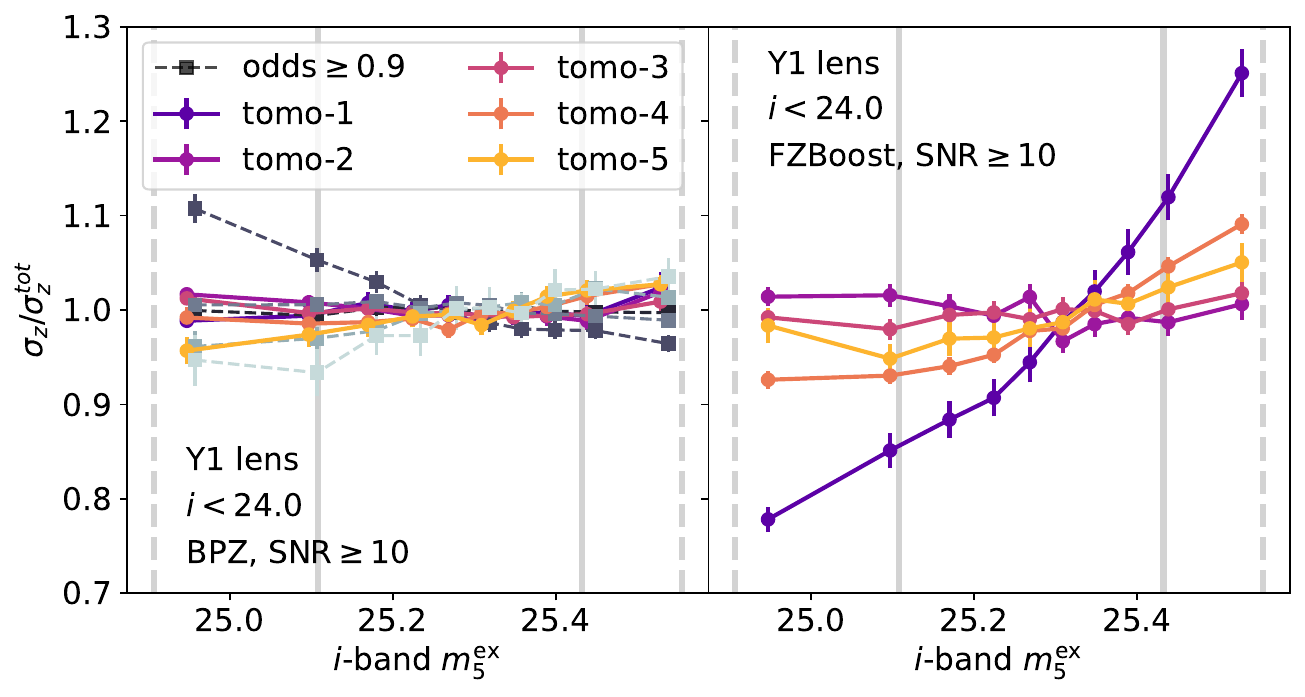}
     \end{subfigure}
      \hfill
     \begin{subfigure}[b]{0.47\textwidth}
         \centering
         \includegraphics[width=\textwidth]{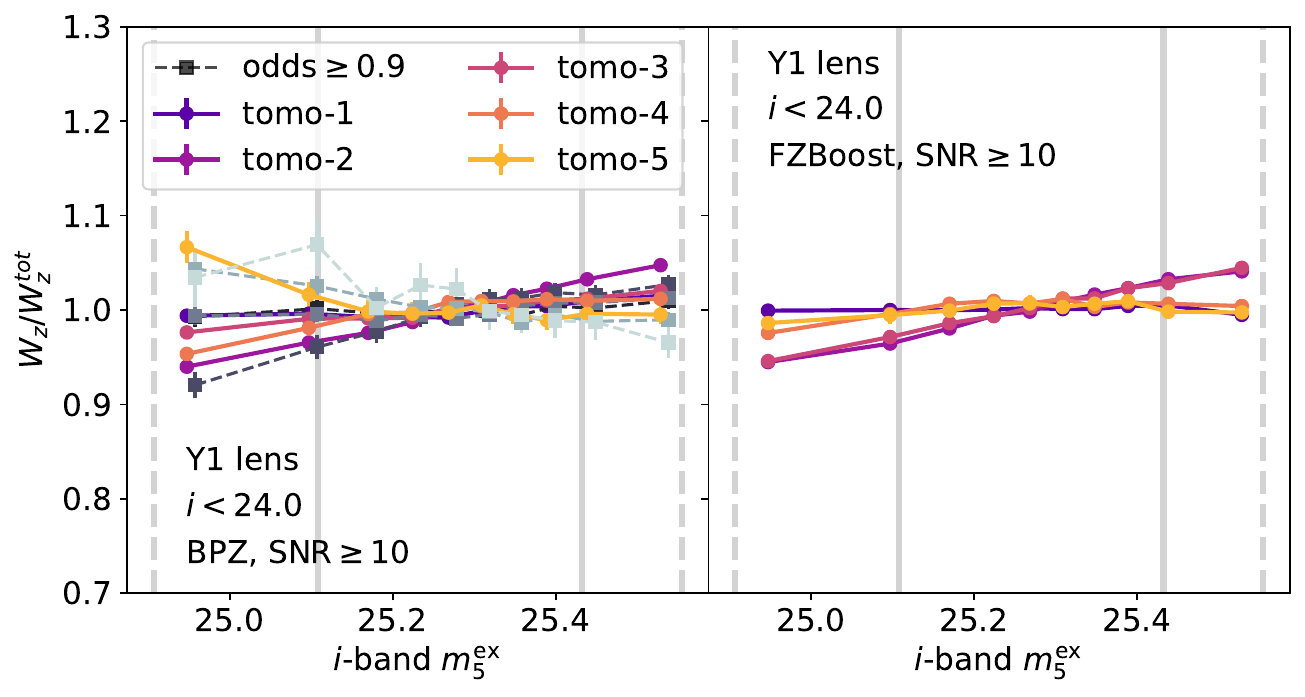}
     \end{subfigure}
     \hfill
     \begin{subfigure}[b]{0.47\textwidth}
         \centering
         \includegraphics[width=\textwidth]{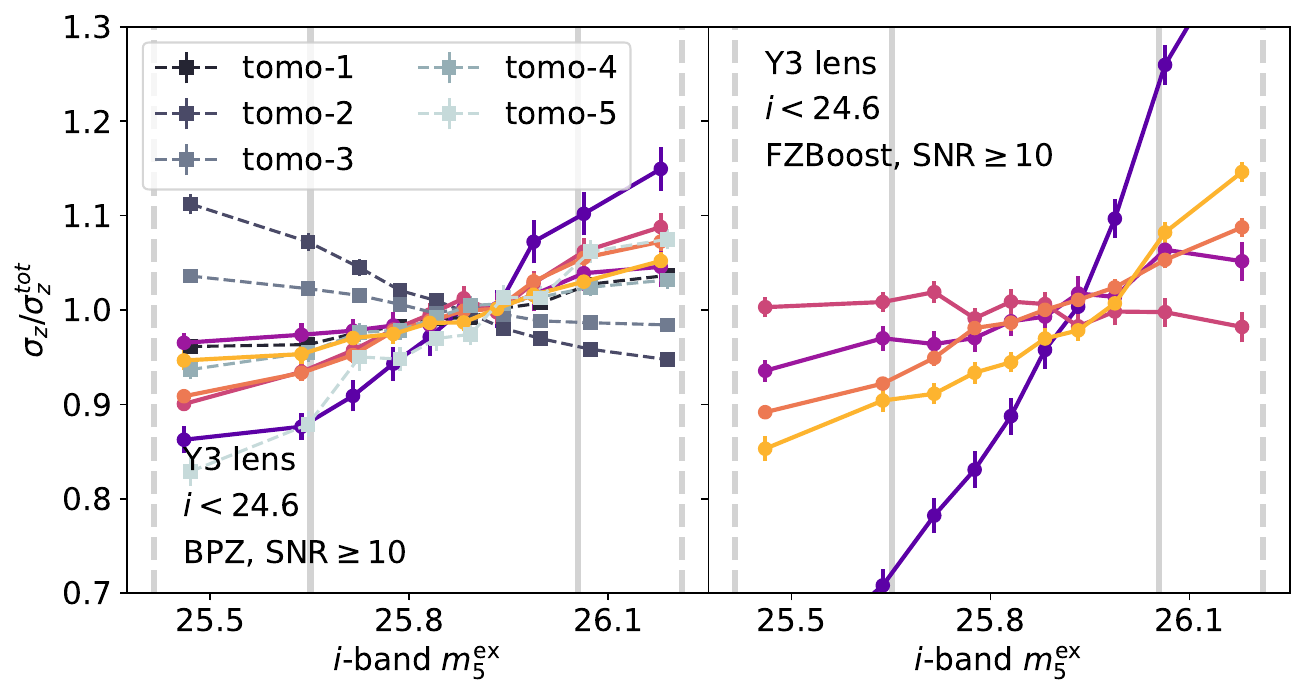}
     \end{subfigure}
     \hfill
     \begin{subfigure}[b]{0.47\textwidth}
         \centering
         \includegraphics[width=\textwidth]{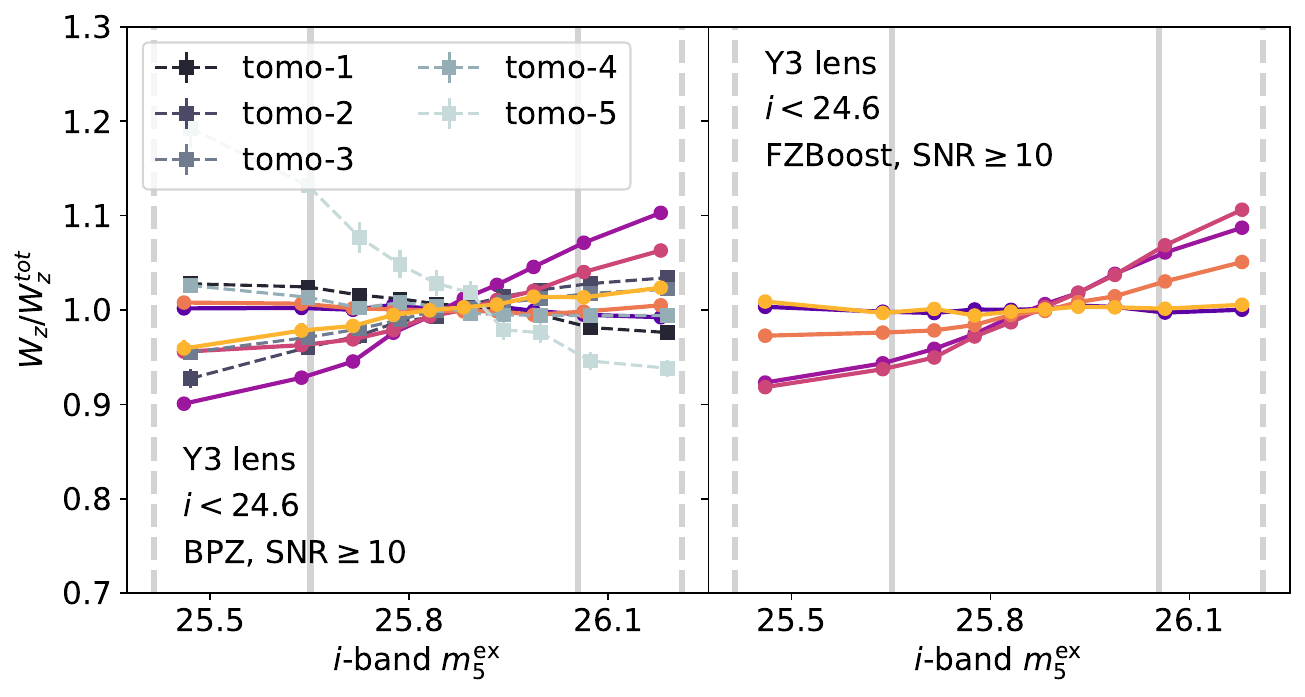}
     \end{subfigure}
     \hfill
     \begin{subfigure}[b]{0.47\textwidth}
         \centering
         \includegraphics[width=\textwidth]{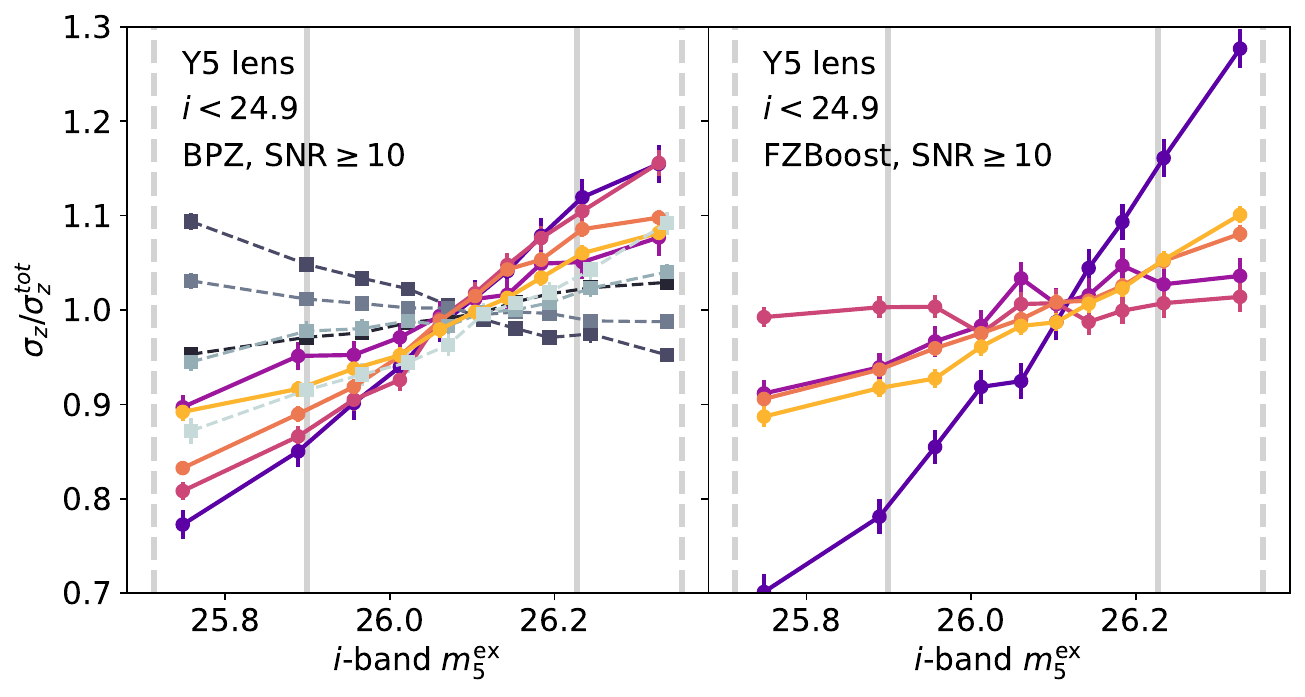}
     \end{subfigure}
     \hfill
     \begin{subfigure}[b]{0.47\textwidth}
         \centering
         \includegraphics[width=\textwidth]{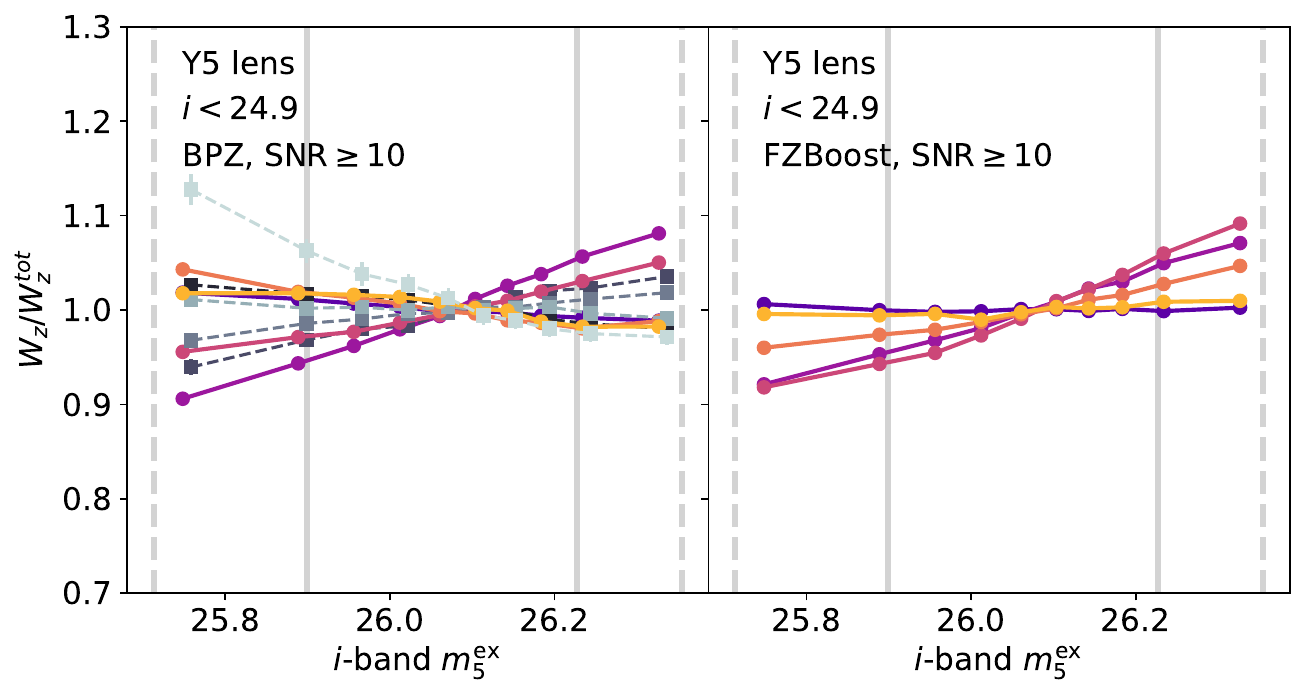}
     \end{subfigure}
        \caption{The relative change in the width of the lens tomographic bin as a function of the $i$-band extinction-corrected coadd depth, $m_5^{\rm ex}$, for Y1, Y3, and Y5. The left two columns show the second moment of the normalised redshift distribution, $\sigma_z$, in each quantile normalized by that of all quantiles combined, $\sigma_z^{\rm tot}$, for each tomographic bin. The right two columns show the LSS diagnostic parameter, $W_z$, as defined in Eq.~\ref{eq: wz}, for each quantile normalised by all quantiles combined $W_z^{\rm tot}$. The left and right panels for each width parameter show results with BPZ and FZBoost respectively. In the case of BPZ, the subsample with selection ${\rm odds}\geq 0.9$ is shown in squares with dashed lines. The vertical solid and dashed lines marks the $1\sigma$ and $2\sigma$ regions of the depth distribution.}
        \label{fig: delta-wz-baselinev2-y1-y5-goldcut-snr-10}
\end{figure*}

\subsection{Number of objects}
\label{sec: Number of objects}

Figure~\ref{fig: ngal-baselinev2-y1-y5-goldcut-snr-10} shows the change in the number of objects, $N_{\rm gal}$, as a function of the $i$-band extinction-corrected coadd depth, $m_5^{\rm ex}$, compared to the overall mean, for lens and source tomographic bins in Y1, Y3, and Y5.
In general, we find an approximately linear increase of number of objects as the $i$-band depth increases, with the higher two redshift bins showing the most extreme variation. For the lower redshift bins, the variation can be $\sim 10\%$ compared to the mean value, whereas for bin 5, the variation can be as large as $\sim 40\%$. The trend does not seem to change much at different observing years. 
This is the result of the $i$-band gold cut and the high SNR selection. The scatter in magnitudes is larger for the shallower sample, hence given a magnitude cut, the shallower sample will have fewer objects. 
At fixed magnitude, the deeper objects have larger SNR, resulting in more faint galaxies surviving the SNR cut.
Given that the gold cut and SNR at given magnitude evolve with depth in the observation year, we expect the trend to be similar across Y1 to Y5. 
It is interesting to see also that per tomographic bin, the trends for baseline BPZ and FZBoost are similar, despite having quite different features in the photo-$z$ vs spec-$z$ plane. The variation between bins 1 - 4 is slightly larger in the BPZ case.
For the BPZ redshifts, the inclusion of the odds selection increases the variation in object number, especially in the highest redshift bin. The steeper slope might be due to the fact that, objects with larger photometric error from the shallower regions are likely to result in a poorer fit, leading to a smaller ${\rm odds}$ value. Hence, the ${\rm odds}\geq0.9$ selection removes more objects from the shallower compared to the baseline case.


\subsection{Mean redshift}
\label{sec: Mean redshift}

Figure~\ref{fig: delta-meanz-baselinev2-y1-y5-goldcut-snr-10} shows the variation in the mean redshift of the tomographic bin, $\langle z \rangle$, as a function of the $i$-band extinction-corrected coadd depth, $m_5^{\rm ex}$, for lens and source samples in Y1, Y3, and Y5. 
In general, $\langle z \rangle$ increases with the $i$-band coadd depth. This is expected as more faint, high redshift galaxies that are scattered within the magnitude cut are included in the deeper sample, resulting in an increased high redshift population.
In general, the slope of this relation is similar across tomographic bins for both lens and source samples, with a variation of $|\Delta z /(1+\langle z \rangle)|\sim0.005 - 0.01$. 
This is not true for bin 5 in the source sample, where the variation with depth is noticeably larger. 
This could be explained by this bin containing objects with the highest $z_{\rm phot}$, which are also most susceptible to scatter in the faint end and outliers in the photo-$z$ estimators. This trend becomes more extreme from Y1 to Y5. By reducing outliers with the BPZ odds cut, the variation in source bin 5 is slightly reduced, although still higher than the nominal level.
There are some difference between the BPZ and FZBoost cases: the slope slightly grows from Y1 to Y5 in the BPZ case, whereas it stays consistent in the FZBoost case, but the two cases converge in Y5.
On the same figure, we mark the DESC SRD requirements for photo-$z$ as a dark grey band at $\Delta z /(1+\langle z \rangle)=\pm0.002$ and a light grey band at $\Delta z /(1+\langle z \rangle)=\pm0.005$. 
The shifts in mean redshift reach the limit of the requirements for Y1, and exceeds the requirement for Y10.


\subsection{Width of the tomographic bin}
\label{sec: Width of the tomographic bin}

Figure~\ref{fig: delta-wz-baselinev2-y1-y5-goldcut-snr-10} shows the change in the tomographic bin width parameters, $\sigma_z$ and $W_z$, as defined in Section~\ref{sec: Metrics for impacts of varying depth} for the lens galaxies as a function of the $i$-band extinction-corrected coadd depth, $m_5^{\rm ex}$, in Y1, Y3, and Y5. 
The width of the tomographic bin can change with depth due to the scatter in the photo-$z$ vs spec-$z$ plane. For example, a deeper sample may have a smaller scatter for the bulk of the sample, but include fainter objects that could result as outliers, resulting a more peaked distribution at the centre with pronounced long tails.

The left two columns of Fig.~\ref{fig: delta-wz-baselinev2-y1-y5-goldcut-snr-10} show the changes in the second moment, $\sigma_z$, for both the BPZ (first column) and FZBoost case (second column).
For BPZ, there is little change in this parameter for Y1 at different depth, but for Y3 and Y5, $\sigma_z$ increases with depth. Including odds selection reduces the trend, and in some cases reverses it. For FZBoost, the trend is similar to BPZ, but bin 1 shows a particularly large variation by as much as $\sim30\%$. 
This is because $\sigma_z$ is sensitive to the entire distribution, not just the peak, and outliers at high redshift can significantly impact this parameter. Fig.~\ref{fig: nz-baselinev3.3-y3-long-logy} shows same $p(z)$ distributions for Y3 in logarithmic scale, where the high redshift outliers are visible. Indeed, one can see an enhanced high-redshift population for bin 1 in the FZBoost case. The odds cut removes most of the outliers, so that $\sigma_z$ is reflecting the change of the peak width with depth, hence giving the reversed trend. 

The right two columns of Fig.~\ref{fig: delta-wz-baselinev2-y1-y5-goldcut-snr-10} show the changes in $W_z$. Given a tomographic bin, a larger $W_z$ means a more peaked redshift distribution, hence a larger clustering signal. 
One can see that $W_z$ is more sensitive to the bulk of the $p(z)$ distribution, as it increases with depth in most bins. We see that the variation in $W_z$ is within 10\% from the mean, with the largest variation coming from bins 2, 3, and 4. The highest and lowest tomographic bins, on the other hand, does not change much, despite their $\sigma_z$ varying significantly with depth. For the BPZ case, adding the additional cut in the odds parameter reduces such trends in general, and the trend in the highest tomographic bin is reversed. 


\section{Impact on the weak lensing \texorpdfstring{$3\times2$}{3x2}pt measurements}
\label{sec: cosmology}

We use the Y3 FZBoost photo-$z$ as an example to showcase the varying depth effects, by propagating the number density and $p(z)$ variation from the previous section into the weak lensing $3\times2$pt data vector. In Section~\ref{sec: Mock maps with varying depth}, we describe how the mock large-scale structure and weak lensing shear maps are constructed with the inclusion of non-uniformity. In Section~\ref{sec: 3x2pt}, we show case the measured $3\times2$pt data vector in both uniform and variable depth case.


\subsection{Mock maps with varying depth}
\label{sec: Mock maps with varying depth}

To construct the mock LSST catalogue, we use one of the publicly available Gower Street Simulations \citep{jeffrey2024dark}. This is a suite of 800 N-body cosmological simulations created using PKDGRAV3 \citep{potter2016pkdgrav3} with various $w$CDM cosmological parameters. The simulation outputs are saved as 101 lightcones in HEALPix format with $N_{\rm side}=2048$ between $0<z<49$. To fill the full sky, the boxes are repeated 8000 times in a $20\times20\times20$ array. For shells $z<1.5$, though, only three replications are required.
We use the particular simulation with $\Lambda$CDM cosmology: $w=-1$, $h=0.70$, $\Omega_m=0.279$, $\Omega_b=0.046$, $\sigma_8=0.82$, and $n_s=0.97$. The dark matter density contrast map, $\delta_m$, is computed using particle counts at $N_{\rm side}=512$ (corresponding to a pixel size of $47.2 \, {\rm arcmin}^2$), and the corresponding lensing convergence map, $\kappa$, is produced with Born approximation using \texttt{BornRayTrace}\footnote{\url{https://github.com/NiallJeffrey/BornRaytrace}} \citep{Bornraytrace2020}. Finally, the shear map, $(\gamma_1,\gamma_2)$ in spherical harmonic space, is produced via
\begin{equation}
    \gamma_{E,\ell m} = \frac{\kappa_{E, \ell m}}{\ell(\ell + 1)\sqrt{(\ell + 2)(\ell - 1)}},
\end{equation}
and we transform $\gamma_{E,\ell m}$ as a spin-2 field, $\gamma_{\ell m}=\gamma_{E,\ell m} + i \gamma_{B,\ell m}$, assuming zero B-mode. For more details see \cite{jeffrey2024dark}.

We construct the lens and source shear maps as follows. In the noiseless case, given a lens (source) redshift distribution, $p_i(z)$, for a tomographic bin $i$, we construct the lens density (source shear) map by $M_i= \sum_j M_j  p_i(z_j)\Delta z_j$, where $j$ denotes the lightcone shells in the Gower street simulation, $M_j$ denotes the map in this particular shell, and $\Delta z_j$ denotes the shell width. 
The noisy maps are generated in the following way.
Lens galaxy counts in tomographic bin $i$ on each pixel $\boldsymbol{\theta}$ are drawn from a Poisson distribution. For a shell $j$, the Poisson mean is $\mu_j(\boldsymbol{\theta})=n_{{\rm gal},j}[1+b\delta_{m,j}(\boldsymbol{\theta})]$, where $b$ is the linear galaxy bias and $n_{{\rm gal},j}=n_{\rm gal}p_i(z_j)\Delta z_j$, with $n_{\rm gal}$ being the average count per pixel in this tomographic bin. Here, we set $b=1$ to avoid negative counts in extremely underdens pixels. 
However, notice that in a magnitude-limited survey, the galaxy bias is typically $b>1$ and evolves with redshift, not to mention the scale-dependence of bias on non-linear scales. 
One approach to sample $b>1$ is to simply set negative counts to zero. However, this may introduce spurious behaviour in the two-point function of the field.
Given the main purpose here is to propagate the systematic effects due to depth only, we justify our choice by prioritizing the precision of the measured two-point statistics compared to theory inputs.
We assume the ensemble-averaged per-component shape dispersion to be $\sigma_e=\left\langle \sqrt{(e_1^2+e_2^2)/2} \right\rangle=0.35$, chosen to roughly match that measured in the Stage III lensing surveys \citep[e.g.][]{2021A&A...646A.129J,2021MNRAS.504.4312G,2022PASJ...74..421L}.
For a tomographic bin $i$, we first assign source counts in the same way as above, resulting in $\hat{n}_{\rm source} (\boldsymbol{\theta})$ galaxies in pixel $\boldsymbol{\theta}$. We then randomly assign shapes drawn from a Gaussian distribution, $\mathcal{N}\sim(0,\sigma_e)$, for each component $\hat{n}_{\rm source} (\boldsymbol{\theta})$ times, and we compute the mean shape noise in each pixel. We end up with a shape noise map, which we then add to the true shear map for each tomographic bin.

To imprint the varying depth effects, we divide the footprint into 10 sub-regions containing the pixels in each of the $i$-band $m_5^{\rm ex}$ deciles, and repeat the above procedure with distinct number density and $p(z)$ for both the lens and source galaxies, according to the findings in previous sections. We do not assign depth-varying shape noise, following the finding in \cite{2021A&A...646A.129J} that the shape noise is only a weak function of depth. 
We also produce the noiseless cases for varying depth. For density contrast, we produce two versions: one with varying $p(z)$ only, and one with additional amplitude modulation $\delta_m + \Delta \delta$, where $\Delta \delta+1=N_{\rm gal}/\bar{N}_{\rm gal}$, as shown in Fig.~\ref{fig: ngal-baselinev2-y1-y5-goldcut-snr-10}. The former is to used isolate the effect of varying $p(z)$ only.

We adopt the cumulative number density of the photometric sample as a function of the $i$-band limiting magnitude given by the DESC SRD:
\begin{equation}
    N(<i_{\rm lim})=42.9(1-f_{\rm mask})10^{0.359(i_{\rm lim}-25)}\, {\rm arcmin}^{-2},
\end{equation}
where $f_{\rm mask}$ accounts for the reduction factor for masks due to image defects and bright stars, and $f_{\rm mask}=0.12$ corresponds to a similar level of reduction in HSC Y1 \citep{thelsstdarkenergysciencecollaboration2021lsst}. 
Hence, substituting $i_{\rm lim}=24.6$ for LSST Y3, the expected total number density is $N(<24.6)=27.1\,{\rm arcmin}^{-2}$.
This is slightly larger but comparable to the HSC Y3 raw number density of $N=22.9 \,{\rm arcmin}^{-2}$ \citep{2022PASJ...74..421L} at a similar magnitude cut of $i_{\rm lim}<24.5$ in the cModel magnitude. 
We estimate the total lens galaxy number density for our sample by $N_{\rm lens}=N(<24.5)f_{\rm LS}$, where $f_{\rm LS}=0.90$ is the ratio between the total number of lens and source samples (averaged over depth bins) from our degraded Roman-Rubin simulation catalogue, hence $N_{\rm lens}=24.4 \,{\rm arcmin}^{-2}$. 
For each lens tomographic bin, we obtain the following mean number density: $3.93, 6.08, 5.66, 5.71, 3.03 \, {\rm arcmin}^{-2}$.
We also explore the case using a MagLim-like lens sample with a much sparser density in Appendix~\ref{sec: maglim}.
For source sample, it is the effective number density 
$n_{\rm eff}$, rather than the raw number density, that determines the shear signal-to-noise. $n_{\rm eff}$ accounts for the down-weighting of low signal-to-noise shape measurements, as defined in e.g. \cite{2012MNRAS.427..146H, 2013MNRAS.434.2121C}. For LSST, $n_{\rm eff}$ is estimated for Y1 and Y10 with different scenarios in Table F1 in the DESC SRD. 
In the case adopted for forecasting, where the shapes are measured in $i+r$ and accounting for blending effect, $n_{\rm eff}$ is $\sim60\%$ of the raw number density for both Y1 and Y10. 
We follow this estimation for Y3, hence adopting $n_{\rm eff}=16.3 \, {\rm arcmin}^{-2}$ for the full source sample, and $3.26 \,{\rm arcmin}^{-2}$ for each tomographic bin. This is comparable, but slightly more sparse compared to HSC Y3, where $n_{\rm eff}=19.9 \, {\rm arcmin}^{-2}$ \citep{2022PASJ...74..421L}.


Meanwhile, we also generate a uniform sample for comparison, in which the number density and $p(z)$ are given by the mean of the depth quantiles.
We assign uniform weights to lens and source galaxies.



\subsection{Weak lensing \texorpdfstring{$3\times2$}{3x2}pt data vector}
\label{sec: 3x2pt}

\begin{figure*}
   \centering
   \includegraphics[width=0.9\textwidth]{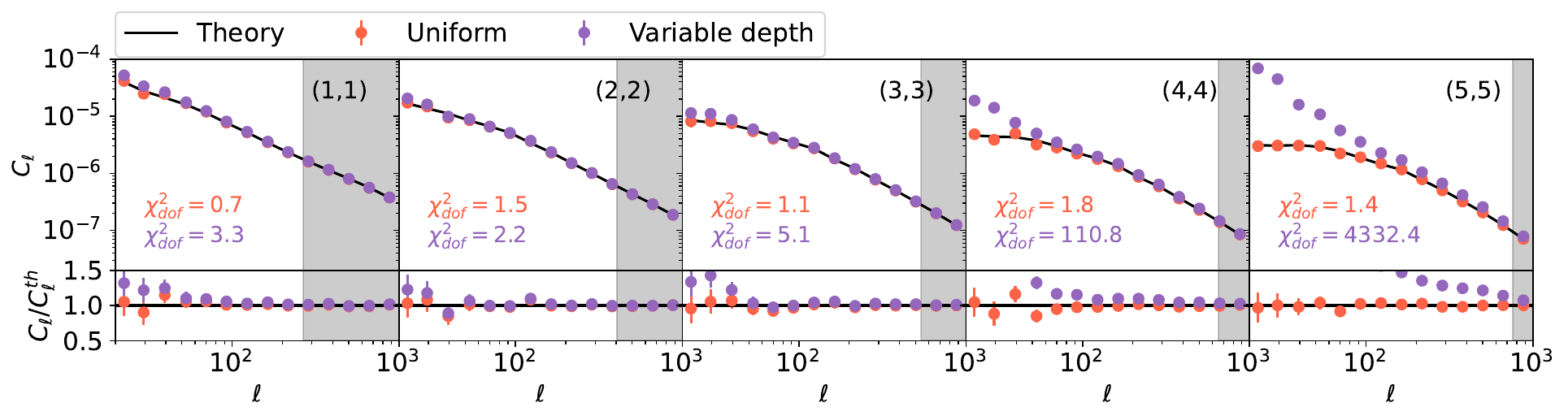}
    \caption{The lens galaxy density angular power spectrum, $C^{\rm gg}_{\ell}$, measured from the mock LSST Y3 data with uniform (red points) and varying depth (purple points). Each panel shows the auto-correlation, $(i,i)$, in each tomographic bin $i$. The lower panels show the ratio between the measurements and the theory (black solid lines), $C^{\rm th}_{\ell}$. The grey area indicates excluded data points from the scale cut corresponding to $k=0.3h {\rm Mpc}^{-1}$. The $\chi^2$ per degree of freedom, $\chi^2_{\rm dof}$, is shown for the uniform and variable depth cases in the lower left corner, computed using a Gaussian covariance assuming spatial uniformity.}
    \label{fig: clgg-y3-fzb}
\end{figure*}

\begin{figure*}
   \centering
   \includegraphics[width=0.9\textwidth]{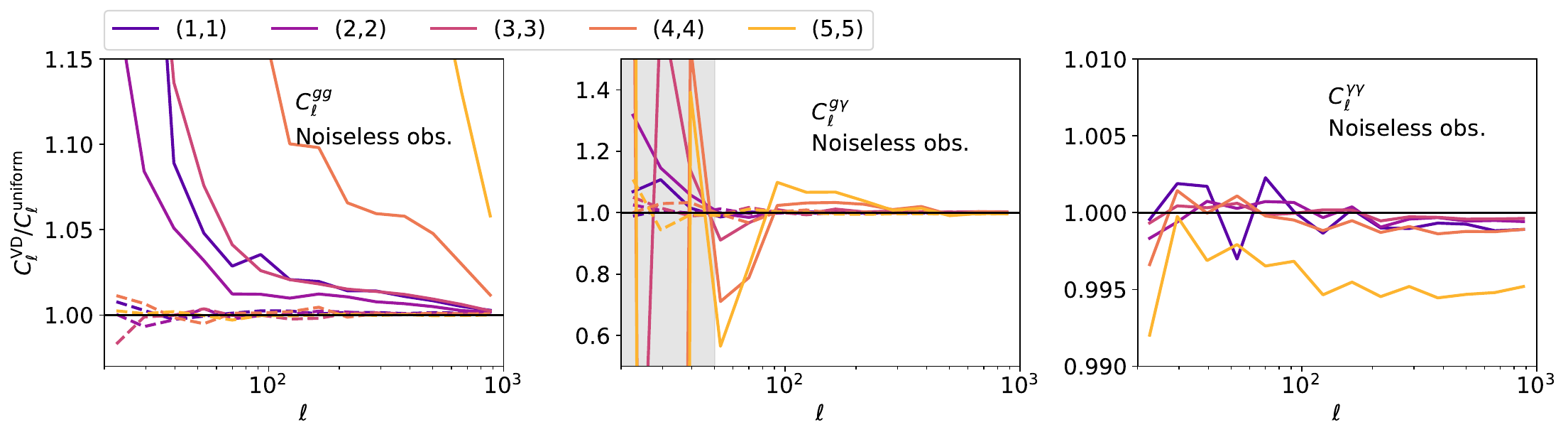}
    \caption{The ratio between noiseless angular power spectra for the varying depth case and the uniform case. The left, middle, and right panels show the ratio for $C_{\ell}^{\rm gg}$, $C_{\ell}^{{\rm g}\gamma}$, $C_{\ell}^{\gamma\gamma}$, respectively. The solid lines indicate the case where both density non-uniformity and varying $p(z)$ are applied to the over-density map, whereas the dashed lines refer to the case where only varying $p(z)$ is implemented. For $C_{\ell}^{{\rm g}\gamma}$ and $C_{\ell}^{\gamma\gamma}$, we only show the diagonal terms, i.e., the combination $(i,i)$ for tomographic bin $i$ for the tracers, for visual clarity. The off-diagonal terms vary within a similar range. In case of $C_{\ell}^{g\gamma}$, the grey region marks $\ell<50$ where measurements are unstable.}
    \label{fig: noiseless-ratio}
\end{figure*}

\begin{figure*}
   \centering
   \includegraphics[width=0.9\textwidth]{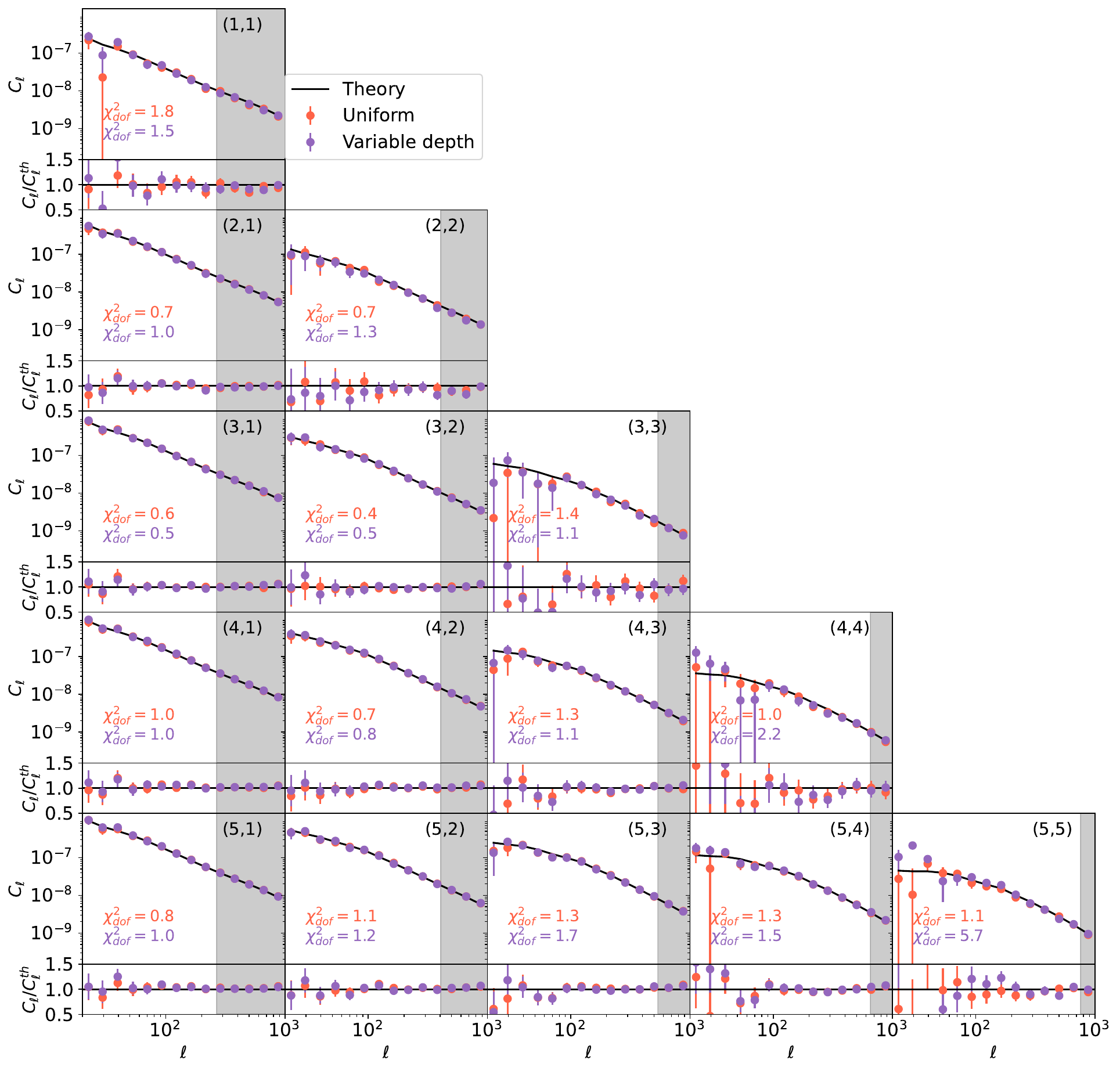}
    \caption{The E-mode of the galaxy-shear angular power spectrum, $C^{{\rm g}\gamma}_{\ell}$, measured from the mock LSST Y3 data with uniform (red points) and varying depth (purple points). Each panel shows the combination, $(i,j)$, for source bin $i$ and lens bin $j$. The lower panels show the ratio between the measurements and the theory (black solid lines), $C^{\rm th}_{\ell}$. The grey area indicates excluded data points from the scale cut corresponding to $k=0.3h {\rm Mpc}^{-1}$ in the lens bin. The $\chi^2$ per degree of freedom, $\chi^2_{\rm dof}$, is shown for the uniform and variable depth cases in the lower left corner, computed using a Gaussian covariance assuming spatial uniformity.}
    \label{fig: clge-y3-fzb}
\end{figure*}

\begin{figure*}
   \centering
   \includegraphics[width=0.9\textwidth]{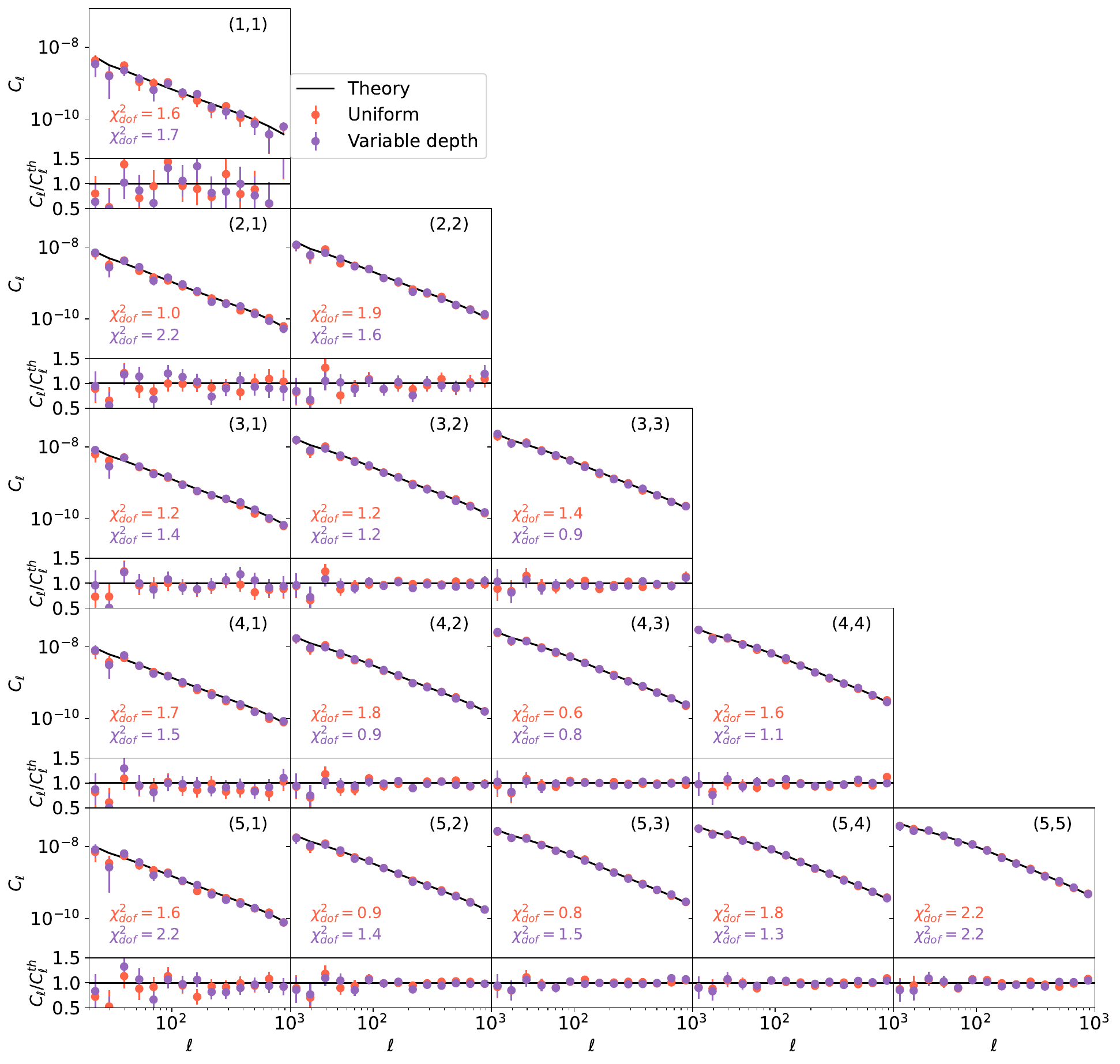}
    \caption{The EE-mode of the shear-shear angular power spectrum, $C^{\gamma\gamma}_{\ell}$, measured from the mock LSST Y3 data uniform (red points) and varying depth (purple points). Each panel shows the source - source combination, $(i,j)$, tomographic bins $i$ and $j$. The lower panels show the ratio between the measurements and the theory (black solid lines), $C^{\rm th}_{\ell}$. The $\chi^2$ per degree of freedom, $\chi^2_{\rm dof}$, is shown for the uniform and variable depth cases in the lower left corner, computed using a Gaussian covariance assuming spatial uniformity.}
    \label{fig: clee-y3-fzb}
\end{figure*}

We use NaMaster \citep{Alonso_2019} to measure the $3\times2$pt data vector in Fourier space: $C_{\ell}^{\rm gg}$, $C_{\ell}^{{\rm g}\gamma}$, and $C_{\ell}^{\gamma\gamma}$ for the lens and source tomographic bins. NaMaster computes the mixing matrix to account for the masking effects, and produces decoupled band powers. The HEALPix pixel window function correction is also applied when comparing the data with input theory.
We adopt 14 $\ell$-bins in range $[20,1000]$ with log spacing. Notice that the maximum $\ell$ is a conservative choice for $C_{\ell}^{\gamma\gamma}$ compared to the DESC SRD, where $\ell_{\rm max}=3000$ is adopted, based on the assumption of improved modelling of non-linearity and baryonic feedback when the LSST data becomes available. Nevertheless, this is sufficient for our purpose to demonstrate the impact of variable depth on relatively large scales. For galaxy clustering and galaxy-galaxy lensing, we apply an additional scale cut at $\ell_{\rm max}=k_{\rm max}\chi(\langle z \rangle)-0.5$ following the DESC SRD, where $k_{\rm max}=0.3h{\rm Mpc}^{-1}$, and $\chi(\langle z \rangle)$ is the comoving distance at the mean redshift $\langle z \rangle $ of the lens tomographic bin.
We generate theory angular power spectra assuming spatial uniformity with the Core Cosmology Library\footnote{\url{https://github.com/LSSTDESC/CCL}} \citep[CCL; ][]{Chisari_2019}. CCL uses HALOFIT \citep{Smith_2003, Takahashi_2012} non-linear power spectrum and Limber approximation when computing the angular power spectra. 
We compute the Gaussian covariance matrix using NaMaster with theoretical data vectors. The covariance includes mask effects, shot-noise, and shape noise power spectra. It should be noted that this is done assuming uniformity. 
In the varying depth case, the true covariance contains extra variance, due to spatial correlation in the noise with the number count. 
Also, the assumption of a purely Gaussian covariance is not completely true. On very large scales, non-Gaussian mode coupling at scales larger than the survey footprint results in a term called super-sample covariance \citep{2014PhRvD..89h3519L}. Here we expect it to be relatively small because of the large sky coverage of LSST. 
On small scales, non-linear structure formation also introduces non-Gaussian terms \citep[e.g.][]{2001ApJ...554...56C}. With the scale cuts adopted in $C_{\ell}^{gg}$ and $C_{\ell}^{g\gamma}$ we expect that such non-Gaussian contribution to be small.

The galaxy clustering angular power spectra measurements, $C_{\ell}^{\rm gg}$, are shown in Fig.~\ref{fig: clgg-y3-fzb}. The tomographic bin number is indicated in the upper right corner as $(i,i)$ for bin $i$. The measurements for the uniform case are shown as red dots, and that for the varying depth case are shown in purple. The data points are shot-noise-subtracted. 
We see a clear difference between the uniform and the varying depth cases at $\ell<100$, and it becomes more significant at higher redshifts. 
The impact at large scales is expected, as the $i$-band coadd depth varies relatively smoothly and the rolling pattern is imposed at relatively large scales. 
The trend with redshifts is also expected, due to two main reasons. Firstly, the slope $d(N_{\rm gal}/\bar{N}_{\rm gal})/dm_5$ increases slightly with redshift, and is significantly larger for bin 5, as shown in the right middle panel of Fig.~\ref{fig: ngal-baselinev2-y1-y5-goldcut-snr-10}. This means that non-uniformity is most severe in these bins. Secondly, the clustering amplitude increases towards lower redshifts due to structure growth, hence the non-uniformity imprinted in $\delta_g$ is less obvious in lower redshift bins. 
In practice, the number density fluctuations are mitigated via the inclusion of the selection weights, $w(\boldsymbol{\theta})$, such that the corrected density field is defined as $ \Tilde{\delta}_g(\boldsymbol{\theta}) = N(\boldsymbol{\theta}) / w(\boldsymbol{\theta}) \bar{N}_w$, where $\bar{N}_w=\sum N(\boldsymbol{\theta}) / \sum w(\boldsymbol{\theta})$ (see e.g. \cite{Nicola_2020}). In addition, these weights will be used to compute the mode coupling matrix and shot noise, such that the varying number density is taken into account in the likelihood analysis.
A more subtle effect is the difference in redshift distribution at different depth. To isolate its impact, we compare the clustering power spectra from the noiseless sample varying $p(z)$ only with that from the noiseless uniform case. The ratio of the measurements are shown as dashed lines in the first panel of Fig.~\ref{fig: noiseless-ratio}. We find that once the non-uniformity in number density is removed, the variation in $p(z)$ does not significantly bias the power spectra, and we recover the uniform case at better than $0.5\%$.

The galaxy-shear and shear-shear power spectra, $C_{\ell}^{{\rm g}\gamma}$ and $C_{\ell}^{\gamma\gamma}$, are shown in Figs.~\ref{fig: clge-y3-fzb} and \ref{fig: clee-y3-fzb}, respectively. The source - lens and source - source combinations are indicated on the upper right as $(i, j)$. In both cases, we only show the non-zero E-modes, and we check that the B-modes are consistent with zero. 
For the galaxy-shear case, measurements from combinations $i<j$ are not shown, because we do not include effects such as magnification or intrinsic alignment, hence these measurements are low signal-to-noise or consistent with zero. 
We see that, overall, the impact of variable depth is much smaller compared to galaxy clustering. 
In the galaxy-galaxy shear measurements, only combination $(5,5)$ shows a significant $\chi^2$ in the variable depth case, and the main deviations is at $\ell<100$. This could be a joint effect where non-uniformity is largest in the highest redshift bin for both lens and source.
There is negligible difference in the shear-shear measurements for all other combinations given the measurement error.
To look at this further, we take the noiseless case and compute the ratio between measurements from the varying depth sample and the uniform sample. We show some examples along the diagonal, i.e., the $(i,i)$ combinations, in the middle and right panels of Fig.~\ref{fig: noiseless-ratio}. The off-diagonal measurements lie mostly within the variation range of the ones shown here. In case of $C_{\ell}^{{\rm g}\gamma}$, we see that deviations are large at low $\ell$ when both density and $p(z)$ is non-uniform (shown as solid line); when the density non-uniformity is removed (shown in dashed line), the results are more consistent within $5\%$. For $C_{\ell}^{\gamma\gamma}$, we see that the largest impact is from the highest tomographic bin reaching up to $0.5\%$. 

These results are consistent with the analytical approach in \cite{2023JCAP...07..044B}, where, in general, the varying depth effect in the redshift distributions is sub-percent and the weak lensing probes are less susceptible to these variations.
Our results are quite different from \cite{2020A&A...634A.104H} (hereafter H20) for KiDS cosmic shear analysis in several aspects. H20 found that the largest impact comes from the sub-pointing, small scales, and for a KiDS-like set-up, the difference between the uniform and variable depth cases is 3\% - 5\% at an angular scale of $\theta=10\,{\rm arcmin}$. Furthermore, the variable depth effect is stronger in lower redshift bins than higher redshift bins. Several differences in the analysis may contribute to these different results. Firstly, the non-uniformity in KiDS is rather different from that considered here: the KiDS footprint consists of many $1\,{\rm deg}^2$ pointings, each having distinctive observing conditions due to that each field only received a single visit. This means that survey properties such as depth are weakly correlated at different pointings. One can write down a scale-dependent function, $E(\theta)$, to specify the probability of a pair of galaxies falling in the same pointing at each $\theta$, and this essentially gives rise to the scale dependence of the variable depth effect in H20. For LSST, the above assumptions are not true, and $E(\theta)$ (if one can write it down) would take a very different form compared with that in KiDS. 
Secondly, due to the single visit, there is a much larger variation in depth, number density, and $\Delta z$ in KiDS compared to this work (tomographic bin centre can shift up to $\Delta z\sim 0.2$ in redshift, as shown in Fig.2 of H20). This means that the variable depth effects in KiDS as explored by H20 is significantly larger compared to this work. 
This also explains their redshift dependence, because for KiDS, the average redshift between pointings varies the most in the lowest redshift bins. Lastly, although our $\ell_{\rm max}$ here corresponds to $\theta \sim 10 \, {\rm arcmin}$. the results are not directly comparable, as H20 conducted the analysis in real space, i.e. $\xi_{\rm \pm}(\theta)$.

To sum up, the largest impact of varying depth comes from galaxy clustering, whereas the impact on weak lensing probes is much smaller. Higher redshift bins are more susceptible due to a higher sensitivity in number density and redshifts with depth. Given the mock LSST Y3 uncertainty, one can clearly detect bias in the power spectrum in galaxy clustering and the galaxy-galaxy shear bin $(4,4)$, while all other combinations do not seem to have detectable impacts. 
Furthermore, once the density non-uniformity is removed, the impact of varying depth is further reduced. There are several ways to mitigate number density variation, such as mode projection \citep[e.g.][]{1992ApJ...398..169R, 2016MNRAS.456.2095E}, template subtraction \citep[e.g.][]{2011MNRAS.417.1350R, 2012ApJ...761...14H}, iterative regression \citep[e.g.][]{2018PhRvD..98d2006E, 2021MNRAS.503.5061W}, and machine learning methods using neural networks \citep{2020MNRAS.495.1613R} and a Self-Organizing Map (SOM) \citep{2021A&A...648A..98J}. 
See \cite{2021MNRAS.503.5061W} for a thorough review. 
Notice that, despite these methods, it is difficult to guarantee a complete removal non-uniformity, and in some cases, clustering signal can also be reduced as a result.
Additional sky cuts to exclude problematic regions can also effectively reduce density variation, at the cost of losing sky coverage.
Finally, for the lens sample, a brighter magnitude cuts can also greatly reduce the variable depth effect (see Appendix~\ref{sec: maglim} for a MagLim-like lens selection), at the cost of sample sparsity.
Nevertheless, non-uniformity in $p(z)$ only seems to be safely averaged out in the 2-point statics measurements.

\subsubsection{Impact on spectroscopic calibration}
\label{sec: Impact on spectroscopic calibration}

\begin{figure*}
   \centering
   \includegraphics[width=0.9\textwidth]{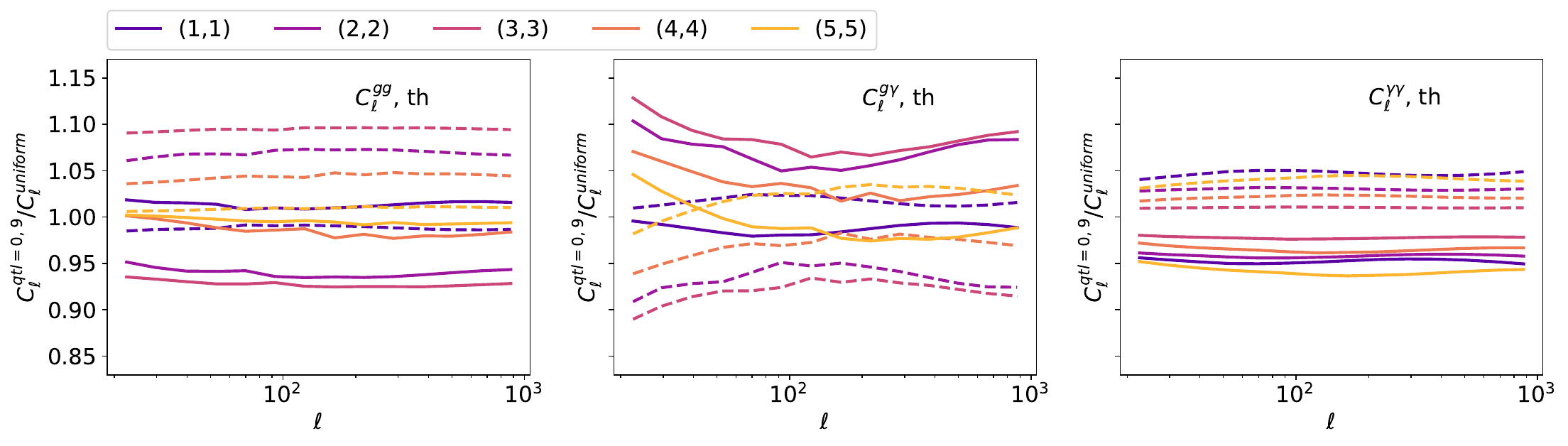}
    \caption{The ratio between the $3\times2$pt theory vectors computed using the $p(z)$ from depth quantiles 0 (shallowest, shown as solid lines) and 9 (deepest, shown as dashed lines), and those computed using the mean $p(z)$. Different colours show different tracer tomographic bin combinations, as indicated in the legend. For $C_{\ell}^{g\gamma}$ and $C_{\ell}^{\gamma\gamma}$, only cases where the tracers are in the same bin are shown for visual clarity, but the other tracer combinations have a comparable variation.}
    \label{fig: q0-q9-ratio-3x2pt-y3-fzb-dirac}
\end{figure*}

Here, we consider another potential source of systematics arising from small spectroscopic calibration fields. Redshift calibration for photometric surveys such as LSST are usually done using small but deep spectroscopic surveys, e.g. C3R2 survey \citep{Masters_2019}. Each field in these surveys has a coverage of a few ${\rm deg}^2$. 
Suppose that a calibration field overlaps with a particularly shallow or deep region, the calibration (e.g. a trained SOM) could cause bias to the overall redshift distribution when it is generalised to the whole field. For example, a SOM trained in a shallow region will contain larger noise, which may increase the scatter for the overall sample. The lack of high redshift, fainter objects in the shallow region could also cause bias when the SOM is applied to objects in deeper regions. 

The specific impact will depend on the calibration method and details of the calibration, which is beyond the scope of this paper. 
Here, we qualitatively assess the impact via the difference in the $3\times2$pt theory vectors computed using the $p(z)$ from a particular quantile and those computed using the mean $p(z)$, as shown in Fig.~\ref{fig: q0-q9-ratio-3x2pt-y3-fzb-dirac}. The solid lines show cases from the shallowest quantile, ${\rm qtl}=0$, and the dashed lines show cases from the deepest quantile, where ${\rm qtl}=9$, highlighting the worst case scenarios. For $C_{\ell}^{g\gamma}$ and $C_{\ell}^{\gamma\gamma}$, only cases where the tracers are in the same bin are shown, but the other lens - source combinations have a comparable variation.
We see that naively taking the $p(z)$ from a quantile and assume it as the $p(z)$ for the full sample can give rise to as much as $10\%$ bias compared to the uniform case.

This effect is reduced by having multiple calibration fields across the LSST footprint. Currently, many of the calibration fields overlaps with the LSST Deept Drilling Field (DDF), which will be much deeper compared to the WFD. Impact of variable depth can then be mitigated via a two-tiered SOM calibration, mapping from the deep to the wide field \citep{2021MNRAS.505.4249M}, and synthetic source injection \citep{Everett_2022}, mimicking the degradation of the deep field objects across the LSST footprint, as done in the DES Y3 analysis. 

\subsection{Impact on cosmological parameters}
\label{sec:results:cosmology}

We further predict the impact of survey non-uniformity on the cosmological analysis by conducting Fisher forecasting. The Fisher forecast estimates the constraints on cosmological parameters by assuming a Gaussian likelihood function, a fiducial cosmology, and a covariance matrix on the data vector \citep{Wasserman2004,Coe2009,Bhandari2021}. In the Bayesian statistics framework, we can write the Fisher Information matrix as 
\begin{equation}\label{eq:fishergauss}
    \mathbb{I}_{ij} = \frac{\partial \bmath{d}^T}{\partial \alpha_i} \, \mathbfss{V} \, \frac{\partial \bmath{d}}{\partial \alpha_j} + \frac{1}{\sigma_{\alpha_i}^2} \delta_{ij},
\end{equation}
where $\bmath{d}$ is the data vector, $\bmath{\alpha}$ is the model parameter vector and $\mathbfss{V}$ is the inverse of the covariance matrix. $\sigma_{\alpha_i}$ is the standard deviation of the Gaussian prior on parameter $\alpha_i$, and $\delta_{ij}$ is the Kronecker delta. We use the Fisher forecast code developed in Zhang et al. \textit{In prep.} 
The covariance matrix is computed by NaMaster using the theoretical angular power spectra generated by CCL, assuming Gaussianity.

\begin{table}
    \centering
    \begin{tabular}{l|c|c}
        Parameter & Fiducial Value & Prior $\sigma$ \\ \hline
        \hline
        Cosmological\\
        $\Omega_m$& 0.279 & 0.15\\
        $\sigma_8$& 0.82 & 0.2\\
        $w_0$     & -1 & 0.8\\
        $w_a$     & 0 & 1.3\\
        $h$       & 0.7 & 0.125\\
        $n_s$     & 0.97 & 0.2\\
        $\Omega_b$& 0.046 & 0.003 \\\hline
        Intrinsic alignment\\
        $A_0$     & 5.92 & 2.5\\
        $\eta_l $ & -0.47 & 1.5\\
        $\eta_h $ & 0.0  & 0.5\\
        $\beta$   &  1.1 & 1.0\\\hline
        Galaxy bias\\
        $b_i$ & 1.0 & 0.9 \\\hline
    \end{tabular}
    \caption{The fiducial value and the Gaussian standard deviation of the prior assumed in the Fisher information matrix for the cosmological and intrinsic alignment parameters as defined in \protect\cite{2017MNRAS.470.2100K}.
    }
    \label{tab:cosmoparams}
\end{table}

We use \textsc{CCL} 
to compute the fiducial data vector of the LSST Y3 $3\times2$pt. 
We use the non-linear intrinsic alignment (NLA) model as in \cite{2017MNRAS.470.2100K}, adopted in the DESC SRD,
to describe the contribution of intrinsic alignments to the data vectors. There are four NLA parameters, namely, the overall intrinsic alignment amplitude, $A_0$, the power-law luminosity scaling, $\beta$, the redshift scaling, $\eta_l$, and the additional high-redshift scaling $\eta_h$. 
The fiducial value and prior of the cosmological and astrophysical parameters are taken from the DESC SRD, as shown in Table~\ref{tab:cosmoparams}. 
The fiducial galaxy bias, $b_i$, of the lens catalogue in each tomographic bin $i$, is set to $1.0$, with a Gaussian standard deviation of $0.9$ and a cut at $b_i<0$.
The contours shown in this section include the statistical uncertainty of the data vector and the marginalized uncertainty over other cosmological and astrophysical parameters described above. 
The contour can be overconfident since it does not marginalize over observational systematic uncertainties, which can include photometric redshift uncertainty, point spread function uncertainty, and multiplicative shear uncertainty. Additionally, the non-Gaussian contributions to the covariance matrix is not taken into account. Non-linear galaxy bias is also not modelled.

Fisher forecasts can be used to predict bias in the parameters given a shift in the data vector. We take the difference between the biased and fiducial $3\times2$pt power spectra, $\bmath{d}^{\textrm{biased}}$ and $\bmath{d}$, respectively, from Section~\ref{sec: 3x2pt}, and use it to calculate the bias in cosmological parameters that the survey non-uniformity induces, 
under the assumption of small, linear changes in $\bmath{d}$ \citep{biasformula2, centroidshift}: 
\begin{equation}
\label{eq:fisher_bias}
    f_b = \mathbb{I}^{-1} \cdot \left( \frac{d \bmath{d}}{ d \bmath{\alpha }}\,   \mathbfss{V}\, \left(\bmath{d}^{\textrm{biased}} -\bmath{d}\right) \right),
\end{equation}
where $\bmath{d}$ is the fiducial $3\times2$pt data vector. The Fisher information matrix used in Eq.~\ref{eq:fisher_bias} is the full $16\times 16$ matrix which includes $11$ cosmological and intrinsic alignment parameters, as well as $5$ galaxy bias parameters, as shown in Table~\ref{tab:cosmoparams}. 

\begin{figure*}
   \centering
   \includegraphics[width=0.99\textwidth]{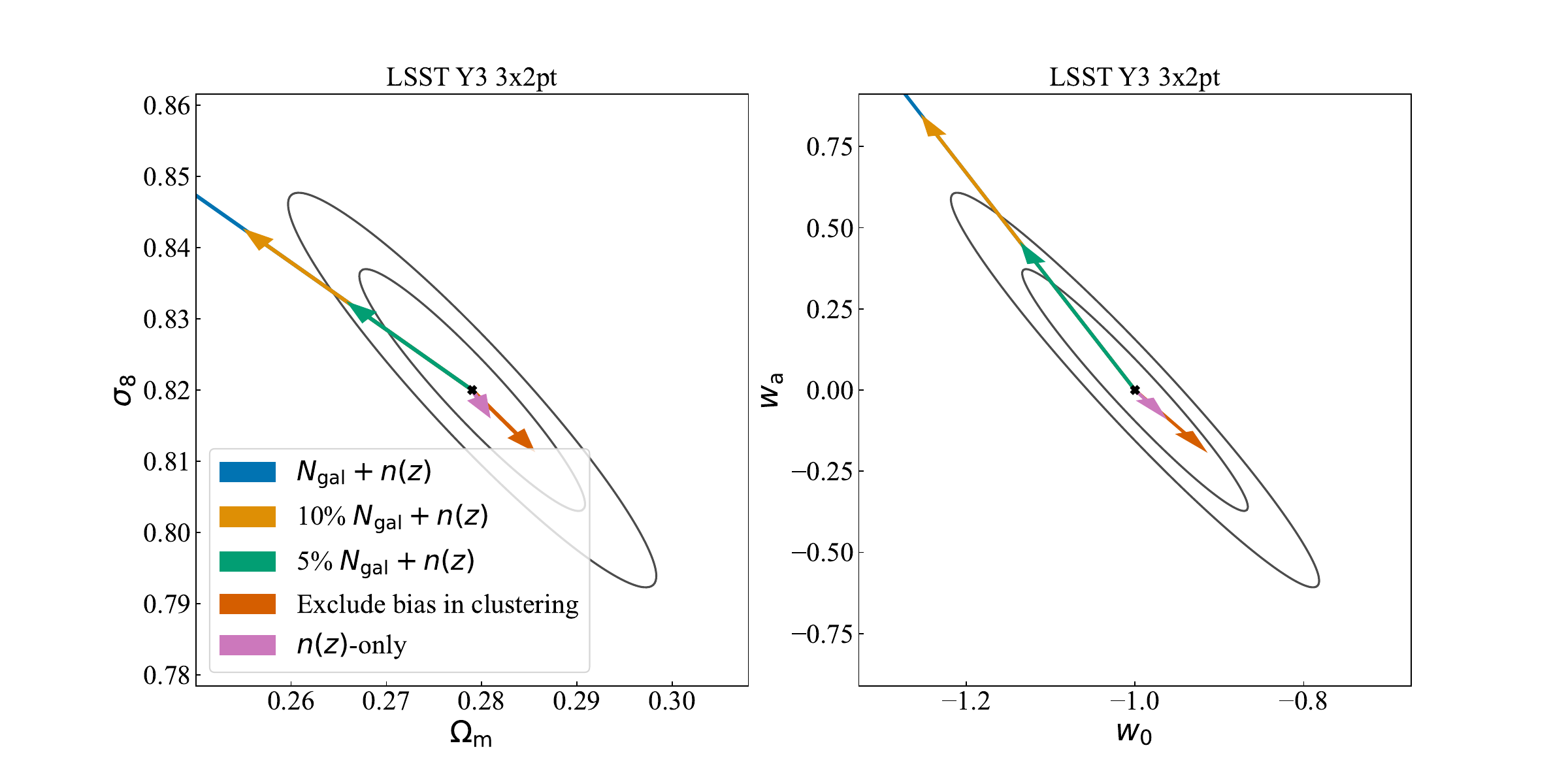}
    \caption{The black ellipse shows the Fisher forecasted $1\sigma$ and $2\sigma$ contour of $\Omega_{\rm m}-\sigma_8$ and $w_0-w_a$, marginalized over 16 parameters as described in Section~\ref{sec:results:cosmology}. The parameter biases induced by survey non-uniformity are given by the vectors in the plot. The blue, orange and green vectors show the biases corresponding to $100\%$, $10\%$, and $5\%$ of both $N_{\rm gal}$ and $n(z)$ non-uniformity. The brick red vector shows the bias corresponding to the $100\%$ case but without the clustering bias. The purple vector shows the bias corresponding to only $n(z)$ non-uniformity.}  
    \label{fig:fisher_forecast}
\end{figure*}

The forecasted impact of non-uniformity on LSST Y3 $3\times2$pt cosmological analysis is shown in Fig.~\ref{fig:fisher_forecast}. When neither non-uniform $N_{\rm gal}$ nor $n(z)$ are modelled in the data vector, the forecasted bias on $\Omega_{\rm m}-\sigma_8$ and $w_0-w_a$ are both on the order of $\sim 20 \sigma$, making the analysis completely unfeasible. Notice that in this case, strictly speaking, the small difference assumption in Eq.~\ref{eq:fisher_bias} breaks down, and so one should take these numbers with caution.
Assuming the non-uniformity residual can be reduced to a level of $10\%$ (orange) and $5\%$ (green), the bias on the cosmological parameters reduces to about $3\sigma$ and $1.5\sigma$, respectively. We observe that the main contributor to the cosmological bias in this case is the galaxy clustering, $C_\ell^{\rm gg}$. When the bias in clustering is set to zero, the overall bias in cosmology is contained within $1\sigma$, shown in brick red.
The cosmological bias when only non-uniformity of $n(z)$ is mis-modelled is negligible, as shown in the purple vector. 

As a result of the Fisher forecast, we recommend the $N_{\rm gal}$ non-uniformity of the LSST $3\times2$pt lens sample should be modeled with less than $3\%$ residual, to ensure an accurate cosmological analysis with bias within $1\sigma$. Otherwise, large-scale modes or high-redshift bins of the galaxy clustering signal must be removed from the data vector to avoid the parts where non-uniformity makes the most significant impact, as also shown in Fig.~\ref{fig: clgg-y3-fzb}.

\section{Conclusions}
\label{sec: Conclusions}

In this paper, we investigated and quantified the impact of spatial non-uniformity due to survey conditions on redshift distributions in the context of early LSST data. We used the Roman-Rubin simulation as the truth catalogue, and degraded the photometry using the LSST error model implemented in the RAIL package. The degradation utilizes the survey condition maps from the OpSim baseline v3.3 for the 1-year, 3-year, and 5-year LSST data. We run BPZ and FZBoost photometric redshift estimators on the degraded sample and use the photo-$z$ mode to separate the samples into five lens and five source tomographic bins. Finally, we apply the LSST Gold selection and a signal-to-noise cut. Taking the extinction-corrected $5\sigma$ coadd depth of the detection band, $i$-band, as the primary source of non-uniformity, we quantify the impact in terms of three measures: the number of objects, the mean redshift of the tomographic bin, and the tomographic bin width. 
We find that:
\begin{itemize}
    \item The number of objects increases with the $i$-band depth in general, and at extreme depth values, the number of objects can vary by a factor of two. The trend is relatively consistent between cases using BPZ and FZBoost, although selecting ${\rm odds}\geq 0.9$ for BPZ amplifies the trend. The largest correlation comes from the highest tomographic bin.
    \item The mean redshift in each bin increases with the $i$-band depth, with a variation of $|\Delta z /(1+\langle z \rangle)|\sim0.005 - 0.01$. The lens samples show a relatively consistent trend across different tomographic bins, whereas for the source sample, the highest tomographic bin shows the largest variation. This reaches the limit of the requirements of $0.005$ for Y1 as listed in the DESC SRD, and exceeds the requirement of $0.003$ for Y10. At extreme depth variations, however, deviation in $\langle z \rangle$ could exceed Y1 requirements.
    \item The width of the lens tomographic bin is measured in terms of $\sigma_z$, which is sensitive to the entire redshift distribution, $p(z)$, and $W_z$, which is sensitive to the peak of $p(z)$, both varying at the level of $10\%$ and slightly increases with year. We find that in general, $\sigma_z$ increases with the $i$-band depth due to fainter objects included in the deeper sample. $W_z$ also increases with the $i$-band depth, due to a more peaked bulk $p(z)$ as a result of higher SNR in deeper samples, although the trend can be reversed in some cases.
\end{itemize}

As emphasized before, results derived for Y3 and Y5 are with particularly large rolling non-uniformity. Hence, the variations shown should be interpreted as an upper-limit for the early Rubin LSST static science. As shown in Appendix~\ref{sec: maglim}, if the final LSST lens selection is similar to the DES Y3 MagLim sample with a bright magnitude cut, then the expected variable depth impact will be milder than shown in our baseline cases.

We took the Y3 FZBoost photo-$z$ as an example to propagate the impact of varying depth to the weak lensing $3\times2$pt measurements. To do this, we used one realization of the Gower Street N-body simulation, and generated lens galaxy maps and source shear maps with spatially varying number density and $p(z)$. We measure the data vector in harmonic space using NaMaster, and also compare them with the theory expectation generated from the Core Cosmology Library. We find that the largest impact is on $C_{\ell}^{\rm gg}$ with the higher redshift bin measurements significantly biased. $C_{\ell}^{{\rm g}\gamma}$ is less sensitive to varying depth effects, although in the source - lens combination $(4,4)$, there is a visible difference at low $\ell$. $C_{\ell}^{\gamma\gamma}$ shows no significant impact in all source - source combinations from varying depth, given the uncertainties in LSST Y3.
Finally, we also investigate cases where we do not include noise in the lens and source maps. 
The difference between uniform and varying depth cases can be up to a few percent for $C_{\ell}^{{\rm g}\gamma}$, and less than $0.5\%$ for $C_{\ell}^{\gamma\gamma}$. Furthermore, by removing the density non-uniformity, and varying $p(z)$ only with depth, one can reduce the bias in $C_{\ell}^{\rm gg}$ and $C_{\ell}^{{\rm g}\gamma}$ to sub-percent level. 
We use a Fisher forecast to assess the impact of non-uniformity on cosmological parameter inference for the $3\times2$pt data vector. We conclude that the mitigation in number density variation is crucial, and for our baseline setup for LSST Y3, this should be controlled below $3\%$.
Therefore, for early LSST analysis, it is crucial to account for the galaxy density variation, but the impact of varying $p(z)$ seems to be negligible. We leave the investigation of an accurate mitigation strategy of the number density variation to future work.

Our current approach has some caveats. 
Firstly, the fidelity of the colour-redshift relation in the Roman-Rubin simulation at $z>1.5$ is questionable. As already mentioned, the strong bifurcation of the blue objects at this high redshift may lead to worse (in the case of BPZ) or overly optimisic (in the case of FZBoost) performance when estimating the photo-$z$.
Secondly, we have adopted an analytic model to obtain the observed magnitudes in each band based on survey conditions. However, in reality, the observed magnitudes and colours also depend on the way they are measured. For example, for extended objects, cModel \citep{2002AJ....124.1810S} and GAaP \citep{2015MNRAS.454.3500K} methods are often applied. Although the photometry will be calibrated, the magnitude error may not be the same for different methods. This could introduce extra scatter in photo-$z$.
Thirdly, we have only tested on two major photo-$z$ estimators, observing some level of differences in the results. For example, compared to BPZ, the FZBoost samples show more consistency between different tomographic bins regarding to the trend with $i$-band depth. Therefore, one should take the result as an order of magnitude estimate of the impact, but the specific trends are likely to differ for different photo-$z$ methods.
Moreover, when propagating the effects to the data vector, we have made some simplifications. We considered a galaxy bias of $b=1$, and did not include systematics such as magnification bias or intrinsic alignments. This choice is to isolate the effect of varying depth on the pure lensing and clustering contribution, but it would be more realistic to include these effects. 
Finally, we have not folded in the effects of blending, i.e. spatially nearby galaxies are detected as one object. This occurs when the surface density is high and the image is crowded, and could be significant for deep photometric surveys such as LSST. 
The level of blending depends on both seeing and depth of the survey, hence, it could correlate with the variable depth effects discussed here.
The impact of blending on photo-$z$ is the inclusion of a small fraction of ill-defined redshifts in the sample, increasing the photo-$z$ scatter. Clustering redshift calibration, which measures galaxy clustering on small scales, can also be affected as these scales are most susceptible to blending.
Moreover, blending can affect shear measurements via e.g. lensing weights, hence introduce impact on galaxy-galaxy lensing and cosmic shear.
As such, \cite{Nourbakhsh_2022} showed that approximately 12\% of the galaxy sample in LSST is unrecognized blends, and can bias $S_8$ measurement from cosmic shear by $2\sigma$. 

Furthermore, so far our results are based on the $p(z)$ of the true redshifts of the sample. In reality, we do not have access to this, and our theory curve will be based on the calibrated redshift distribution $p_c(z)$, which itself can be impacted by non-uniformity based the calibration method. 
For example, in many weak lensing surveys, a Self-Organizing Map (SOM) is used to calibrate redshifts by training on a photometric sub-sample with spectroscopic counterparts \citep{2020A&A...637A.100W,2021MNRAS.505.4249M}. By taking sub-samples from a small calibration field (typically of a few square degrees) located in a particularly shallow region could result in a trained SOM that captures different magnitudes, redshifts, and SNR than that from a deep region, as quanlitatively shown in Section~\ref{sec: Impact on spectroscopic calibration}.
One remedy may come from calibration using clustering redshifts, which takes advantage of galaxy clustering of the target sample with a spectroscopic sample, spliced in thin redshift bins \citep{2020A&A...642A.200V, 2022MNRAS.510.1223G,2023MNRAS.524.5109R}. The nonphysical variation with depth will drop out in this method, giving unbiased estimate of $p(z)$.

We have only explored the impact of variable depth on two-point statistics here, but there could be potential impact on statistics beyond two-point. For example, for weak lensing shear, a similar effect in manifestation is source clustering, where the number density of source galaxies $n(\hat{\theta}, z)$ is correlated with the measured shear $\gamma(\hat{\theta})$ for a given direction $\hat{\theta}$ on the sky, because source galaxies are themselves clustered. Impact of source clustering is negligible in two-point statistics for Stage III surveys, but is detected significantly in several higher order statistics in the DES Y3 data \citep{2024MNRAS.527L.115G}. Given that the variable depth effect also modulates $n(\hat{\theta}, z)$ (hence imprinting a fake `source clustering'), there may be non-negligible impact on higher order statistics with LSST.
We leave these explorations to future work.

\section*{Author contribution statements}
\textit{QH}: Contributed to the conceptualization, data curation, formal analysis, and writing of the draft. Contributed to the development of RAIL.\\
\textit{BJ}: Contributed to the conceptualization, funding acquisition, project administration, and revisions of the text.\\
\textit{EC}: Contributed to RAIL software core functionalities used in the analysis. Minor contributions to revisions of text.\\
\textit{JFC}: Contributed to the development of RAIL and the \texttt{photerr} model.\\
\textit{PL}: Contributed to the development of the Roman-Rubin Diffsky simulation.\\
\textit{AIM}: Contributed to RAIL software core functionalities used in the analysis. Minor contributions to revisions of text.\\
\textit{SS}: Contributed to RAIL software used in the analysis, namely the BPZ and FlexZBoost algorithms used in estimation, along with general software development. Minor contributions to revisions of text.\\
\textit{ZY}: Contributed to the development of RAIL and the \texttt{photerr} model; provided reviewing comments for the manuscript.\\
\textit{TZ}: Conducted the cosmological forecast for the paper. Contributed to the development of RAIL, including the LSST error model, and RAIL's core API; provided reviewing comments for the manuscript.\\

\section*{Acknowledgements}

QH and BJ are supported by STFC grant ST/W001721/1 and the UCL Cosmoparticle Initiative.
This paper has undergone internal review in the LSST Dark Energy Science Collaboration. 
The authors thank the internal reviewers, Boris Leistedt and Markus Rau, for their thorough and insightful comments. 
This work also benefited from helpful comments by Federica Bianco, Pat Burchat, Andrew Hearin, Eve Kovacs, Ofer Lahav, Rachel Mandelbaum, Andrina Nicola, and Peter Yoachim.
The DESC acknowledges ongoing support from the Institut National de 
Physique Nucl\'eaire et de Physique des Particules in France; the 
Science \& Technology Facilities Council in the United Kingdom; and the
Department of Energy, the National Science Foundation, and the LSST 
Corporation in the United States.  DESC uses resources of the IN2P3 
Computing Center (CC-IN2P3--Lyon/Villeurbanne - France) funded by the 
Centre National de la Recherche Scientifique; the National Energy 
Research Scientific Computing Center, a DOE Office of Science User 
Facility supported by the Office of Science of the U.S.\ Department of
Energy under Contract No.\ DE-AC02-05CH11231; STFC DiRAC HPC Facilities, 
funded by UK BEIS National E-infrastructure capital grants; and the UK 
particle physics grid, supported by the GridPP Collaboration.  This 
work was performed in part under DOE Contract DE-AC02-76SF00515.
JFC acknowledges support from the U.S. Department of Energy, Office of Science, Office of High Energy Physics Cosmic Frontier Research program under Award Number DE-SC0011665.
AIM acknowledges the support of Schmidt Sciences.

We acknowledge the use of arXiv and ADS for referneces, the use of Python libraries and software mentioned in the main text for data analysis and plotting, and Overleaf for the writing of this paper.

\section*{Data Availability}

 The methodology of generating mock LSST photometry with observing conditions is included in the RAIL pipeline\footnote{\url{https://github.com/LSSTDESC/rail_pipelines/tree/main/src/rail/pipelines/examples/survey_nonuniformity}}.
 The mock galaxy catalogues with varying depth effect are available upon reasonable request.



\bibliographystyle{mnras}
\bibliography{main} 




\appendix

\section{Comparison of LSST Error Model on DC2}
\label{sec: Calibration of LSST error model}


The $5\sigma$ depth per visit, $m_5$, depends on a set of observing conditions in the following way \citep{2019ApJ...873..111I}: 
\begin{equation}
    \begin{aligned}
    m_5  = & C_m + 0.50 (m_{\rm sky}-21) + 2.5\log_{10}(0.7/\theta_{\rm eff}) + \\
     & 1.25\log_{10}(t_{\rm vis}/30)-k_m(X-1),
     \label{eq: mag err}
\end{aligned}
\end{equation}
where $C_m$ is a constant that depend on the overall throughput of the system, $m_{\rm sky}$ is the sky brightness in AB mag arcsec$^{-2}$, $\theta_{\rm eff}$ is the seeing in arcsec, $t_{\rm vis}$ is the exposure time in seconds,
$k$ is the atmospheric extinction coefficient, and $X$ is the airmass. 
The default values of the parameters in the above equation per band are given in Table 2 in \cite{2019ApJ...873..111I}. The magnitude error for $N$-years observation is computed by $\sigma/Nn_{\rm vis}$, where the mean number of visits per year $n_{\rm vis}$ can be derived from Table 1 in \cite{2019ApJ...873..111I}.

In this Appendix, we compare the LSST error model with the Rubin OpSim output as well as the Data Challenge 2 \citep[DC2; ][]{2021ApJS..253...31L} dr6 magnitude error. We perform our tests on the specific OpSim version \texttt{minion\_1016}, and we use the 5-year observing conditions including: coadd $5\sigma$ point source depth (\texttt{CoaddM5}), single-visit $5\sigma$ point source depth (\texttt{fiveSigmaDepth}), sky brightness (\texttt{filtSkyBrightness}), and number of visits (\texttt{Nvisits}).

\begin{figure*}
    \includegraphics[width=\textwidth]{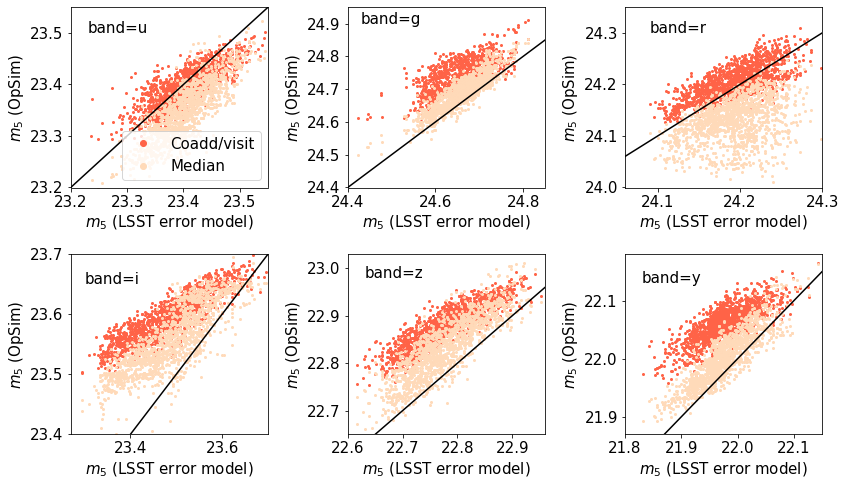}
    \caption{Comparison of the $5\sigma$ PSF limiting magnitude computed from Eq.~\ref{eq: mag err} with the OpSim output: the median $m_5$ over 5 years (bright pink) and the coadded $5\sigma$ depth converted to the equivalent of per visit (red). The $m_5$ computed from Eq.~\ref{eq: mag err} utlises the median sky brightness ($m_{\rm sky}$), median airmass ($X$), and median seeing ($\theta_{\rm eff}$), and other parameters are set to the default value in \protect\cite{2019ApJ...873..111I}.}
    \label{fig: lsst-error-model-minion-1016-m5}
\end{figure*}

We begin by checking Eq.~\ref{eq: mag err} using OpSim MAF maps over the DC2 footprint. The various survey conditions $m_{\rm sky}$, $\theta_{\rm eff}$, and $X$ are taken as the median values over the 5-year period, and other parameters $C_m$, $t_{\rm vis}$, and $k_m$ are taken as the default values from \cite{2019ApJ...873..111I}. The results from Eq.~\ref{eq: mag err} are the $5\sigma$ PSF magnitude limit in each band per visit, and we compare it with two quantities: the median $5\sigma$ depth map, and the equivalent per-visit depth from the coadded map: $m_5=m_5^{\rm coadd}-2.5\log(\sqrt{N_{\rm vis}})$, where $N_{\rm vis}$ is the number of visits at each pixel. The results are shown in Fig.~\ref{fig: lsst-error-model-minion-1016-m5}. We see that in general, $m_5$ predicted by Eq.~\ref{eq: mag err} tends to be brighter than that from OpSim, and the difference is larger considering the coadd depth than the median depth. It seems that except for $i$-band which has a slightly different slope from unity, the difference in all other bands can be fixed by introducing a correction to $C_m$. For example, for the median $m_5$ case, the shifts needed are $\delta C_m=\{ -0.053, 0.032, -0.063, 0.070, 0.057, 0.027 \}$ for $ugrizy$, respectively.

\begin{figure*}
    \includegraphics[width=\textwidth]{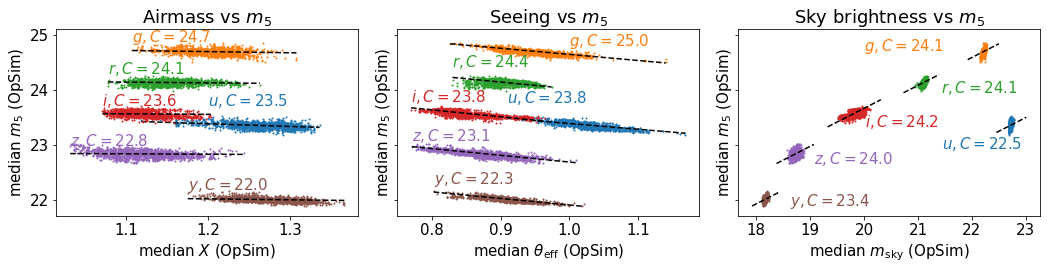}
    \caption{The relation between $m_5$ and other survey conditions using the LSST error model. We show the comparison between the data points from OpSim for each of the six LSST bands, and the relation from the LSST error model using the default parameters as black dashed lines, with a fitted constant $C$. We see that the LSST error model captures the correlation between $m_5$ and the underlying survey conditions well. The different colours correspond to different LSST filters, as indicated in the texts next to the data points in the same colour.}
    \label{fig: lsst-error-model-minion-1016-X-theta-msky}
\end{figure*}

We also explicity check whether the dependence of the airmass, seeing, and sky brightness are as expected in Eq.~\ref{eq: mag err} with the default parameters. This is shown in Fig.~\ref{fig: lsst-error-model-minion-1016-X-theta-msky}. In all these exercises, we test whether the dependencies of the particular survey condition with $m_5$ on the ensemble pixels, fixing all other dependence to a constant $C$ which we fit to the ensemble. We see that the airmass and seeing are well captured by Eq.~\ref{eq: mag err}, although the dependence of $m_5$ on airmass is weak. The sky brightness relation is less well captured by Eq.~\ref{eq: mag err} especially for $u$ and $g$. In general, however, we conclude that in absence of a depth map, one can estimate the unbiased $m_5$ for Rubin observation conditions using Eq.~\ref{eq: mag err} with a modification of the $C_m$ parameters for each band.

\begin{figure*}
    \includegraphics[width=\textwidth]{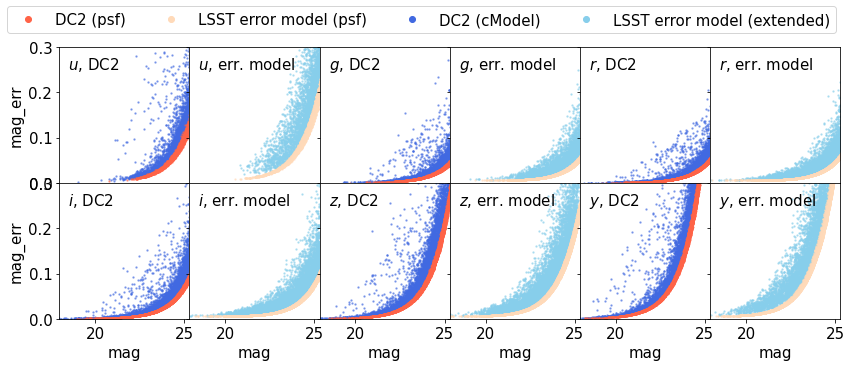}
    \caption{Comparison of the magnitude error as a function of magnitude in each of the six LSST bands between the DC2 dr6 catalogue and the LSST error model. The red and pink points show the PSF magnitude errors, whereas the dark and light blue points show that of the extended errors compared with the DC2 cModel magnitudes. The coadd $5\sigma$ depth from OpSim is used to compute the magnitude errors.}
    \label{fig: lsst-error-model-magerr-dc2}
\end{figure*}

We then check Eq.~\ref{eq: mag err 1} and \ref{eq: mag err 2} with the DC2 DM catalogue, where the magnitude errors are obtained through the detection pipeline, thus supposed to be more realistic. In this case, we directly adopt the coadded depth as $m_5$. We also compute in the low SNR limit (Eq.~\ref{eq: low snr}) which allows us to check the fainter magnitudes. For the extended magnitude errors, we compare with the CModel magnitudes in DC2. 
This is shown in Fig.~\ref{fig: lsst-error-model-magerr-dc2}. We see that there is reasonable agreement for the PSF megnitude errors in most bands, except for the $u$-band, where the LSST error model predicts larger error compared to that measured in DC2. However, it is also noticeable that the DC2 error seems to be underestimated when comparing the observed magnitude to the truth. It is also noticeable that the LSST error model also predicts consistently larger error at the bright end.
When we add the extended error from the size of the galaxy (Eq.~\ref{eq: extended err}), we find that the scatter of the magnitude error at fixed magnitude is quite a bit larger than that measured by the cModel in DC2. 
Both the PSF magnitude error and the scatter for the extended error in DC2 can be matched by the LSST error model by simple scaling of the PSF magnitude error by a constant for each band, as well as scaling the galaxy size $a_{\rm gal}, b_{\rm gal}$. We emphasise that due to the known issues in the DC2 catalogue, we do not calibrate the LSST error model to DC2 in our analysis. However, it is worth bearing in mind what the differences are, and that one needs to calibrate the model with the real data.

\section{Comparison of Roman-Rubin galaxy colour with BPZ templates}
\label{sec: bpz templates}

\begin{figure*}
   \centering
   \includegraphics[width=0.6\textwidth]{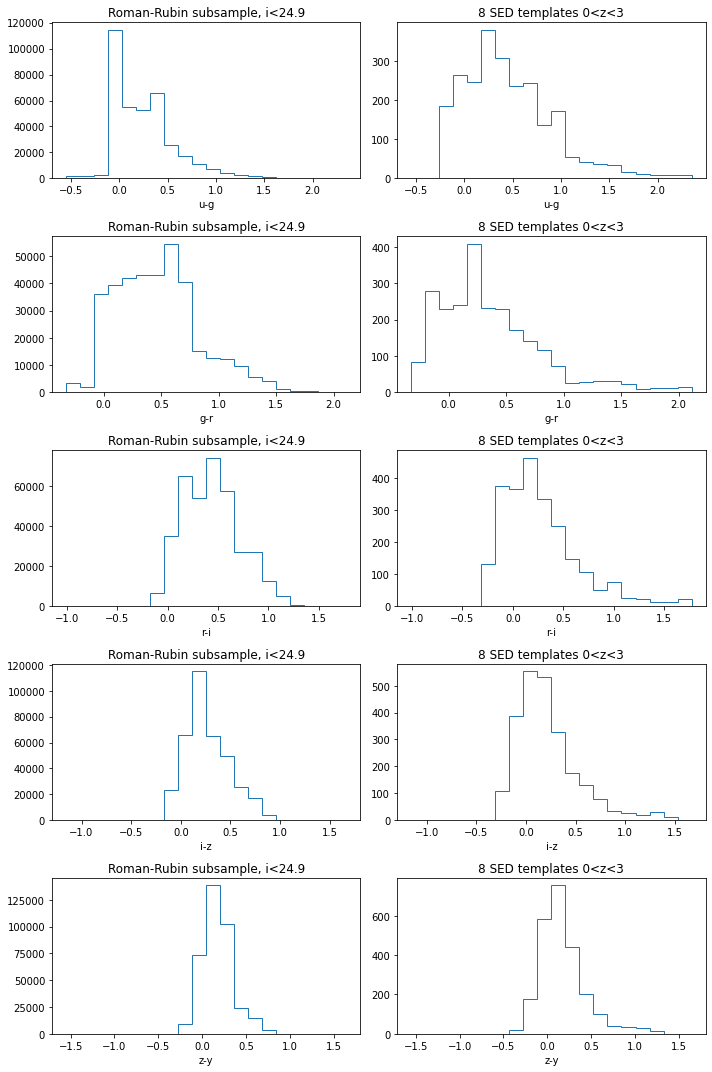}
    \caption{Colours in the LSST filters, for the Roman-Rubin (DiffSky) galaxies with $i < 24.9$ (left),
and that derived from the SED templates used in BPZ in the redshift range $0 < z < 3$ (right).}
    \label{fig: diffsky_bpz_sed_colors}
\end{figure*}

We show the coverage of BPZ templates adopted in this paper for the Roman-Rubin (DiffSky) simulation galaxy colours. We obtain template magnitudes in the LSST six-band filters by integrating each SED templates with the corresponding filter curves, with the template shifted in redshift range $0<z<3$. We then compare the five colour distributions of the resultant templates with that of the Roman-Rubin galaxies ($i < 24.9$, corresponding to the Y5 Gold cut). The results are shown in Fig.~\ref{fig: diffsky_bpz_sed_colors}. We see that the colour ranges of the simulated galaxies are captured by the BPZ templates used.

\section{Lens and source tomographic bin details}
\label{sec: Lens and source tomographic bin details}

\begin{figure*}
   \centering
   \includegraphics[width=\textwidth]{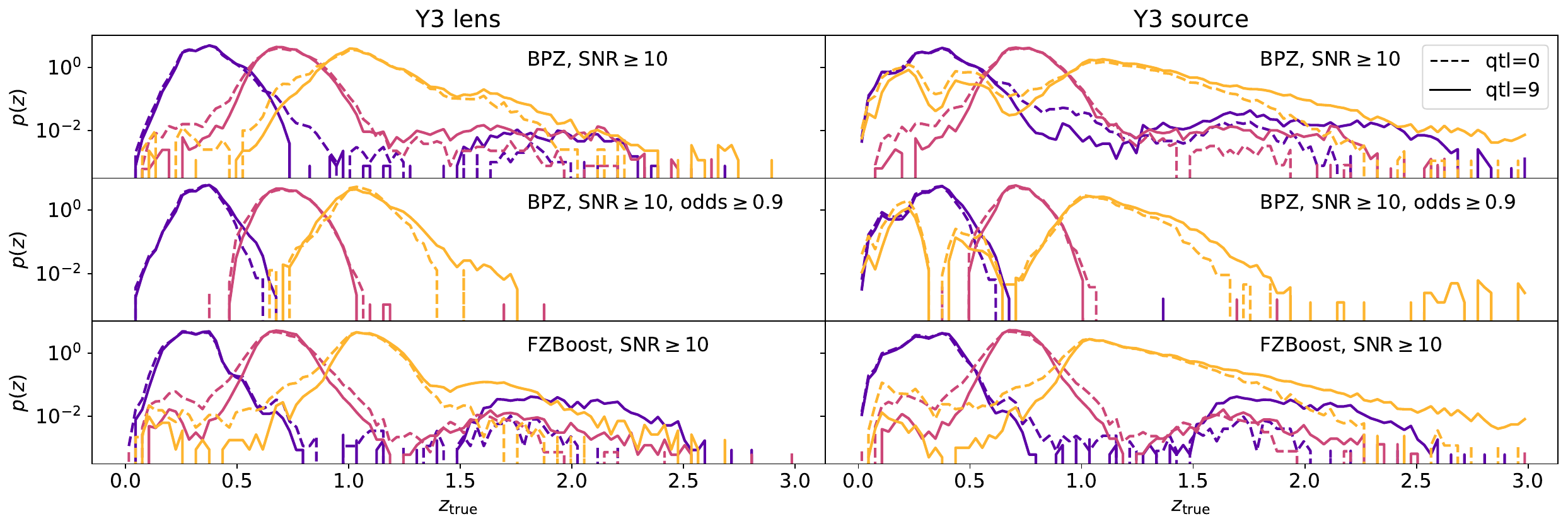}
    \caption{Tomographic redshift distribution of the Y3 sample for ${\rm qtl=0}$ (solid lines) and ${\rm qtl=9}$ (dashed lines) in log scale. Only tomographic bins 1, 3, and 5 are shown for visual clarity. The log scale highlights the tails towards high redshifts in each bin, which significantly impact the second moment of the distribution, $\sigma_z$ for each case.}
    \label{fig: nz-baselinev3.3-y3-long-logy}
\end{figure*}

This section includes some supplementary information for the lens and source tomographic bins for the mock photometry sample, as discussed in Section~\ref{sec: Tomographic bins}.

Figure~\ref{fig: nz-baselinev3.3-y3-long-logy} shows a similar plot as Fig.~\ref{fig: nz-baselinev2-y2-goldcut-snr-10}, but with the $y$-axis in logarithmic scale, and extended to $z=3$. Only tomographic bins 1, 3, and 5 are shown for visual clarity. This scaling enhances the small, high-redshift population for both lens and source galaxies. 

\begin{table*}
 \caption{The summary statistics on photo-$z$ performance for BPZ and FZBoost at the 10\% shallowest $i$-band coadd depth (${\rm qtl}=0$) and deepest depth (${\rm qtl}=9$) for the 1-year, 3-years, and 5-years mock LSST data, as shown in Fig.~\ref{fig: photo-$z$ vs spec-z}. Defining $\Delta z = (z_{\rm phot}-z_{\rm true})/(1+z_{\rm true})$, the summary statistics are: median bias, defined as the median of $\Delta z$, STD, defined as the standard deviation of $\Delta z$, the normalized Median Absolute Deviation, defined as $\sigma_{\rm NMAD}=1.48 {\rm Median}(|\Delta z|)$, and outlier fraction, defined as the fraction of sample with $|\Delta z|>0.15$. Both cases for full sample without cuts and for the high signal-to-noise sample with ${\rm SNR}\geq10$ are shown. For BPZ, we also show the selection with ${\rm odds}\geq0.9$.}
 \label{tab: stats on photoz}
 \begin{tabular}{llcccccccc}
 \hline
 Sample & & \multicolumn{4}{|c|}{${\rm qtl}=0$} & \multicolumn{4}{|c|}{${\rm qtl}=9$}\\
  \hline
   & & Median bias & STD & $\sigma_{\rm NMAD}$ & Outlier fraction & Median bias & STD & $\sigma_{\rm NMAD}$ & Outlier fraction\\
  \hline
  Y1 BPZ & Full & $-0.011$ & $0.411$ & $0.0772$ & $20.1$\% & $-0.011$ & $0.404$ & $0.0634$ & $15.5$\%  \\
   &  ${\rm SNR}\geq 10$ & $-0.005$ & $0.444$ & $0.0632$ & $14.2$\% & $-0.009$ & $0.409$ & $0.0585$ & $12.6$\% \\
   &  ${\rm odds}\geq 0.9$ & $-0.001$ & $0.446$ & $0.0431$ & $5.8$\% & $-0.006$ & $0.388$ & $0.0401$ & $4.5$\% \\
   \hline
   Y1 FZBoost & Full & $0.008$ & $0.122$ & $0.0479$ & $7.2$\% & $-0.006$ & $0.082$ & $0.0371$ & $4.2$\%  \\
   & ${\rm SNR}\geq 10$ & $0.008$ & $0.072$ & $0.0410$ & $3.1$\% & $-0.004$ & $0.065$ & $0.0351$ & $2.4$\% \\
  \hline
  Y3 BPZ & Full & $-0.013$ & $0.380$ & $0.0770$ & $20.9$\% & $-0.011$ & $0.369$ & $0.0613$ & $14.3$\%  \\
   & ${\rm SNR}\geq 10$ & $-0.009$ & $0.399$ & $0.0612$ & $13.9$\% & $-0.010$ & $0.368$ & $0.0586$ & $12.1$\%\\
   & ${\rm odds}\geq 0.9$ & $-0.005$ & $0.388$ & $0.0407$ & $4.7$\% & $-0.008$ & $0.337$ & $0.0421$ & $4.2$\%  \\
   \hline
    Y3 FZBoost & Full & $0.005$ & $0.145$ & $0.0408$ & $8.3$\% & $-0.004$ & $0.079$ & $0.0257$ & $4.2$\%  \\
    & ${\rm SNR}\geq 10$ & $0.005$ & $0.089$ & $0.0326$ & $3.1$\% & $-0.003$ & $0.065$ & $0.0247$ & $2.7$\% \\
  \hline
  Y5 BPZ & Full & $-0.013$ & $0.371$ & $0.0774$ & $21.3$\% & $-0.011$ & $0.353$ & $0.0666$ & $15.9$\%  \\
   & ${\rm SNR}\geq 10$ & $-0.009$ & $0.384$ & $0.0620$ & $14.2$\% & $-0.009$ & $0.350$ & $0.0633$ & $13.5$\% \\
   & ${\rm odds}\geq 0.9$ & $-0.006$ & $0.372$ & $0.0403$ & $4.8$\% & $-0.007$ & $0.330$ & $0.0442$ & $4.4$\% \\
   \hline
    Y5 FZBoost & Full & $0.004$ & $0.137$ & $0.038$ & $8.3$\% & $-0.003$ & $0.08$ & $0.0256$ & $4.7$\% \\
    & ${\rm SNR}\geq 10$ & $0.004$ & $0.086$ & $0.0305$ & $3.4$\% & $-0.003$ & $0.068$ & $0.0244$ & $3.2$\% \\
  \hline
 \end{tabular}
\end{table*}

\begin{table*}
 \caption{The mean values of the metrics across all depth quantiles for samples with BPZ redshifts. The metrics per tomographic bin include number of galaxies $\bar{N}_{\rm gal}$ and mean redshift $\langle z \rangle$. For lens galaxies, we compute two additional metrics regarding to the width of the tomographic bin: the second moment $\sigma_z$ and the LSS diagnostic $W_z$ defined in Eq.~\ref{eq: wz}. Gold cut in the respective year and ${\rm SNR}\geq10$ are applied to all samples, and a case with ${\rm odds}\geq0.9$ is also included for comparison.}
 \label{tab: mean metrics bpz}
 \begin{tabular}{ll ccccc ccccc}
  \hline
  Sample & Metric & \multicolumn{5}{|c|}{${\rm SNR}\geq10$} & \multicolumn{5}{|c|}{${\rm SNR}\geq10, {\rm odds}\geq0.9$}\\
  \hline
   & & Bin 1 & Bin 2 & Bin 3 & Bin 4 & Bin 5 & Bin 1 & Bin 2 & Bin 3 & Bin 4 & Bin 5\\
  \hline
  Y1 lens & $\bar{N}_{\rm gal}$ & $28796.5$ & $32021.9$ & $35387.0$ & $24288.8$ & $8463.0$ & $7635.1$ & $6158.4$ & $10002.0$ & $6221.3$ & $1875.5$\\
   & $\langle z \rangle$ & $0.362$ & $0.537$ & $0.714$ & $0.882$ & $1.050$ & $0.337$ & $0.549$ & $0.716$ & $0.890$ & $1.056$\\
   & $\sigma_z$ & $0.098$ & $0.123$ & $0.110$ & $0.145$ & $0.172$ & $0.068$ & $0.098$ & $0.087$ & $0.089$ & $0.083$\\
   & $W_z$  & $3.216$ & $2.797$ & $2.785$ & $2.535$ & $1.945$ & $3.453$ & $3.453$ & $3.311$ & $2.919$ & $3.062$\\
  \hline
  Y1 source & $\bar{N}_{\rm gal}$  & $30534.5$ & $30534.5$ & $30534.5$ & $30534.6$ & $30534.7$ & $7042.0$ & $7042.0$ & $7042.0$ & $7042.0$ & $7042.0$\\
   & $\langle z \rangle$ & $0.350$ & $0.482$ & $0.661$ & $0.813$ & $0.883$ & $0.310$ & $0.508$ & $0.678$ & $0.817$ & $0.845$\\
  \hline
  Y3 lens & $\bar{N}_{\rm gal}$ & $38988.1$ & $39330.9$ & $50523.5$ & $44645.0$ & $21978.0$ & $13904.9$ & $12239.9$ & $22208.5$ & $14705.3$ & $5822.9$\\
   & $\langle z \rangle$ & $0.373$ & $0.541$ & $0.728$ & $0.912$ & $1.067$ & $0.345$ & $0.550$ & $0.721$ & $0.895$ & $1.064$\\
   & $\sigma_z$ & $0.123$ & $0.135$ & $0.127$ & $0.166$ & $0.185$ & $0.069$ & $0.093$ & $0.084$ & $0.098$ & $0.103$\\
   & $W_z$ & $3.208$ & $3.123$ & $2.872$ & $2.514$ & $2.147$ & $3.602$ & $3.517$ & $3.361$ & $2.977$ & $2.952$\\
  \hline
  Y3 source & $\bar{N}_{\rm gal}$ & $48205.8$ & $48205.8$ & $48205.8$ & $48205.5$ & $48203.5$ & $15578.9$ & $15579.0$ & $15579.0$ & $15579.0$ & $15579.0$\\
   & $\langle z \rangle$ & $0.384$ & $0.557$ & $0.755$ & $0.954$ & $1.033$ & $0.332$ & $0.577$ & $0.726$ & $0.866$ & $0.973$\\
  \hline
  Y5 lens & $\bar{N}_{\rm gal}$  & $44162.9$ & $43177.3$ & $58004.0$ & $54826.6$ & $31379.0$ & $17002.9$ & $15398.3$ & $28564.3$ & $19690.5$ & $8850.2$\\
   & $\langle z \rangle$ & $0.383$ & $0.545$ & $0.736$ & $0.931$ & $1.085$ & $0.348$ & $0.549$ & $0.723$ & $0.898$ & $1.064$\\
   & $\sigma_z$ & $0.156$ & $0.151$ & $0.160$ & $0.198$ & $0.208$ & $0.071$ & $0.090$ & $0.083$ & $0.102$ & $0.111$\\
   & $W_z$ & $3.136$ & $3.267$ & $2.894$ & $2.394$ & $2.087$ & $3.920$ & $3.683$ & $3.423$ & $2.945$ & $2.982$\\
  \hline
  Y5 source & $\bar{N}_{\rm gal}$ & $59745.5$ & $59747.1$ & $59746.1$ & $59745.0$ & $59725.6$ & $20746.7$ & $20746.8$ & $20746.7$ & $20746.8$ & $20744.1$\\
   & $\langle z \rangle$ & $0.417$ & $0.594$ & $0.804$ & $1.028$ & $1.150$ & $0.345$ & $0.599$ & $0.751$ & $0.904$ & $1.019$\\
  \hline
 \end{tabular}
\end{table*}

\begin{table*}
 \caption{Same as Table~\ref{tab: mean metrics bpz}, but for FZBoost redshifts. All samples have ${\rm SNR}\geq 10$.}
 \label{tab: mean metrics fzb}
 \begin{tabular}{llccccc}
  \hline
  Sample & Metric & Bin 1 & Bin 2 & Bin 3 & Bin 4 & Bin 5\\
  \hline
  Y1 lens & $\bar{N}_{\rm gal}$ & $31054.7$ & $38964.9$ & $33717.0$ & $29871.8$ & $9104.6$\\
   & $\langle z \rangle$ & $0.327$ & $0.515$ & $0.706$ & $0.865$ & $1.083$\\
   & $\sigma_z$  & $0.119$ & $0.103$ & $0.114$ & $0.118$ & $0.108$\\
   & $W_z$ & $4.236$ & $3.098$ & $3.202$ & $3.242$ & $3.719$\\
  \hline
  Y1 source & $\bar{N}_{\rm gal}$ & $30534.6$ & $30534.5$ & $30534.5$ & $30534.5$ & $30534.7$\\
   & $\langle z \rangle$ & $0.291$ & $0.466$ & $0.623$ & $0.779$ & $1.030$\\
  \hline
  Y3 lens & $\bar{N}_{\rm gal}$ & $34498.5$ & $52662.6$ & $51564.7$ & $52858.0$ & $27561.2$\\
   & $\langle z \rangle$ & $0.320$ & $0.500$ & $0.702$ & $0.894$ & $1.093$\\
   & $\sigma_z$   & $0.183$ & $0.116$ & $0.118$ & $0.135$ & $0.144$\\
   & $W_z$  & $4.220$ & $3.263$ & $3.385$ & $3.061$ & $3.053$\\
  \hline
  Y3 source & $\bar{N}_{\rm gal}$ & $48205.9$ & $48205.8$ & $48205.8$ & $48205.9$ & $48203.1$\\
   & $\langle z \rangle$ & $0.328$ & $0.531$ & $0.718$ & $0.893$ & $1.213$\\
  \hline
  Y5 lens & $\bar{N}_{\rm gal}$ & $37286.1$ & $57641.9$ & $59009.5$ & $63214.3$ & $39344.7$\\
   & $\langle z \rangle$ & $0.339$ & $0.510$ & $0.709$ & $0.905$ & $1.104$\\
   & $\sigma_z$ & $0.194$ & $0.136$ & $0.122$ & $0.146$ & $0.142$\\
  & $W_z$ & $4.150$ & $3.332$ & $3.428$ & $3.037$ & $2.908$\\
  \hline
  Y5 source & $\bar{N}_{\rm gal}$  & $59747.0$ & $59747.1$ & $59747.0$ & $59747.2$ & $59721.0$\\
   & $\langle z \rangle$ & $0.350$ & $0.567$ & $0.764$ & $0.971$ & $1.342$\\
  \hline
 \end{tabular}
\end{table*}

Table~\ref{tab: stats on photoz} shows the summary statistics on photo-$z$ performance for BPZ and FZBoost at the 10\% shallowest $i$-band coadd depth (${\rm qtl}=0$) and deepest depth (${\rm qtl}=9$) for the 1-year, 3-years, and 5-years mock LSST data. The summary statistics are: median bias, standard deviation (STD), normalized Median Absolute Deviation (NMAD), and outlier fraction.
Tables~\ref{tab: mean metrics bpz} and \ref{tab: mean metrics fzb} show the mean values of the various metrics over the depth quantiles, given the gold cut adjusted for each year. The metrics include mean galaxy number $\bar{N}_{\rm gal}$ and mean redshift of the tomographic bin $\langle z \rangle$ for both lens and source samples, and additionally the width metrics $\sigma_z$ and $W_z$ for lens samples. In the BPZ case, we include an additional case where we select objects with ${\rm odds}\geq0.9$.

\section{Variation with other survey properties}
\label{sec: Variation with other survey properties}

In the main analysis, we investigated the trend of galaxy number and redshift distribution as a function of the $i$-band coadd depth. We considered the $i$-band depth to be most impactful because it is the detection band, and fluxes in all other bands are measured with forced photometry based on the $i$-band detection. However, other survey properties can also be important. For example, $u$-band is important for the quality of photo-$z$ estimation, so extreme variation in the $u$-band depth could cause additional scatter. The effective seeing could be another important factor, which directly impact the noise-to-signal for extended objects. 
We investigate the variation of galaxy number density and photo-$z$ properties with these other survey properties in this section.

\begin{table*}
\scriptsize
    \centering
   \caption{The mean and standard deviation of all other survey condition maps used for degradation in this work in 10 quantiles of $i$-band depth, as shown in Table~\ref{tab: i-band quantiles}. This particular example shows the case for year 3, but year 1 and year 5 follow a similar trend. There is a strong correlation between these survey conditions and the $i$-band depth.}
    \label{tab: sub-prop}
   \begin{tabular}{llcccccccccc} 
       \hline
        Prop. & band & \multicolumn{10}{|c|}{$i$-band $m_5^{\rm ex}$ quantiles} \\
        \hline
        & & 0 & 1 & 2 & 3 & 4 & 5 & 6& 7& 8& 9\\ 
        \hline 
        $m_5^{\rm ex}$ & $u$ &  $24.00\pm0.24$ & $24.19\pm0.20$ & $24.30\pm0.19$ & $24.40\pm0.20$ & $24.49\pm0.19$ & $24.57\pm0.18$ & $24.63\pm0.16$ & $24.68\pm0.15$ & $24.76\pm0.15$ & $24.91\pm0.14$ \\
        & $g$ &  $25.66\pm0.18$ & $25.82\pm0.16$ & $25.91\pm0.16$ & $25.99\pm0.15$ & $26.06\pm0.15$ & $26.13\pm0.13$ & $26.18\pm0.12$ & $26.22\pm0.12$ & $26.31\pm0.14$ & $26.49\pm0.13$ \\
        & $r$ &  $25.88\pm0.13$ & $26.05\pm0.09$ & $26.13\pm0.09$ & $26.20\pm0.08$ & $26.26\pm0.08$ & $26.31\pm0.07$ & $26.36\pm0.07$ & $26.41\pm0.07$ & $26.49\pm0.08$ & $26.62\pm0.07$\\
        & $z$ &  $24.80\pm0.13$ & $24.95\pm0.09$ & $25.02\pm0.09$ & $25.08\pm0.08$ & $25.13\pm0.08$ & $25.18\pm0.07$ & $25.22\pm0.07$ & $25.27\pm0.07$ & $25.35\pm0.07$ & $25.47\pm0.07$ \\
        & $y$ &  $23.89\pm0.15$ & $24.03\pm0.11$ & $24.10\pm0.11$ & $24.16\pm0.11$ & $24.21\pm0.10$ & $24.25\pm0.09$ & $24.28\pm0.09$ & $24.32\pm0.10$ & $24.40\pm0.09$ & $24.50\pm0.08$ \\
        \hline
        $\theta_{\rm FWHM}^{\rm eff}$ & $u$ & $1.25\pm0.11$ & $1.20\pm0.12$ & $1.17\pm0.12$ & $1.15\pm0.12$ & $1.13\pm0.12$ & $1.12\pm0.13$ & $1.10\pm0.12$ & $1.10\pm0.12$ & $1.12\pm0.13$ & $1.05\pm0.10$\\
        & $g$ &  $1.07\pm0.12$ & $1.02\pm0.13$ & $1.00\pm0.14$ & $1.00\pm0.14$ & $0.99\pm0.13$ & $0.97\pm0.13$ & $0.96\pm0.12$ & $0.96\pm0.12$ & $0.96\pm0.11$ & $0.89\pm0.09$ \\
        & $r$ &  $1.04\pm0.09$ & $0.98\pm0.09$ & $0.96\pm0.09$ & $0.94\pm0.09$ & $0.92\pm0.08$ & $0.91\pm0.07$ & $0.90\pm0.07$ & $0.90\pm0.06$ & $0.89\pm0.06$ & $0.85\pm0.05$ \\
        & $i$ &  $1.01\pm0.08$ & $0.95\pm0.07$ & $0.94\pm0.08$ & $0.93\pm0.08$ & $0.91\pm0.07$ & $0.90\pm0.07$ & $0.88\pm0.06$ & $0.87\pm0.06$ & $0.87\pm0.06$ & $0.83\pm0.04$ \\
        & $z$ &  $1.00\pm0.07$ & $0.97\pm0.07$ & $0.96\pm0.07$ & $0.95\pm0.07$ & $0.94\pm0.07$ & $0.93\pm0.07$ & $0.92\pm0.07$ & $0.91\pm0.07$ & $0.91\pm0.07$ & $0.86\pm0.05$ \\
        & $y$ &  $1.02\pm0.07$ & $0.97\pm0.06$ & $0.96\pm0.06$ & $0.95\pm0.06$ & $0.94\pm0.06$ & $0.93\pm0.06$ & $0.93\pm0.06$ & $0.92\pm0.06$ & $0.91\pm0.05$ & $0.88\pm0.04$ \\
        \hline
   \end{tabular}
\end{table*}

Table~\ref{tab: sub-prop} summarizes the mean and standard deviation of the coadd depth in the other five LSST bands and the median effective seeing for Y3 survey properties from Rubin OpSim baseline v3.3. The other years show a similar trend, although Y1 has a larger scatter. We see that there is a strong correlation between the $i$-band depth and these other survey properties. On average, a deeper $i$-band quantile also contains deeper coadd depth in all other five bands, as well as a smaller median effective seeing, with more scatter in the latter.  


\begin{figure*}
    \includegraphics[width=\textwidth]{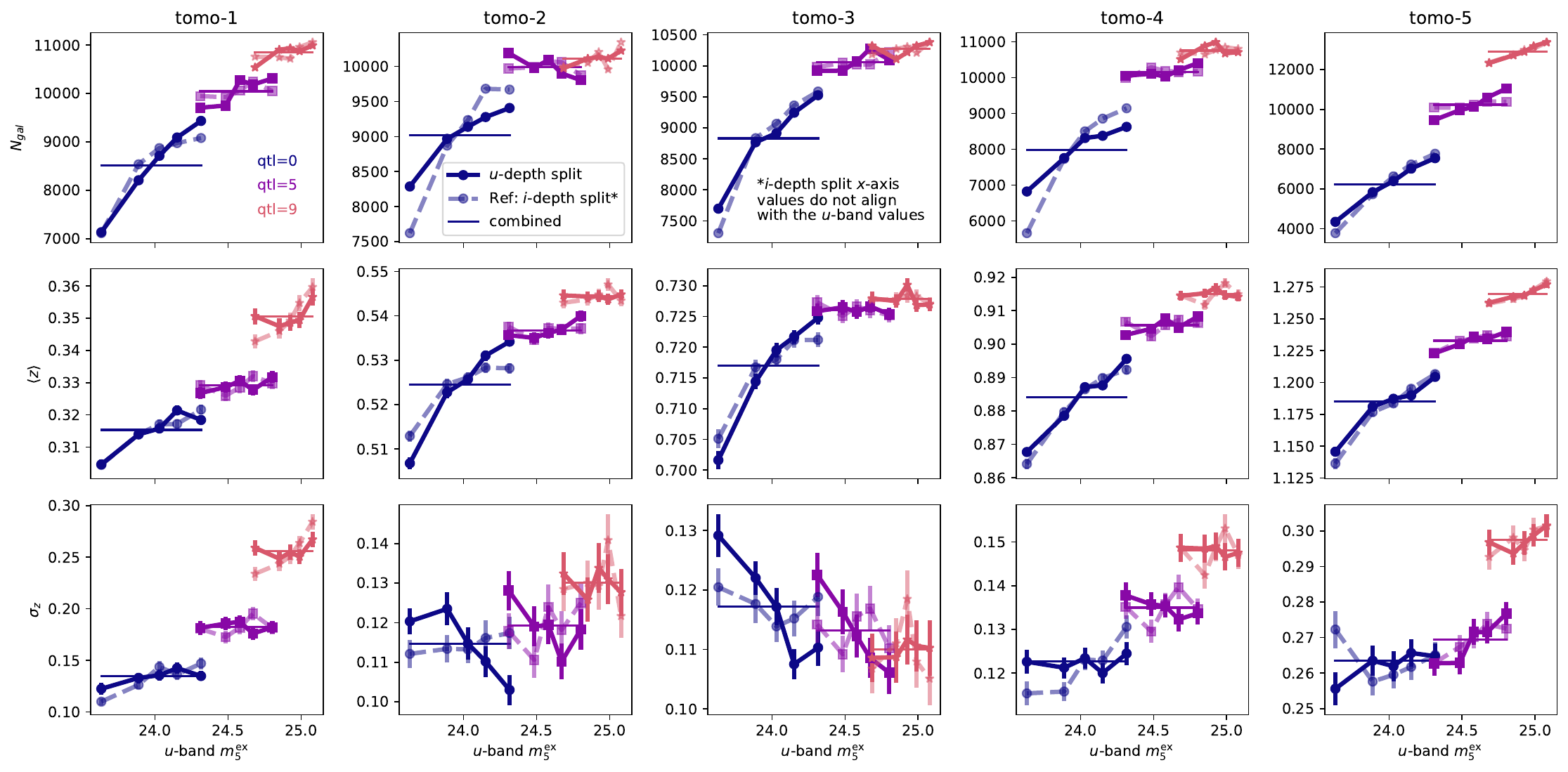}
    \caption{The variation of the number of objects, $N_{\rm gal}$, the mean redshift, $\langle z \rangle$, and the standard deviation, $\sigma_z$, of the Y3 source tomographic bins, as a function of the extinction-corrected $u$-band coadd depth, $m_5^{\rm ex}$. The $u$-band depth bins are determined by 5 quantiles sub-dividing each of the $i$-band quantiles used in the main analysis. Examples shown here are for the $i$-band quantiles 0 (dark blue), 5 (purple), and 9 (pink). The tomographic bins are split by FZBoost photo-$z$. The horizontal lines indicate the combined values as shown in Fig.~\ref{fig: ngal-baselinev2-y1-y5-goldcut-snr-10} - \ref{fig: delta-wz-baselinev2-y1-y5-goldcut-snr-10}. The faint, dashed lines indicate a reference case where the sub-division is done for 5 quantiles in the $i$-band depth. Notice that the $i$-band split case is only over-plotted here to provide a visual comparison of the level of fluctuations, but its actual $x$-axis values do \textit{not} align with those on the figure, which are for the $u$-band depth.}
    \label{fig: subprop-ExgalM5_u-y3-fzb-source}
\end{figure*}

\begin{figure*}
    \includegraphics[width=\textwidth]{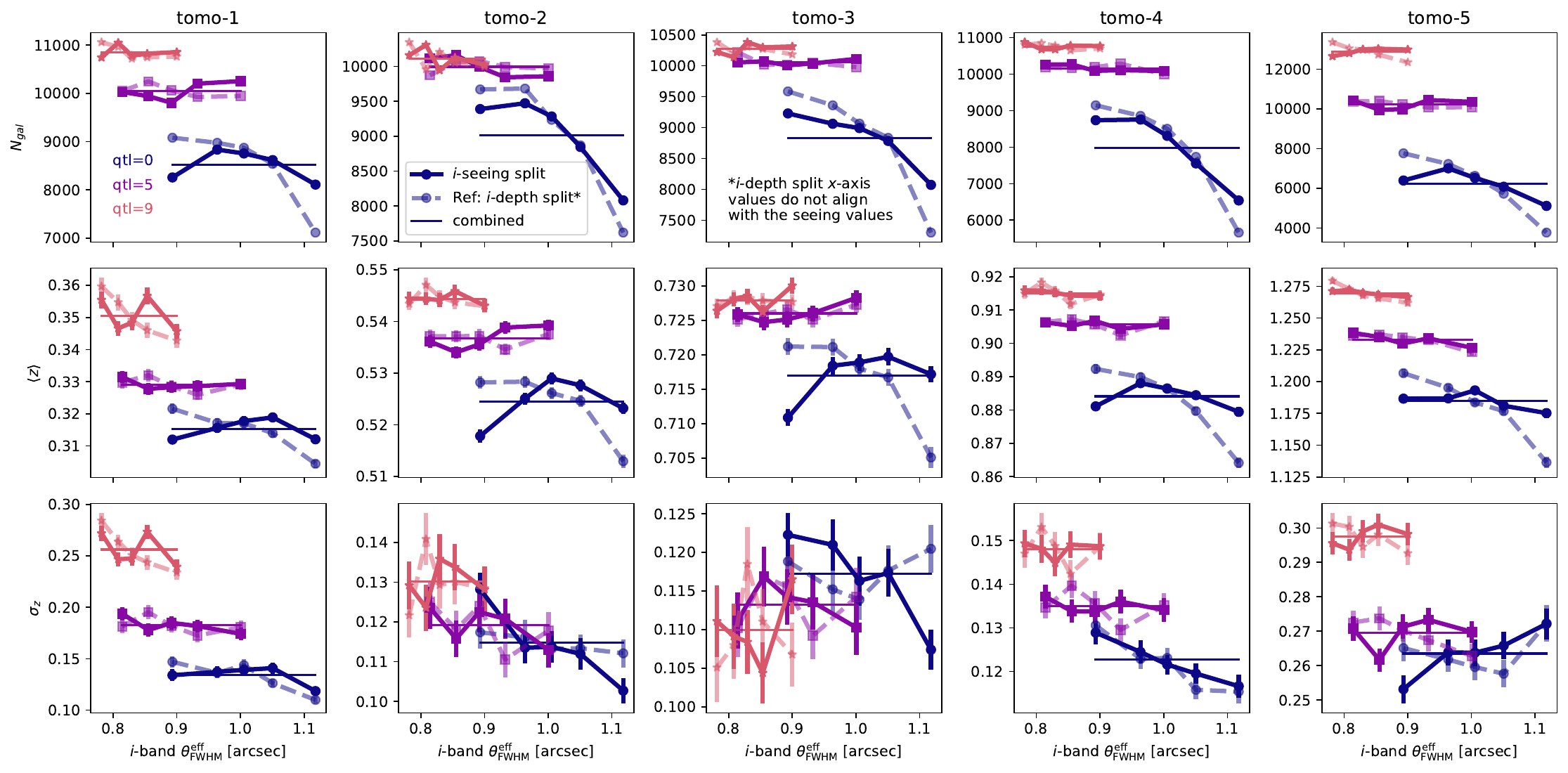}
    \caption{Same as Fig.~\ref{fig: subprop-ExgalM5_u-y3-fzb-source}, but for sub-division in the $i$-band median effective seeing, $\theta_{\rm FWHM}^{\rm eff}$. Notice here that the reference $i$-band depth sub-quantiles, as indicated by the faint dashed lines, are flipped, i.e., the first quantile in $i$-band depth is over-plotted on top of the last quantile in the $i$-band seeing. This is because, on average, a deeper coadd depth corresponds to a smaller seeing angle.}
    \label{fig: subprop-Median_seeingFwhmEff_i-y3-fzb-source}
\end{figure*}

To check the dependences of other survey properties, we sub-divide each of the $i$-band deciles into five sub-quantiles of another survey property (such as depth in another band), and check the variation of the metrics, i.e. number of objects $N_{\rm gal}$, mean redshift $\langle z \rangle$, and width of the redshift bin $\sigma_z$, with these properties. As a reference, we also compute and compare the variation with sub-quantiles of the $i$-band depth itself. 
In this section, we show two representative examples for source tomographic bins determined by FZboost photo-$z$: the sub-quantiles in coadd $u$-band depth and the $i$-band seeing, for the fainest, median, and deepest $i$-band deciles: ${\rm qtl}=0, 5, 9$. 
In the results presented here, we over-plot the variation from the $i$-band depth sub-quantiles (as faint, dashed lines) on top of that from the other survey properties (as solid lines), for visual comparison. That is, one can read off the level of fluctuation from the deepest and shallowest $u$-band depth sub-bin, for example, and compare it with that from the deepest and shallowest $i$-band depth sub-bin.
It should be noted, however, that these reference $i$-band split cases have a \textit{different} actual $x$-axis values from those shown in the plots.

The results are shown in Fig.~\ref{fig: subprop-ExgalM5_u-y3-fzb-source} and~\ref{fig: subprop-Median_seeingFwhmEff_i-y3-fzb-source} respectively. We see that in general, these trends are consistent with the $i$-band depth fluctuation for all three metrics: the deeper (smaller) the depth (seeing), the more objects included in the sample, the higher the mean redshift of the tomographic bin, and the larger $\sigma_z$. 
Also, ${\rm qtl}=0$ has a significantly larger variation compared to ${\rm qtl}=9$ in most cases. 
Compared to the trends in the $i$-band depth sub-bins, we see that the $N_{\rm gal}$ variations are always less strong for other properties. This is understood as selections are primarily taken in $i$-band. The $\langle z \rangle$ variations for the $u$-band tightly follows the $i$-band, although the first bin can have slightly larger fluctuations. For seeing, on the other hand, the trend is quite different for ${\rm qtl}=0$, where the smallest seeing does not always correspond to a higher mean redshift. This could happen because the seeing is not as well correlated with depth - there are more scatter in the coadd depth and seeing at the faint end. Finally, the variation in $\sigma_z$ seems to be relatively minor in most cases.

From these exercises, we see that within each $i$-band decile, the number of objects and $p(z)$ properties can still change significantly with other survey properties such as $u$-band depth and $i$-band seeing. Meanwhile, given that these quantities are also quite tightly correlated, we expect that a lot of these variations are also due to the co-variation of the $i$-band depth. Hence, our main analysis, by splitting into the $i$-band quantiles, should capture the level of variations of the metrics. However, if one wishes to apply this method in e.g. forward modeling, then co-variation of all bands need to be taken into account.

\section{MagLim-like lens sample}
\label{sec: maglim}

\begin{figure}
    \centering
    \includegraphics[width=\linewidth]{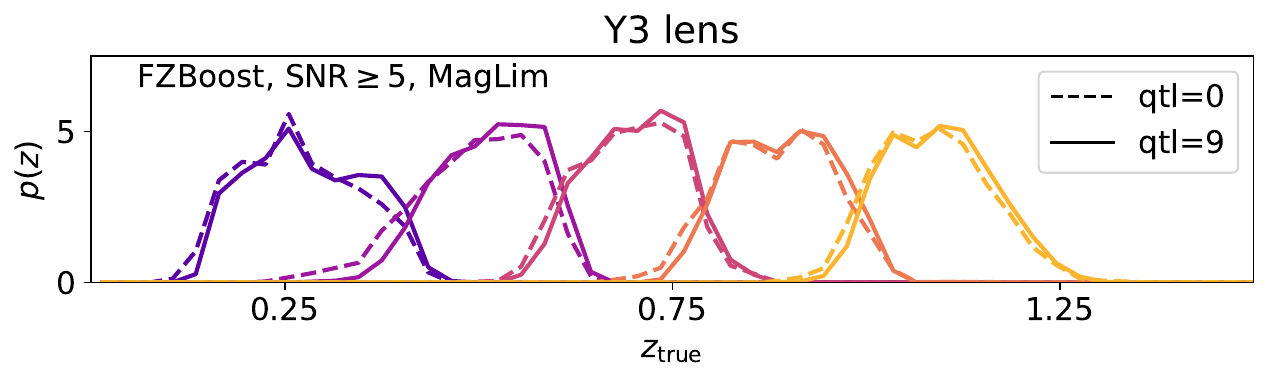}
    \caption{True redshift distribution of the LSST Y3 MagLim lens sample, split in tomographic bins as defined in the DESC SRD. The MagLim cuts and the tomographic edhes are determined using the mode of FZBoost redshifts. The sample has also been applied a cut with ${\rm SNR}\geq5$. The dashed lines show samples degraded using the shallowest 10\% pixels in $i$-band coadd depth (${\rm qtl}=0$), and the solid lines show those from the deepest 10\% (${\rm qtl}=9$).}
    \label{fig:nz maglim}
\end{figure}

\begin{figure*}
     \centering
      \hfill
     \begin{subfigure}[b]{0.24\textwidth}
         \centering
         \includegraphics[width=\textwidth]{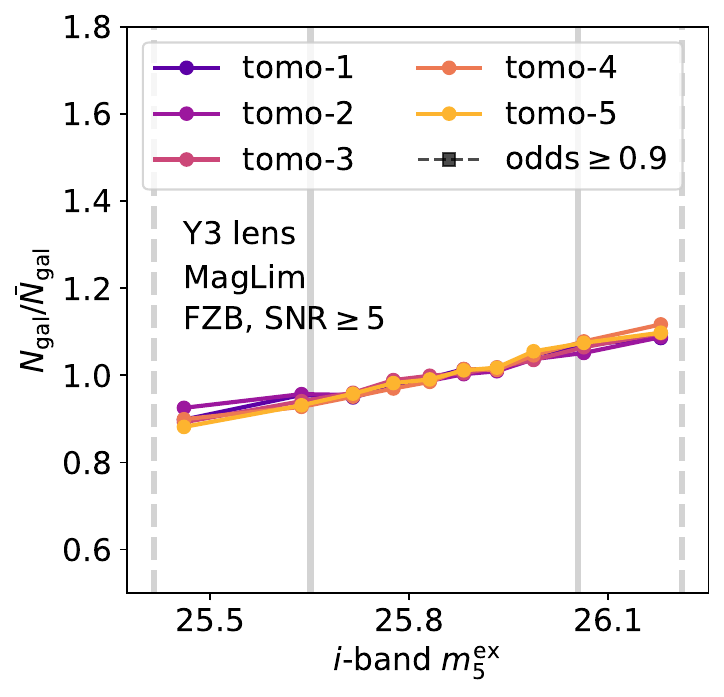}
         \subcaption[]{$N_{\rm gal}$ vs depth.}
     \end{subfigure}
     \begin{subfigure}[b]{0.25\textwidth}
         \centering
         \includegraphics[width=\textwidth]{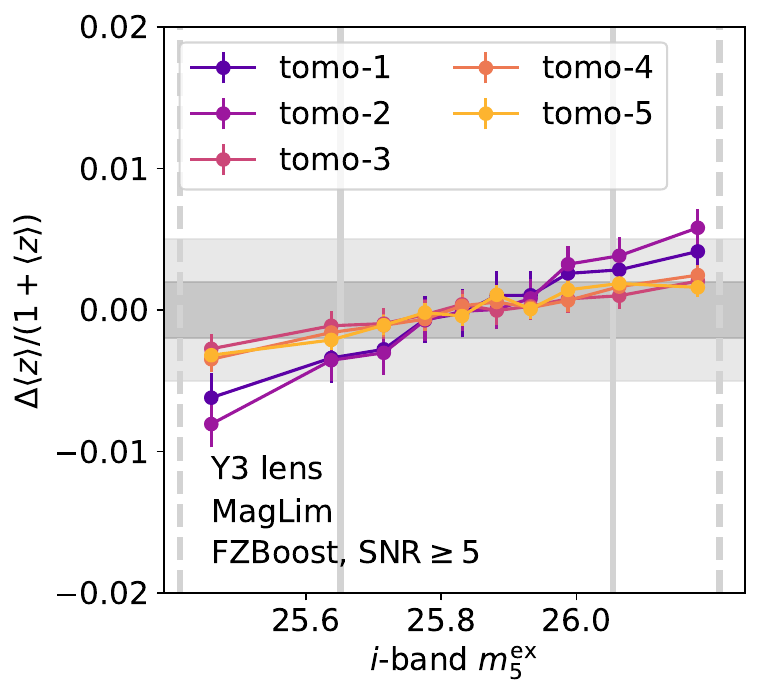}
         \subcaption[]{Mean redshifts vs depth.}
     \end{subfigure}
     \begin{subfigure}[b]{0.24\textwidth}
         \centering
         \includegraphics[width=\textwidth]{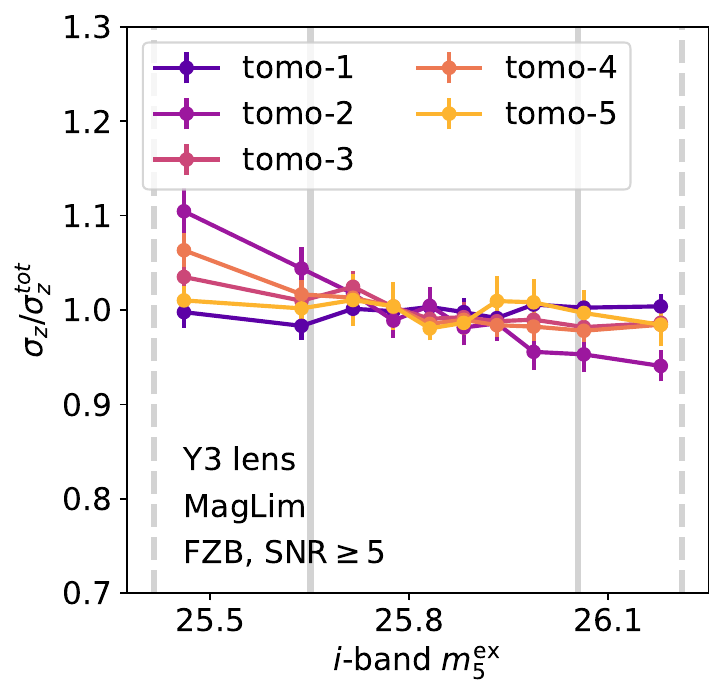}
         \subcaption[]{Width $\sigma_z$ vs depth.}
     \end{subfigure}
     \begin{subfigure}[b]{0.24\textwidth}
         \centering
         \includegraphics[width=\textwidth]{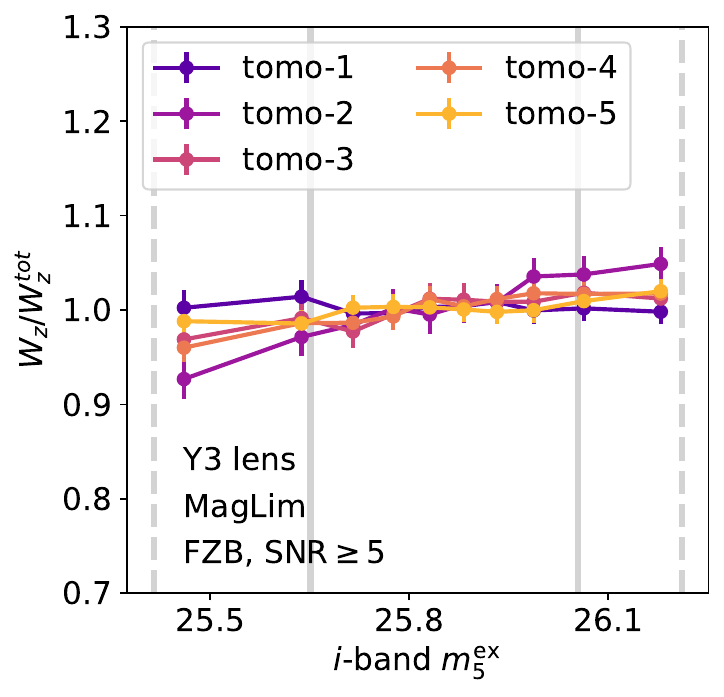}
         \subcaption[]{LSS-relevant width $W_z$ vs depth.}
     \end{subfigure}
        \caption{Metrics for impact of variable depth for the LSST Y3 MagLim lens sample, split in five tomographic bins.
        (a). The fractional change in number of galaxies, $N_{\rm gal}/\bar{N}_{\rm gal}$ in tomographic bins as a function of the $i$-band extinction-corrected coadd depth, $m_5^{\rm ex}$; (b). The scaled shifts in mean redshift, $\Delta \langle z \rangle/ (1+\langle z \rangle)$ as a function of $m_5^{\rm ex}$; (c). The fractional change in second moment of the redshift distribution, $\sigma_z/\sigma_z^{\rm tot}$, as a function of $m_5^{\rm ex}$; (d). The fractional change in the LSS-related kernel, $W_z/W_z^{\rm tot}$, as a function of $m_5^{\rm ex}$.
        The MagLim cuts and the tomographic bins edges are determined using the mode of FZBoost redshifts, and the sample has an $i$-band ${\rm SNR}\geq5$. The vertical solid and dashed lines marks the $1\sigma$ and $2\sigma$ regions of the depth distribution.}
        \label{fig: depth maglim}
\end{figure*}

\begin{figure*}
   \centering
   \includegraphics[width=0.9\textwidth]{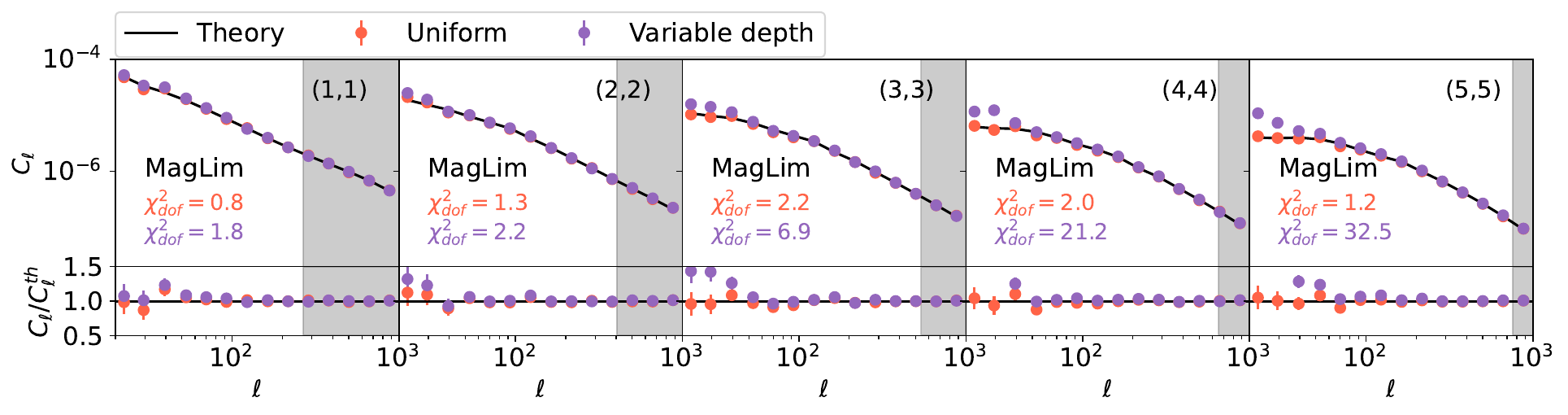}
    \caption{Similar to Fig.~\ref{fig: clgg-y3-fzb}, but for the LSST Y3 MagLim lens sample.}
    \label{fig: clgg-y3-fzb maglim}
\end{figure*}

In this section, we explore the impact of variable depth on a lens sample selected with the DES Y3 MagLim cuts \citep{2022PhRvD.106j3530P}. Because this sample has a brighter cut, we relax the $i$-band signal-to-noise limit to ${\rm SNR\geq 5}$. The sample is selected with 
\begin{equation}
  17<i<4 z_{\rm phot}+18,   
\end{equation}
where we use the FZBoost mode redshift as $z_{\rm phot}$. This cut reduces the number of lens sample significantly compared to our fiducial case, resulting in a total sample size of $3.67\%$ of the baseline (Gold cut) lens sample. The true redshift distribution of each tomographic bin is shown in Fig.~\ref{fig:nz maglim}, where the dashed lines show those from the shallowest quantile, and the solid lines show those from the deepest. Notice that the distribution is less smooth due to the sparsity of the sample. Overall, thanks to the bright cut, the redshift distribution for each bin has a smaller tail compared to the baseline case, especially for the highest redshift bin. 

Figure~\ref{fig: depth maglim} shows the metrics for the variable depth, namely, the galaxy number, mean redshift, and width of the tomographic bin, as a function of the $i$-band depth. The panels (a) - (d) has the same style as, and should be compared to Fig.~\ref{fig: ngal-baselinev2-y1-y5-goldcut-snr-10} - \ref{fig: delta-wz-baselinev2-y1-y5-goldcut-snr-10}. Again, we see a significantly milder, but visible, trend of these metrics with depth, owing to the bright magnitude cut. This shows that the variable depth effect can be greatly reduced, but not completed removed, by introducing a bright cut at the cost of sample size. 

Figure~\ref{fig: clgg-y3-fzb maglim} shows the effect propagated to the galaxy clustering two-point data vector, $C_{\ell}^{gg}$. We followed the same procedure as in Section~\ref{sec: 3x2pt}, and set the number density in each bin to be $0.135, 0.117, 0.156, 0.219, 0.267$ arcmin$^{-2}$ to account for the reduction in the overall number density compared to the fiducial case. The impact of variable depth on $C_{\ell}^{gg}$ is also significantly reduced, especially for $(4,4)$ and $(5,5)$. However, the impact is not negligible still at $\ell<100$. 


\bsp	
\label{lastpage}
\end{document}